\documentclass{aa}  

\usepackage{graphicx}
\usepackage{txfonts}
\usepackage[usenames,dvipsnames]{xcolor}
\usepackage{gensymb}
\usepackage{url}
\usepackage{multirow}
\usepackage[normalem]{ulem}

\begin{document}

   \title{Exoplanet weather and climate regimes with clouds and thermal ionospheres}

   \subtitle{A model grid study in support of large-scale observational campaigns}

   \author{Christiane Helling \inst{1,2,3}
          \and
          Dominic Samra \inst{1,2}
          \and 
          David Lewis \inst{1,2}
          \and
          Robb Calder \inst{2}
          \and
          Georgina Hirst \inst{2}
          \and 
          Peter Woitke \inst{1,2}
          \and
          Robin Baeyens \inst{4}
          \and
          Ludmila Carone \inst{2}
           \and
          Oliver Herbort \inst{1,2,5}
          \and 
          Katy L. Chubb \inst{2}}

   \institute{
      Space Research Institute, Austrian Academy of Sciences, Schmiedlstrasse 6, A-8042 Graz, Austria\\
   \email{Christiane.Helling@oeaw.ac.at}
   \and 
   Centre for Exoplanet Science, School of Physics \& Astronomy, University of St Andrews, North Haugh, St Andrews, KY169SS, UK
         \and
         Institute for Theoretical Physics and Computational Physics, Graz University of Technology, Petersgasse 16
8010 Graz
         \and
             Institute of Astronomy, KU Leuven, Celestijnenlaan 200D, 3001, Leuven, Belgium
          }

   \date{Received September 15, 1996; accepted March 16, 1997}

  \abstract
   {Gaseous exoplanets are the targets that enable us to explore fundamentally our understanding of planetary physics and chemistry.   With observational efforts moving from the discovery into the characterisation mode, systematic campaigns that cover large ranges of global stellar and planetary parameters will be needed to disentangle the diversity of exoplanets and their atmospheres that all are affected by their formation and evolutionary paths. Ideally, the spectral range includes the high-energy (ionisation) and the low-energy (phase-transitions) processes as they carry complementary information of the same object.}
   {We aim to uncover cloud formation  trends and globally changing chemical regimes into which gas-giant exoplanets may fall due to the host star's effect on the thermodynamic structure of their atmospheres. We aim to examine the emergence of an ionosphere as indicator for potentially asymmetric magnetic field effects  on these atmospheres. We aim to provide input for exoplanet missions like JWST, PLATO, and Ariel, as well as potential UV missions ARAGO, PolStar or POLLUX on LUVOIR.}
  {Pre-calculated 3D GCMs for M, K, G, F host stars  are the input for our kinetic cloud  model for the formation of nucleation seeds, the growth to macroscopic cloud particles and their evaporation, gravitational settling, element conservation and gas chemistry.}
   {Gaseous exoplanets fall broadly into three classes: i) cool planets  with homogeneous cloud coverage, ii) intermediate temperature planets with asymmetric dayside cloud coverage, and iii) ultra-hot planets without clouds on the dayside. 
   {   In class ii),} the dayside cloud patterns are shaped by the wind flow and irradiation. Surface gravity and planetary rotation have little  effect. For a given effective temperature, planets around K dwarfs are rotating faster compared to G dwarfs leading to larger cloud inhomogeneities in the fast rotating case. Extended atmosphere profiles suggest the formation of mineral haze in form of metal-oxide clusters (e.g. (TiO$_2$)$_{\rm N})$.
    }
   {The dayside cloud coverage is the tell-tale sign for the different planetary regimes and their resulting weather and climate appearance. Class (i) is representative of planets with a very homogeneous cloud particle size and material compositions  across the globe (e.g., HATS-6b, NGTS-1b), classes (ii, e.g., WASP-43b, HD\,209458b) and (iii, e.g., WASP-121b, WP0137b) have a large day/night divergence of the cloud properties. The C/O ratio is, hence, homogeneously affected in class (i), but asymmetrically in class (ii) and (iii). The atmospheres of class (i) and (ii) planets are little affected by thermal ionisation, but class (iii) planets exhibit a deep ionosphere on the dayside. Magnetic coupling will therefore affect different planets differently 
   and will be more efficient on the more extended, cloud-free dayside. How the ionosphere connects atmospheric mass loss  at the top of the atmosphere with deep atmospheric layers need to be investigated to coherently interpret high resolution observations of ultra-hot planets.  }


   \maketitle

\begin{figure*}[ht]
    \includegraphics[width=20pc]{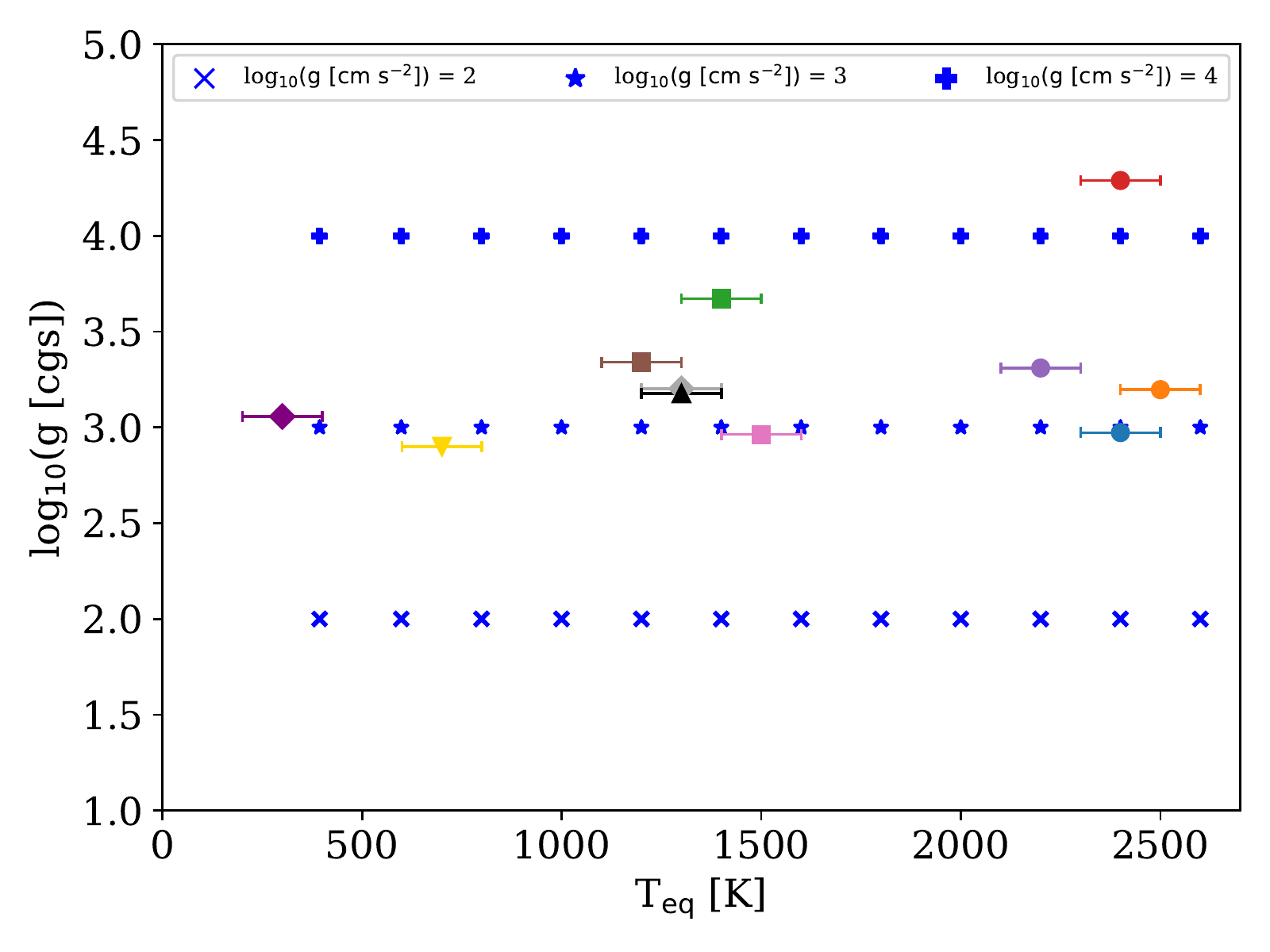}
    \includegraphics[width=20pc]{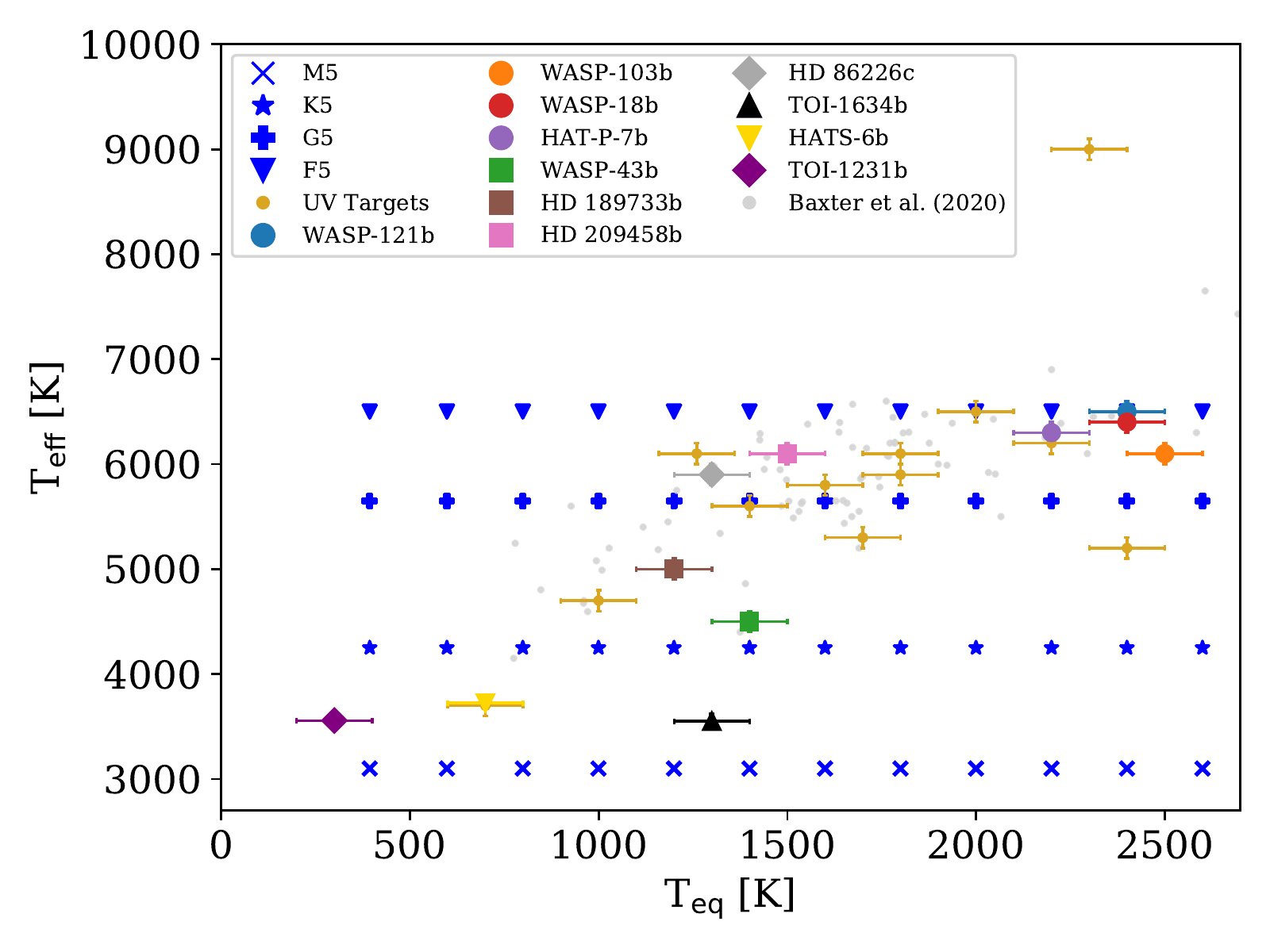}\\
    \includegraphics[width=20pc]{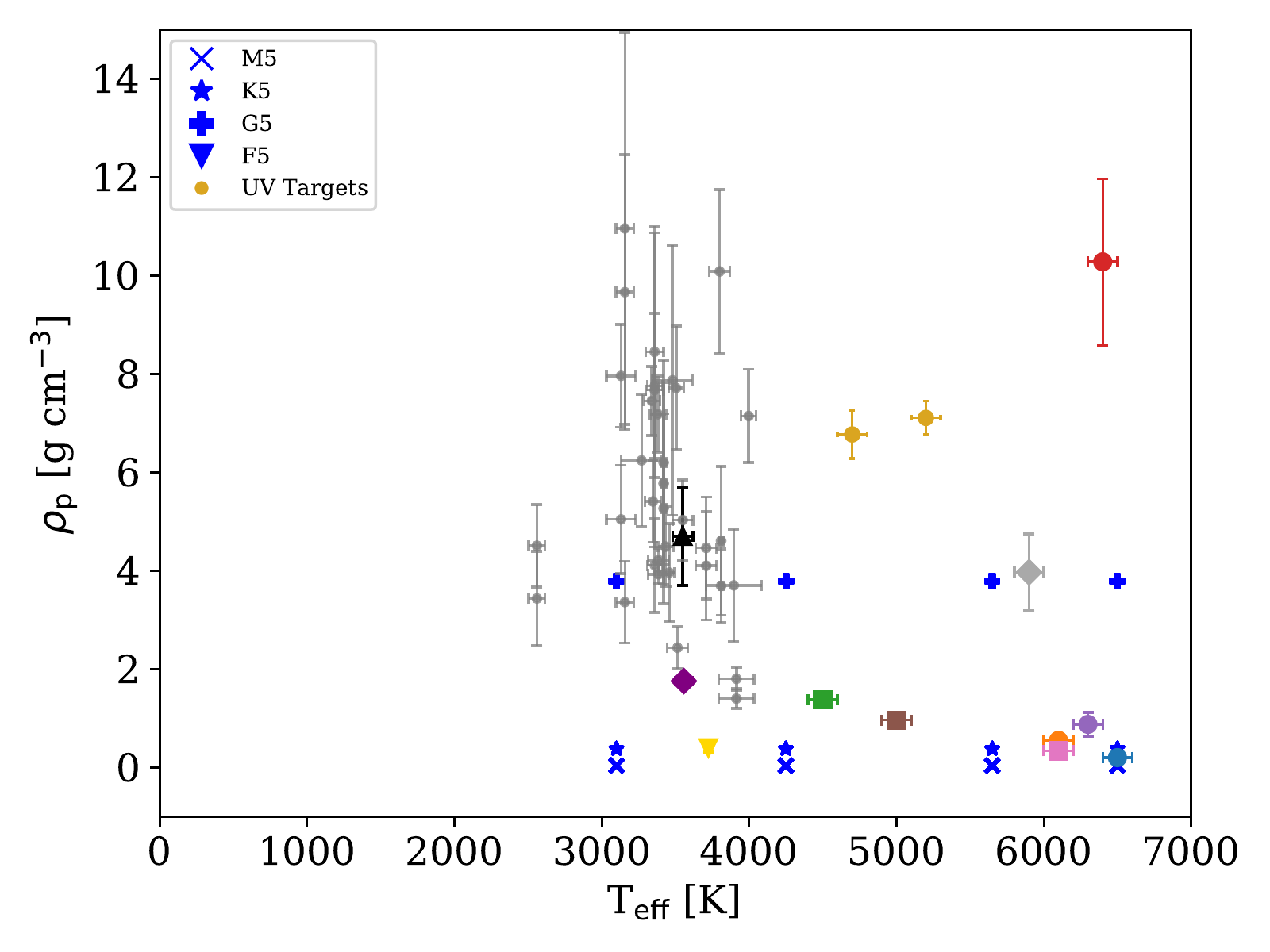}
    \includegraphics[width=20pc]{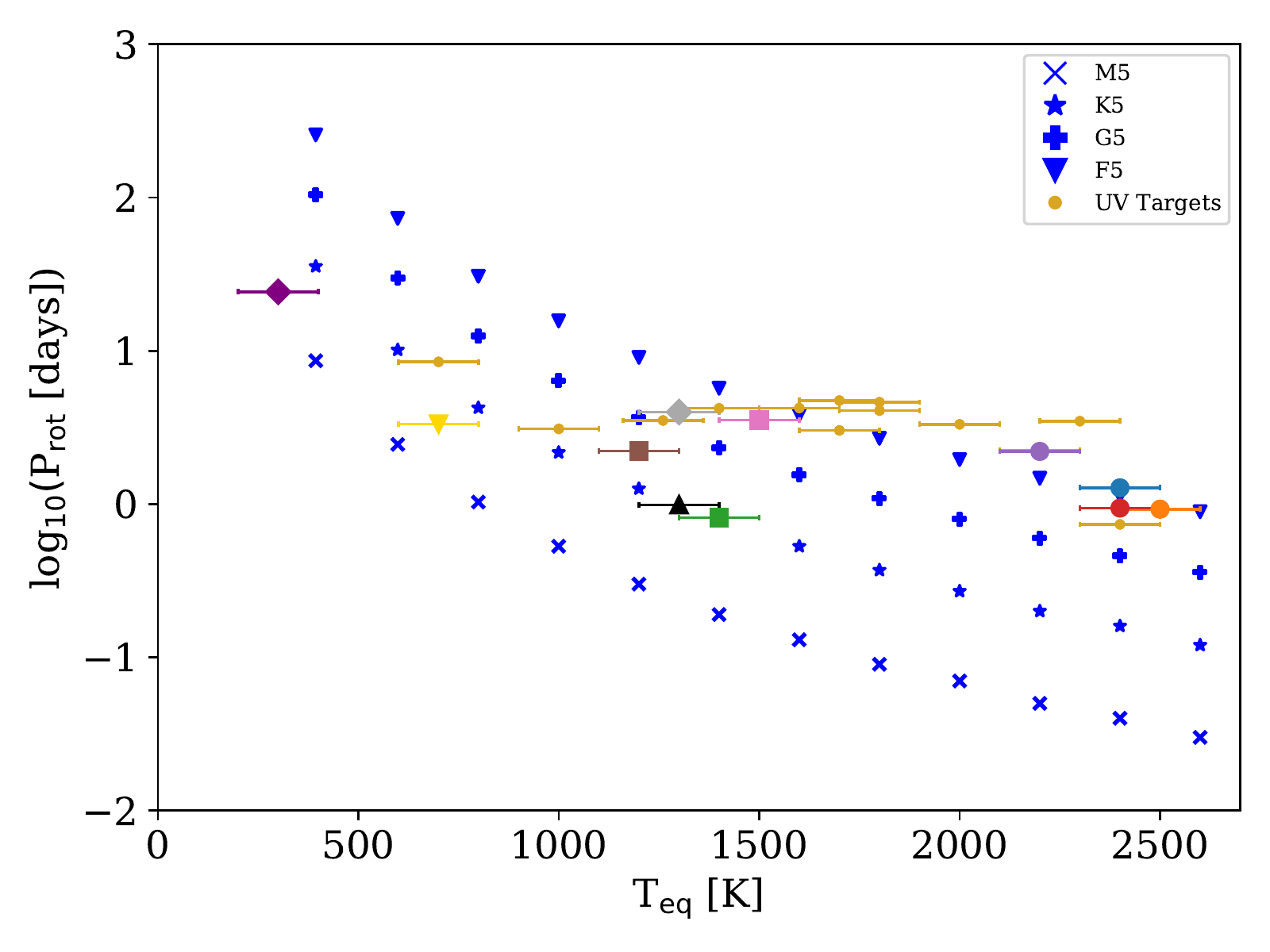}
    
    \caption{System and global parameters for selected planets in comparison to the 3D GCM model grid parameter space from \citet{2021MNRAS.tmp.1277B}. All model planets are tidally locked such that $P_{\rm rot}$=$P_{\rm orb}$. Selected planets include hot (WASP-43b, HD 189733b, HD 209458b) and ultra-hot (HAT-P-7b, WASP-18b, WASP-103b, WASP-121b) Jupiters, sub-Neptunes (HD 86226c, TOI-1634b), a Neptune-like planet (TOI-1231b), and a warm Saturn (HATS-6b).
    We also include the sample of hot and the ultra-hot Jupiters from Table 1 in \citet{2020A&A...639A..36B} in the T$_{\rm eff}$ vs T$_{\rm eq}$ plot (top right) as smaller light grey points 
     and a sample of small M-Dwarf orbiting planets from exoplanet.eu
    in the $\rho_{\rm P}$ vs T$_{\rm eff}$ plot (bottom left) as grey points with error bars.
    Potential UV targets (dark golden dots with error bars) are included for comparison (Tables~\ref{t:UV1},~\ref{t:UV2}).
    \textbf{Notes:} The equilibrium temperature given in the literature have been rounded to the next 100\,K.\\    \textbf{References:} \textit{WASP-103 b}: \citet{Gillon14}; \textit{WASP-121 b}: \citet{10.1093/mnras/stw522}; \textit{WASP-18 b}: \citet{Hellier2009}, \citet{Sheppard17};  \textit{WASP-43 b}: \citet{2012A&A...542A...4G}; \textit{HAT-P-7 b}: \citet{vaneylen13}, \citet{pal09}; \textit{HD 209458 b}: \citet{2008ApJ...677.1324T}; \textit{HD 189733 b}: \citet{2008ApJ...677.1324T}; \textit{HD 86226 c}: \citet{teske2020tess}; \textit{TOI-1634 b}: \citet{2021arXiv210312790C}; \textit{HATS-6 b}: \citet{2015AJ....149..166H}; \textit{TOI-1231b}: \citet{2021arXiv210508077B}; 
    The rotational periods for the GCM models are taken from \citet{2021MNRAS.tmp.1277B}.
    }
   \label{fig:global_properties}
\end{figure*}

\section{Introduction}
\label{section:Intro}

\begin{figure*}
   \includegraphics[width=0.965\textwidth]
    {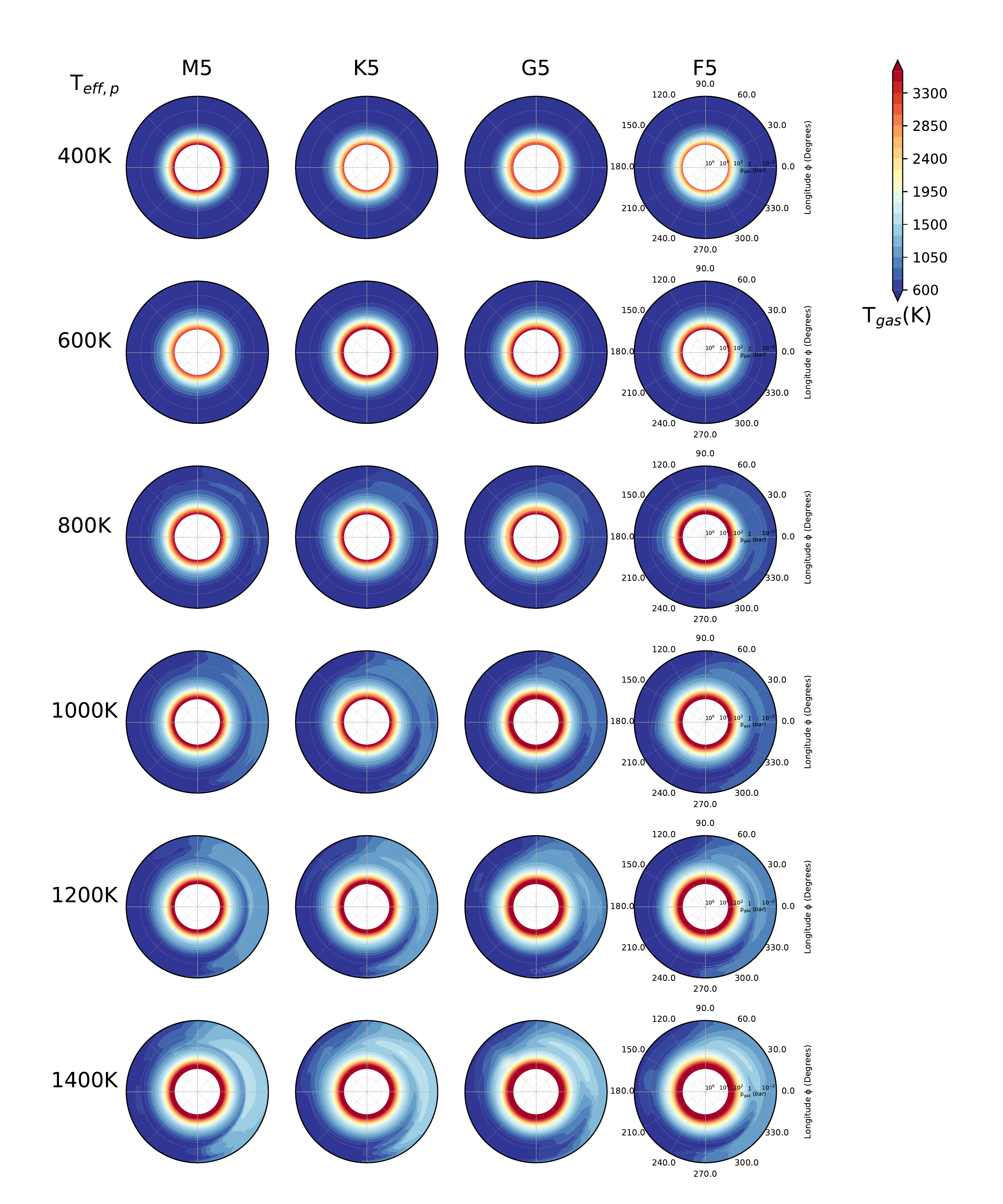}
    \caption{Equatorial temperature maps: 2D cut through the equatorial plane ($\theta = 0\degree$) showing the (T$_{\rm gas}$, p$_{\rm gas}$) distribution for each 3D GCM model (\citealt{2021MNRAS.tmp.1277B}). The planetary effective temperature ranges is T$_{\rm eff} = 400\,\ldots\, 1400$ K shown for all stellar types and $\log_{10}(g\,{\rm [cgs]})$ = 3.}
   \label{fig:global_slice_plots1}
\end{figure*}

\begin{figure*}
   \includegraphics[width=0.965\textwidth]
    {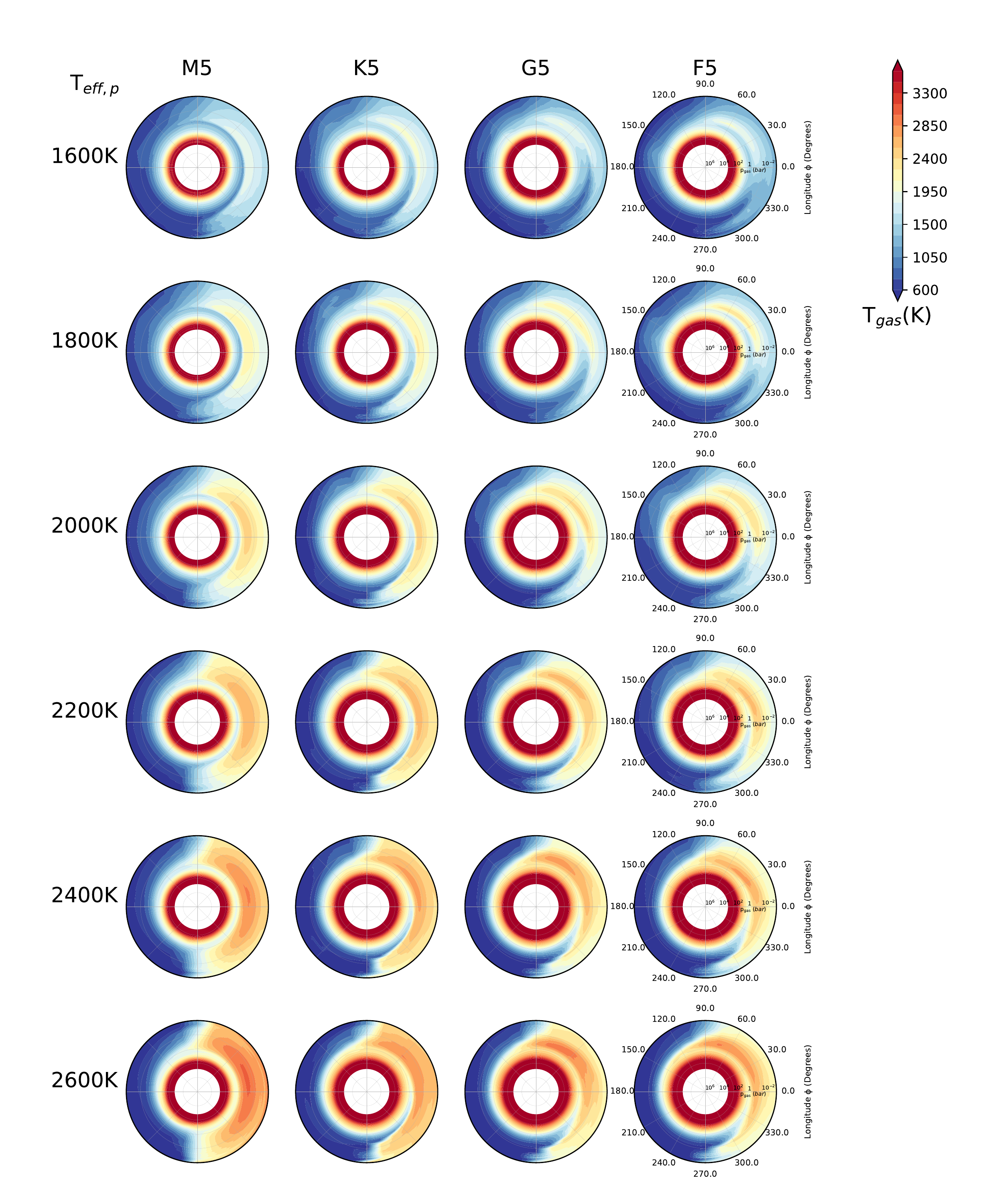}
    \caption{Equatorial temperature maps: 2D cut through the equatorial plane ($\theta = 0\degree$) showing the (T$_{\rm gas}$, p$_{\rm gas}$) distribution for each 3D GCM model (\citealt{2021MNRAS.tmp.1277B}). The planetary effective temperature ranges from T$_{\rm eff} = 1600\,\ldots\, 2600$ K  for all stellar types and $\log_{10}(g\,{\rm [cgs]})$ = 3.}
   \label{fig:global_slice_plots2}
\end{figure*} 

The diversity of exoplanets known so far\footnote{\url{http://exoplanet.eu}} calls for concerted modelling efforts in order to optimally access the information content of the observational data. Mission concepts for deciphering rocky exoplanets and, in the ideal case, an exo-Earth (\citealt{2020arXiv200106683G,2021arXiv210404824T,quanz2021atmospheric}) are being developed and a fleet of exoplanet missions is under development at ESA\footnote{\url{https://sci.esa.int/web/exoplanets}, \url{https://exoplanets.nasa.gov/discovery/exoplanet-catalog/}} and at NASA\footnote{\url{https://exoplanets.nasa.gov/exep/about/missions-instruments/}}.  The planets that can be studied with today's observational facilities in considerable detail (i.e. atmosphere characterisation)  are planets  that orbit close to their host star, like most of the known gas-giants outside the solar system. The number of directly imaged planets, i.e.~those orbiting at a substantial distance from their host star, is increasing due to massive observational efforts, for example with SPHERE at the VLT (\citealt{2021A&A...651A..71L}). Unless in discovery mode, observations will need to focus on specific objects to maximise the outcome of their instruments, for example for JWST, 
or 
future missions
like PLATO and in the UV (e.g. PolStar \cite{scowen2021polstar}; ARAGO \cite{2019BAAS...51c.219N}; POLLUX at LUVOIR~\citep{18BoNeGo}). Ariel, however, will allow to observe a large ensemble of gas planets and thus to move beyond single observations towards the study of 1000 transiting exoplanets and thus to comparative planetology for exoplanets. With JWST, comparative observations of several gas dominated exoplanets will be possible, for example,  comparison between hot Jupiters of different equilibrium temperatures and planetary rotation orbiting K and G dwarf stars.

Modelling efforts allow  to span ranges of global parameters (\citealt{2016ApJ...828...22P,2021MNRAS.501...78P,2021MNRAS.tmp.1277B}) that are wider than what instrument target lists can afford  and are therefore valuable tools to put the individually observed objects into context as demonstrated in Fig.~\ref{fig:global_properties}. They are  necessary to provide context for future ensemble studies.  Questions to explore include the effect of stellar evolution on their planetary companions but also in-depth studies of  the effect of host star radiation field on cloud formation and the formation of a thermal ionosphere (i.e. a region of a sufficiently high number of charges {  that enables plasma behaviour, for example by coupling to the ambient magnetic field (\citealt{2015MNRAS.454.3977R,2021A&A...648A..80H}})), as well as the effect of different planetary rotation on the wind flow and thus on cloud formation.  The study of cloud formation is not only key to understand the atmospheric chemistry that will be observed with space observatories like  JWST, PLATO, Ariel, LUVOIR  (e.g. \citealt{2020A&A...642A..28M}) but also with ground-based telescopes like the VLT and the ELTs. Understanding cloud formation has gained further momentum due to the role that cloud particles may play in aerial biospheres (\citealt{2017ApJ...836..184Y,2021Univ....7..172S}).

We address the question of how cloud formation is affected by the global parameters like planetary effective temperature and rotation period by utilizing a grid of 48 3D General Circulation Models (GCMs) that includes M, K, G and F stars as planetary host stars. Key properties, like nucleation rate and particle sizes, are selected to study how cloud formation and the resulting global distribution of clouds change with changing global parameters of the star-planet system
(global temperatures, type of host star;
Sect.~\ref{section:Cloud_regimes}). We supplement this part of our grid study by a catalogue which contains the complete set of cloud (nucleation rate, mean particles sizes, material properties, dust-to-gas ratios) and derived gas (C/O ratio, degree of ionisation, mean molecular weight) properties for all the models included in this study  (\citealt{Lewis2022}).  We follow this up by presenting integrated properties that help to discern correlated cloud property trends (Sect.~\ref{section:Cloud_regimes}). Sect~\ref{sec:3cases} studies three selected cases that are representative of cloud formation regimes  as exoplanet tell-tale signs for  weather and possible climate regimes. This specific study is followed up by addressing the effect of the outer boundary of our computational domain on our results (Sect~\ref{section:extrapolation_gas_and_cloud_results}) which leads us to suggest the formation of mineral hazes in form of metal-oxide clusters in the atmospheric region of local pressure $< 10^{-8}$ bar.  Sect~\ref{section:observational_implications} presents observational implications in terms of the Transmission Spectroscopy Metric TSM, $p(\tau(\lambda)=1)$-levels, wavelength-dependent albedo. This paper presents all results for the planetary $\log_{10}(g\,{\rm [cgs]})$)=3.0.
In this first cloud-grid study we focus on the interplay between 3D wind flow and temperatures for different planetary rotations and how they affect equilibrium chemistry and cloud formation. Disequilibrium chemistry will only become important  for effective temperatures smaller than 1400~K and high radiation fields. Here we address formation of mineral cloud particles from collisions dominated gases where equilibrium chemistry holds well. The C/N/O/H non-equilibrium will have little effect on our results \citep{helling2020mineral}.

\section{Approach}
\label{section:Approach}
 
\subsection{3D GCM input}
\label{section:3D_GCM}
We utilise the grid of 3D GCMs for tidally-locked irradiated planetary atmospheres from \cite{2021MNRAS.tmp.1277B}. The grid spans host stars of spectral types M5, K5, G5 and F5, which have T$_{\rm eff} = 3100, 4250, 5650, 6500 \rm{K}$ respectively, and planetary effective temperatures of T$_{\rm eff, P} = 400\, ...\, 2600$K  (in 200 K spacing). While the grid of \cite{2021MNRAS.tmp.1277B} also varies the gravity,  the present study addresses the $\log_{10}(g\,{\rm [cgs]})$ = 3 models only. All model  planets are assumed to have the same radius of 1.35~R$_{\rm Jup}$, and  a constant mean molecular weight of $\mu=2.3$ is assumed for the atmospheric modelling.

The atmospheric circulation in the 3D climate models utilised here is driven using parametrized radiative transfer (Newtonian relaxation) towards radiative-convective equilibrium \citep{Carone2020, 2021MNRAS.tmp.1277B}. As equilibrium state, chemical equilibrium abundances with solar metallicity and C/O ratio have been assumed.  
This Newtonian-relaxation approach results in a computationally efficient 3D model of the atmospheric circulation, which qualitatively agrees with those produced by self-consistent GCMs. As such, its use enables large grid studies of the 3D climate. For an in-depth comparison between a Newtonian-relaxed model and a GCM with self-consistent radiative coupling, see \cite{2022arXiv220209183S}.

The planetary rotation is determined by its orbital period under the assumption of synchronous rotation (P$_{\rm rot}$ = P$_{\rm orb}$) and ranges from $P=0.03 - $ 256~days. For all except the longest orbital periods, the assumption of synchronous rotation is valid based on timescale arguments \citep{2021MNRAS.tmp.1277B}. 
For a given planetary equilibrium temperature, M and K dwarf planets orbit closer to their host stars than planets around G and F stars. As such planets around cooler stars, e.g.~NGTS-10b (\citealt{2020MNRAS.493..126M}), HATS-6b (\citealt{2015AJ....149..166H}), or WASP-43b, are fast rotators, which may impact  the planet's climate. One result of fast rotation is an increased day-night temperature contrast (\citealt{Carone2020,2021MNRAS.tmp.1277B}), which sets the GCMs for these planets apart from models for more typical hot Jupiter (e.g.~\citealt{Parmentier2018,Drummond2018a,mendonca2018b}). Large gas planets around M dwarfs like HATS-6b are rare and challenge the present understanding of planet formation \citep[e.g.][]{Kennedy2008,Morales2019,Bayliss2018}. Thus, such planets pose interesting targets for future characterization with JWST. The very short period corner of the parameter space may also be used to represent gas planets that orbit white dwarfs, for example WD 1856+534 b with an orbital period of 1.4 days, or brown dwarfs - white  dwarfs pairs (e.g., WD 0137-349B with P= 0.0803 days, \citealt{2020MNRAS.496.4674L}).

For the grid study conducted in this paper, we extract 48 1D (T$_{\rm gas}$, p$_{\rm gas}$, $\vec{v}(x,y,z)$) profiles per 3D GCM atmospheric model.
The sampled latitudes are $\theta = 0^{\degree}$ (the equator) and $\theta = 45^{\degree}$, and the sampled longitudes span $\phi = -165^{\degree} {~\rm to~} 180^{\degree}$ in $15^{\degree}$ spacing 
The morning and evening terminators are given by $\phi = 270^{\degree}$ and $\phi = 90^{\degree}$ respectively.

In a small fraction of vertical temperature profiles extracted from the GCMs, spurious unphysical variations have been found, which are likely related to numerical artifacts, possibly due to the parameterized radiative forcing. Such profiles have been omitted from our cloud analysis. In the end, we included 85\% of the G star profiles and 86\% of the M star profiles in our investigations, as well as 100\% of the 1D atmosphere profiles for the K and the F star planets.

We further note that 3D GCMs exhibit notoriously slow temperature evolution in the deep atmosphere \citep{Carone2020, Wang2020, 2022arXiv220209183S}. This has largely been mitigated in the grid of \citet{2021MNRAS.tmp.1277B} by starting from a hot initial adiabatic temperature profile, but some 3D GCMs still are not fully converged for p$_{\rm gas}>100$~bar, leading to locally oscillating (T$_{\rm gas}$, p$_{\rm gas}$) structures in these innermost atmospheric regions. {   The main results in this paper will not be altered by these uncertainties since the high gas pressures do stabilise the cloud particles over a substantial temperature range at these deep atmospheric layers (see for example, Fig. 14 in \citealt{2021A&A...649A..44H}) which sits deep in the optically thick part of the atmosphere.} WASP-43b is one example where cloud formation reaches deep into these inner atmospheric regimes (\citealt{2021A&A...649A..44H}).
We include these unconverged regions nevertheless in order to explore how deep inside the atmosphere clouds could form {   for the different global parameters covered in our grid.}

\paragraph{   The grid's global parameters corners:}
The model grid spans a {   large range of global parameters that has not yet been filled with existing exoplanets equivalents completely. Two corners of the grid's parameter space are therefore pointed out  that may appear as unrealistic at a first glance for the time being: }

1) Planets with $P<35$d can safely be assumed to be tidally locked. Longer  orbital periods of tidally locked planets may occur nevertheless for older systems, for example Mercury has a period of 90d. One should, however, expect a spin evolution for exoplanets during their migration through the planet forming disks. {   For example, } brown dwarfs undergo a spin down evolution which may align them with planets  (\citealt{2018ApJ...859..153S}).

2)  Giant gas planets with log(g) = 2 [cgs] = 100 cm\,s$^{-2}$ have not been discovered so far. Nevertheless, there is a very small sub-class of objects, so-called super-puffs with very small bulk densities: For example, the Kepler-51 planets with orbital periods of 45, 85 and 130 days and with densities below  0.1 g/cm3 (\citealt{2014ApJ...783...53M,2020AJ....159...57L}). While the 3D GCMs for very low surface gravity exist, they are not included in the present cloud study.

\begin{figure*}
{\ }\\*[-1.0cm]
   \includegraphics[width=0.965\textwidth]
    {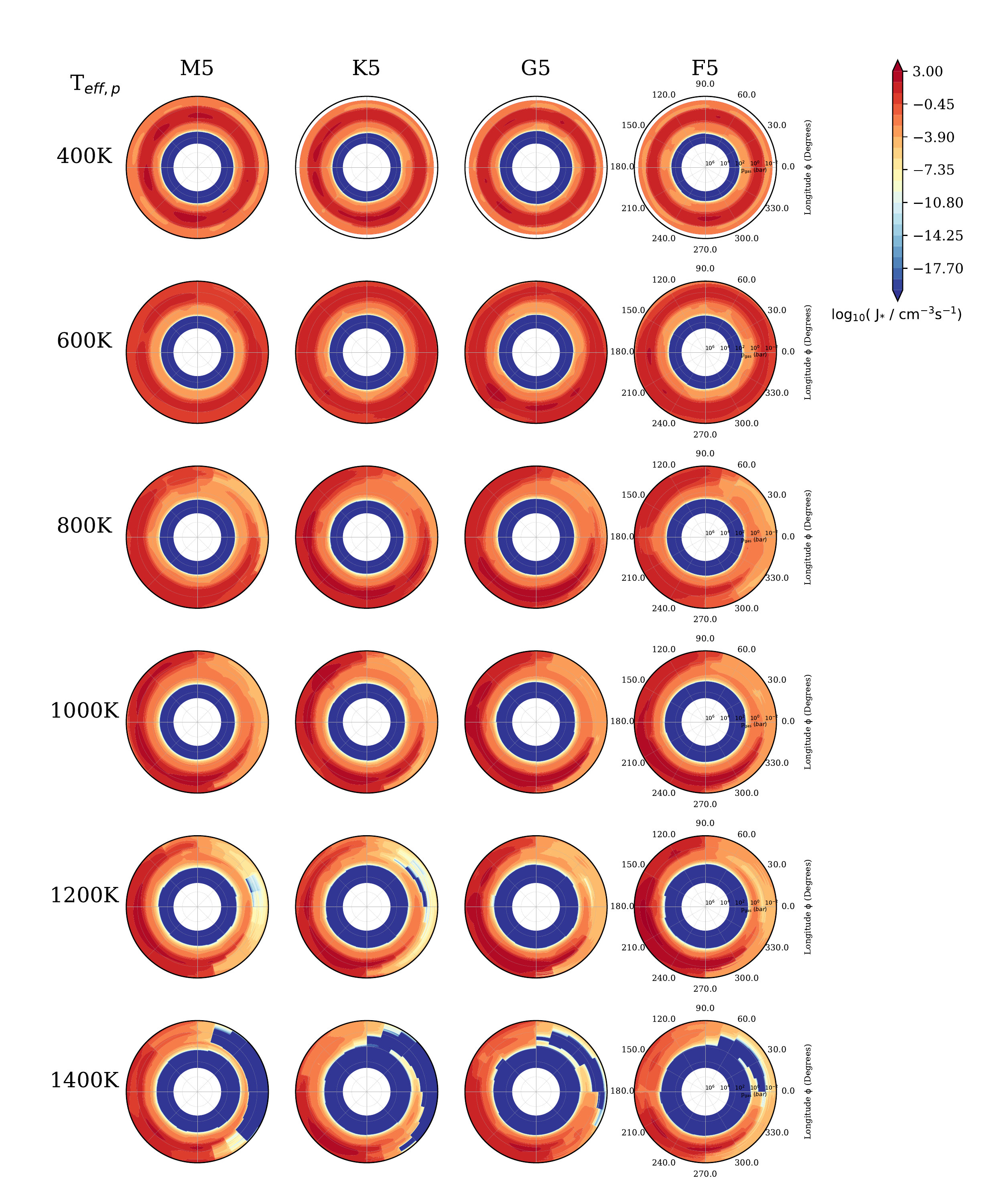}
    \caption{Total nucleation rate, $ J_{*} =\sum_{\rm i}J_{\rm i}$ [cm$^{-3}$ s$^{-1}$], i=TiO$_2$, SiO, NaCl, KCl,  in the 2D  equatorial plane ($\theta = 0\degree$) for the 1D profiles extracted from the 3D GCM models for log$_{10}$g = 3 [cgs]. The same cold model atmospheres like in Fig.~\ref{fig:global_slice_plots1} are depicted.
    In this grid, the $T_{\rm eff, P}=1400$K orbiting an K or G star is representative of NGTS-10b/WASP-43b and 
    HD 209458b respectively. The $T_{\rm eff, P}=1200$K orbiting the K dwarf may represent HD 189733b, $T_{\rm eff, P}=1600$K orbiting the K star for WASP-63b.}
   \label{fig:global_slice_plots_logg3_nucleation_1}
\end{figure*}

\begin{figure*}
{\ }\\*[-0.5cm]
   \includegraphics[width=0.965\textwidth]
    {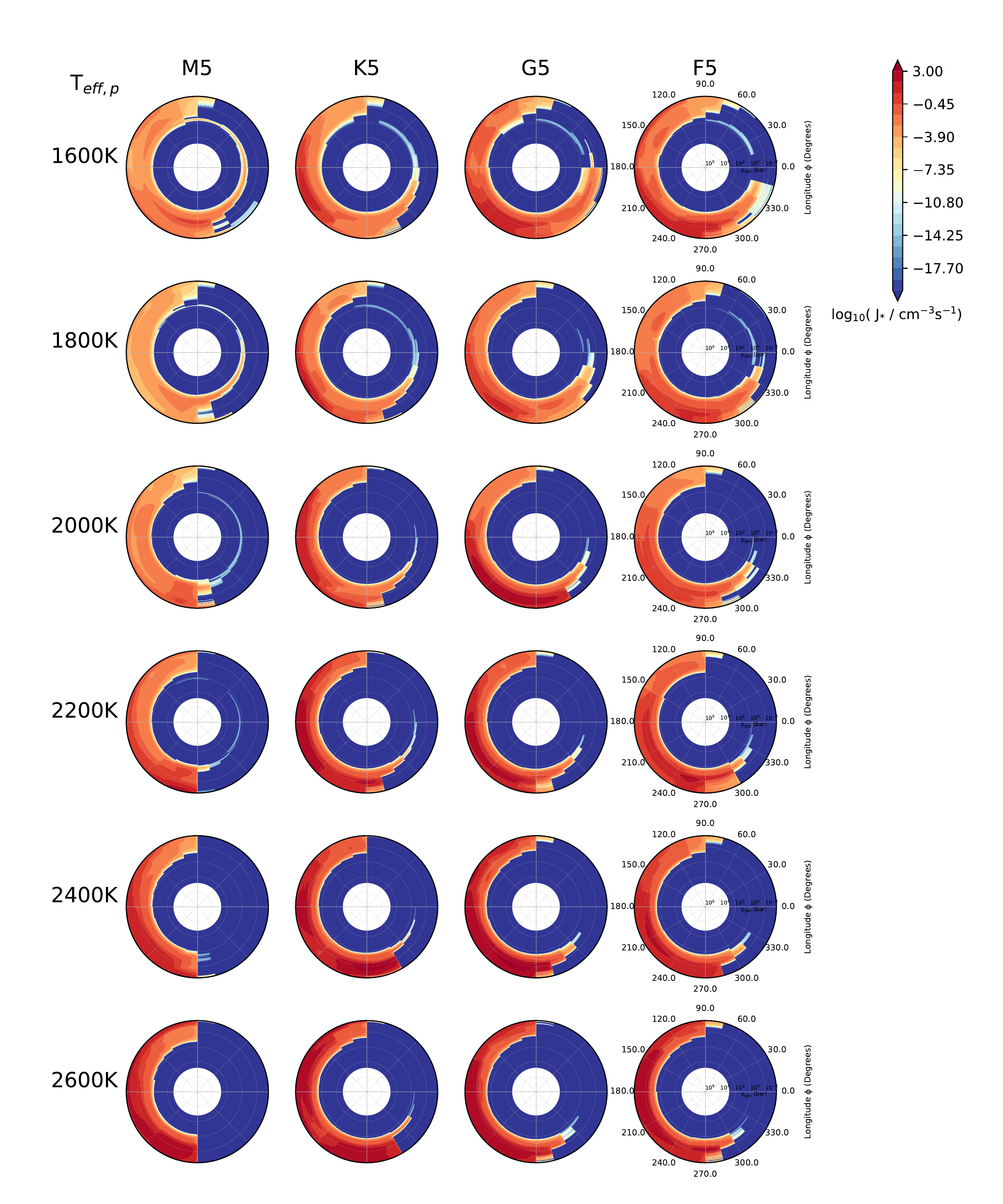}
    \caption{Total nucleation rate, $ J_{*} =\sum_{\rm i}J_{\rm i}$ [cm$^{-3}$ s$^{-1}$], i=TiO$_2$, SiO, NaCl, KCl,  in the 2D  equatorial plane ($\theta = 0\degree$) for the 1D profiles extracted from the 3D GCM models for log$_{10}$g = 3 [cgs]. The same  warm model atmospheres like in Fig.~\ref{fig:global_slice_plots2} are depicted.}
   \label{fig:global_slice_plots_logg3_nucleation_2}
\end{figure*}

\subsection{Cloud formation and gas-phase modelling}
\label{section:CF_and_gas_phase}

{   Cloud formation has become an important piece of physico-chemistry for exoplanet research and various groups worked on understanding cloud formation in the context of atmospheric environments. Model approaches have been compared in \cite{2008MNRAS.391.1854H} and  within the atmosphere modelling context summarised in  \cite{2018ApJ...854..172C} and \cite{2020RAA....20...99Z}. A discussion of the cloud formation models in the exoplanet community as well as the approach applied in the plant-forming disk community compared to the modelling approach used here can be found in the recent paper of \cite{2022A&A...663A..47S}. }

The local gas-phase abundances are calculated assuming chemical equilibrium by applying {\sc GGChem} which is part of our cloud formation code. {   The gas phase is assumed to be in chemical equilibrium throughout the simulation. This, however, does not imply phase equilibrium for the condensate species considered for cloud formation. We addressed the small differences that may be imposed by kinetic gas-phase effects on the gas composition in the cloud forming  regions in \cite{2020A&A...635A..31M}.} Out of the total set of elements considered for the gas-phase, only 11 elements  (Mg, Si, Ti, O, Fe, Al, Ca, S, K, Cl, Na) participate in the bulk growth of the cloud particles and only 6 (Ti, Si, O, K, Cl, Na) participate in the formation of cloud condensation nuclei. Within our kinetic cloud formation approach, we treat the formation of 4 nucleation species (TiO$_2$, SiO, KCl, NaCl) that form the cloud condensation nuclei and determine the total nucleation rate ($J_*$ [cm$^{-3}$s$^{-1}$]). We use the modified classical nucleation theory (e.g., see \citealt{helling2013RSPTA,2018A&A...614A.126L}) the results of which we compare for TiO$_2$ to a Monte Carlo approach treating individual cluster collisions (\citealt{2021A&A...654A.120K}).
The total nucleation rate  determines the number of cloud particles that are forming locally and which grow to macroscopic sizes by the condensation of 16 materials through 132 gas-surface growth reactions. The cloud particle sizes are expressed as local mean cloud particle radii $\langle a\rangle$ [$\mu$m] (\citealt{Woitke2003,Woitke2004,2004A&A...423..657H,Helling2006,2008A&A...485..547H}). For a recent review, see  \cite{2022arXiv220500454H}.

We do present our results in terms of surface averaged cloud particle radii which is more representative of their effect of the local opacity (see Sect.~\ref{amean}). In total, we are solving 31 ODEs in order to model the formation of cloud particles as a sequence of nucleation, surface growth/evaporation, gravitational settling, element replenishment and element conservation. The undepleted gas is assumed to be of solar composition. We have also undertaken an update of our evaporation routine. The updated modelling of the vertical mixing is described below.


\subsection{Treatment of vertical mixing}
\label{section:vertical_mixing}

Atmospheric transport processes remain challenging in combination with cloud formation modelling. Within a hydrodynamic framework, there is advective but also diffusive transport. Both components, gas and cloud particles,  will move with the same velocity if the cloud particles are frictionally coupled to the gas phase. Cloud particles move with different velocities than the gas if an additional force, like gravity, causes a frictional decoupling. Gravitational decoupling is treated as a consistent part of our kinetic cloud model (\citealt{Woitke2003}). Hydrodynamic transport processes that cause a vertical transport, however, are either parameterized \citep[e.g.][]{parmentier20133d,2019A&A...631A..79H,2021A&A...649A..44H,2021MNRAS.504.2783S} or derived from the hydrodynamic velocity field as described in Appendix~\ref{s:diffmix}.
In this paper, we apply two different approaches for two different domains:

a) Within the 3D GCM computational domain ($p_\textrm{gas} > 10^{-4}$~bar): The cloud modelling within the computational domain  of the whole grid of 3D models applies a different treatment of the vertical mixing source term than in previous papers (\citealt{2019A&A...631A..79H,2021A&A...649A..44H}), calculating the standard deviation based on adjacent grid cells. The details are outlined in Appendix~\ref{s:diffmix}. The respective mixing time scale is calculated from Eq.~\ref{final}.

b) Above the 3D GCM computational domain ($p_\textrm{gas} < 10^{-4}$~bar; Sect.~\ref{section:extrapolation_gas_and_cloud_results}): No information about the local velocity fields are available outside the computational domain of the 3D GCMs. Hence, we adopt  the final value for the original non-extrapolated profile for  the vertical velocity within the extrapolated regime to such that the velocity is constant. We demonstrate the validity of the hydrodynamic assumptions within the extrapolated atmosphere profiled in Appendix ~\ref{section:hydrodynamics_validity}. The hydrodynamic assumption would break down at p$_{\rm gas}\approx 10^{-9}$~bar if the collisional processes within the atmospheric gas were only due to neutral molecules (here: H$_2$). This threshold moves to pressures as low as p$_{\rm gas}\approx 10^{-15}$~bar if the atmospheric gas is ionised (also \citealt{Debrecht2020}). The increased degree of ionisation in exoplanet (and brown dwarf) atmospheres has been demonstrated in \cite{Barth2021} due to photochemical effects as well as due to Lyman continuum ionisation by the interstellar radiation field (\citealt{2018A&A...618A.107R}).

\section{Results}
\label{section:Results}

The leading aim of this study is to identify global cloud formation  trends (Sect.~\ref{section:Cloud_regimes}) and globally changing chemical regimes (Sect.~\ref{section:chemical_regimes}) depending on the  global parameters of the star-planet system with view of upcoming space missions like JWST, PLATO and Ariel, but also for potential missions in the UV. The respective mission host stars are covered by the model grid that is utilised here. We concentrate on the  stellar effective temperature and the  orbital period as global system  parameters.

The orbital period does determine the planetary effective (or global) temperature and the stellar spectral type is represented by the stellar effective temperature. An overview of the parameter range can be found in Figs.~\ref{fig:global_properties}-\ref{fig:global_slice_plots2}. A secondary aim is to provide a first insight regarding the potential of magnetic coupling by investigating the general trend of thermal ionisation with global parameters in comparison to the cloud location (Sect.~\ref{section:degree_of_ionisation}). We hope to stimulate follow-up  studies in magnetic coupling effects beyond the assumption of an ideal MH flow that assumes a constant coupling for changing thermodynamic conditions.

Our 3D grid study supports the transition found between hot and ultra-hot Jupiters for $T>1800$~K  \citep[e.g.][]{2020A&A...639A..36B, 
2019NatAs...3.1092K,Showman2020Review,Zhang2020Review,Parmentier2021}: The dayside of ultra-hot planets are cloud-free (Fig. \ref{fig:global_slice_plots2}), which leads to a low bond albedo (and thus efficient dayside irradiation). Further, the day-to-nightside heat circulation is very inefficient in the 3D GCMs \citep[see also][]{Perna2012,2017ApJ...835..198K}, leading to strong horizontal temperature gradients (Fig. \ref{fig:global_slice_plots1}). Low bond albedo at the dayside and inefficient heat circulation in combination lead to particularly large day-to-nightside emission differences. We note, however, that faster rotators, that is, the M and K dwarf planets are prone to have an even more pronounced day-to-nightside dichotomy in temperatures and cloudiness than slower rotators, that is the G and F planets. We note here that we do  not have cloud-feedback on the temperatures that may lead to even less heat circulation in ultra-hot Jupiters \citep{Parmentier2021}. However, \citet{Parmentier2021} used a simpler cloud model than used in this study. A future study that will also incorporate cloud-feedback will show if this effect can be reproduced also with the microphysical cloud model. Further assumptions may alter the exact temperature threshold between hot and ultra-hot Jupiters since 3D atmosphere modelling did require further assumption to enable the simulations. One such assumption is the mean molecular weight which we  address in Sect.~\ref{section:mean_molecular_weight}.

\subsection{Host star trends of planetary (T$_{\rm gas}$, p$_{\rm gas}$)-structures}
\label{section:Tp-struc_trends}

A summary of the change in the 3D (T$_{\rm gas}$, p$_{\rm gas}$)-structures demonstrates first trends that will translate into trends  in the global cloud structure of these planetary atmospheres. A detailed analysis is presented in \cite{2021MNRAS.tmp.1277B} and we focus on global trends only. Figures~\ref{fig:global_slice_plots1} \& ~\ref{fig:global_slice_plots2} show the thermal structure of the equatorial plane ($\theta$=0$^{\circ}$) for all the log(g)=3 [cgs] models, and how the local atmosphere temperature changes when the planet orbits closer to
its host star (increasing T$_{\rm eff, P}$).
 
All models with T$_{\rm eff, P}\leq800{\rm K}$ have a horizontally/zonally/longitudinally homogeneous temperature distribution. The day-night asymmetry emerges at T$_{\rm eff, P}=800{\rm K}$ and is more pronounced for the faster rotating stellar types.  For hotter planets ($T_{\rm eff, P} \geq 1600$~K), advection of cooler air from the nightside onto the dayside can be seen, extending across the evening terminator at 1~bar. This structure is more extended for the F and the G star models, and it does not appear at all for the M star planets. 
 The JWST targets NGTS-10b, HD~209458b (pink square), HD~189733b  (brown square) and WASP-63b have equilibrium temperature between 1200\,K and 1600\,K orbiting G and K stars, hence represent the transitional regime from homogeneous temperature distributions to pronounced day/night temperature differences (e.g. HAT-P-7b (purple filled circle)) in the 3D GCM grid utilized here. 
The super-Saturn HATS-6b (yellow triangle) that orbits an  M1V host star (d=148.4 ($\pm$ 3.3) pc) at a distance of  a=0.03623 AU with a period of P=3.325 d (M$_{\rm P}$=0.319 $M_{\rm Jup}$, R$_{\rm P}$=0.998 $R_{\rm Jup}$) would be represented by the model of T$_{\rm eff, P}$=600\,K, $g=10$m\,s$^{-1}$ and M-type host star. A homogeneous temperature structure can, hence,  be expected for HATS-6b.

\subsection{Cloud regimes with changing global system parameters}
\label{section:Cloud_regimes}

The formation of clouds is triggered by a gas-phase transition leading to the formation of newly formed cloud condensation nuclei (nucleation), unless meteoritic dust re-condenses or the planet under consideration has a rocky surface from which sand particles are transported into the atmosphere.  The chemical processes that lead the nucleation process are only partially known (e.g., \citealt{2021A&A...654A.120K} and references therein) and extensive studies are ongoing.  The nucleation process is key to where clouds can form and it is determined by the local thermodynamic conditions. It is therefore important to be able to determine where in the atmosphere nucleation occurs with which efficiency  as this is already indicative for changing cloud regimes with changing global stellar-planetary parameters (Sect.~\ref{ss:nuc}). The nucleation rate determines the number of cloud particles that eventually make up the whole cloud (and their distribution), hence, it also influences the size of the cloud particles. Due to element conservation (mass conservation), it is reasonable to generally expect large cloud particles in regions of low nucleation efficiency (Sect.~\ref{amean}). 

\begin{figure*}
    \includegraphics[width=20pc]{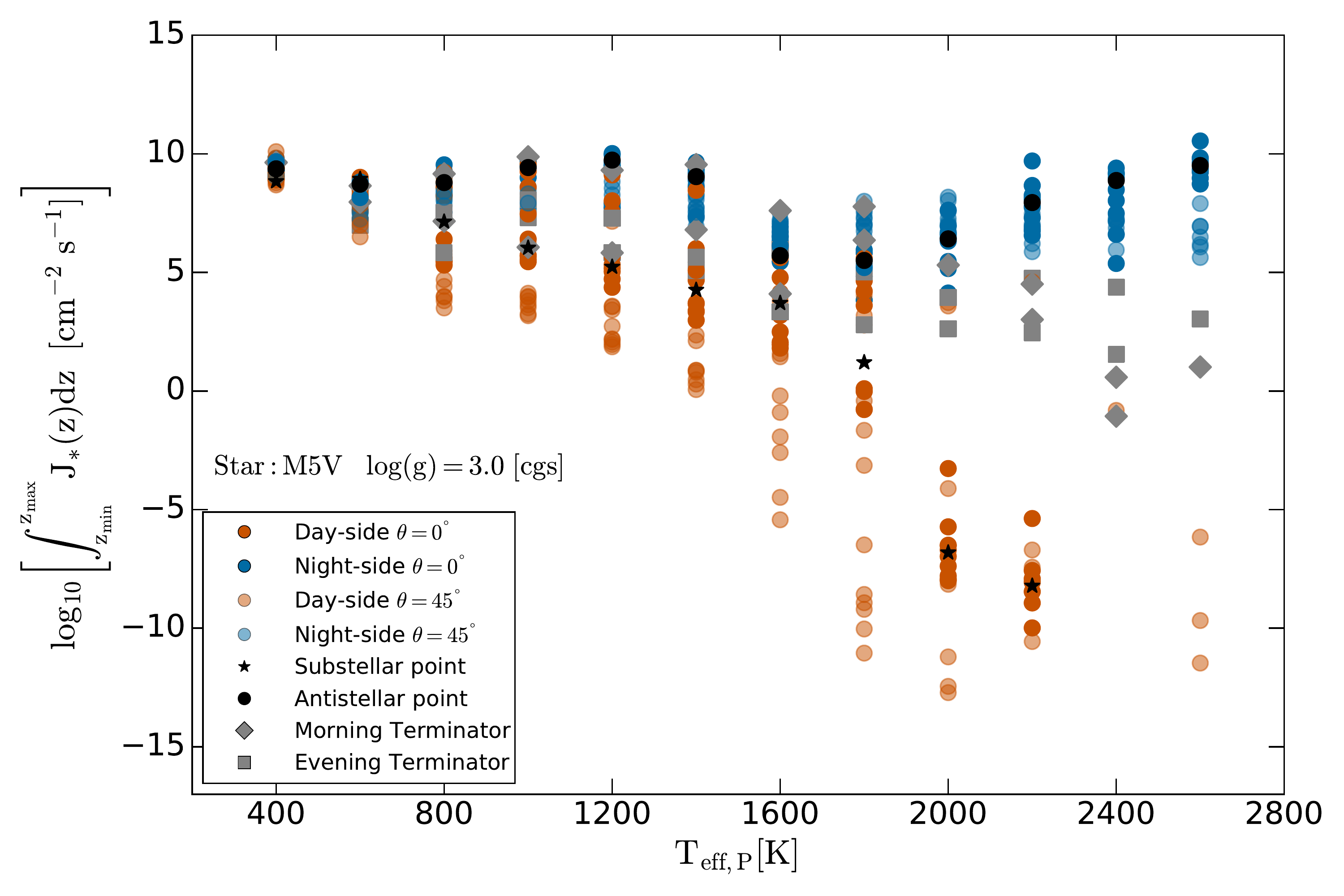}
    \includegraphics[width=20pc]{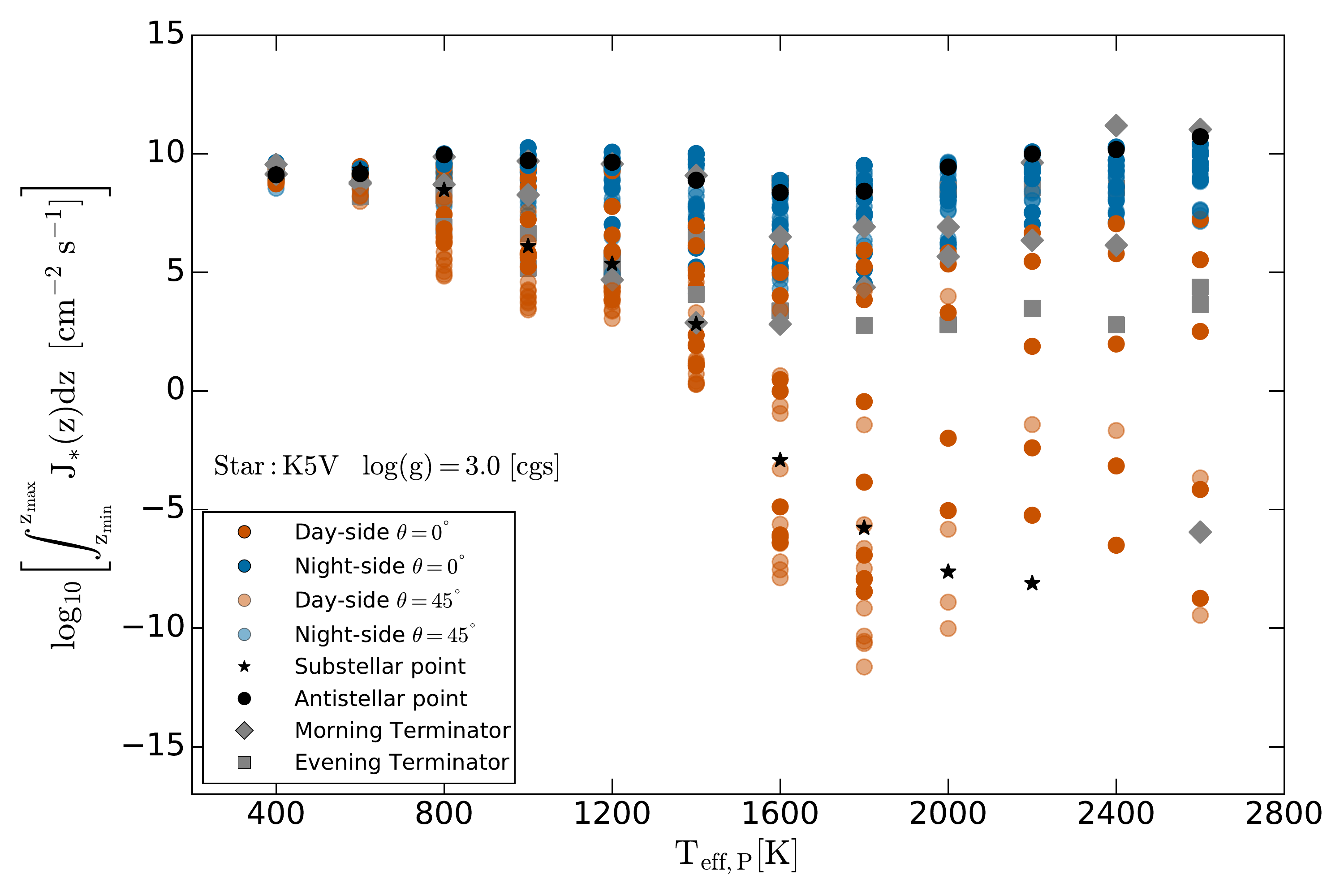}\\
    \includegraphics[width=20pc]{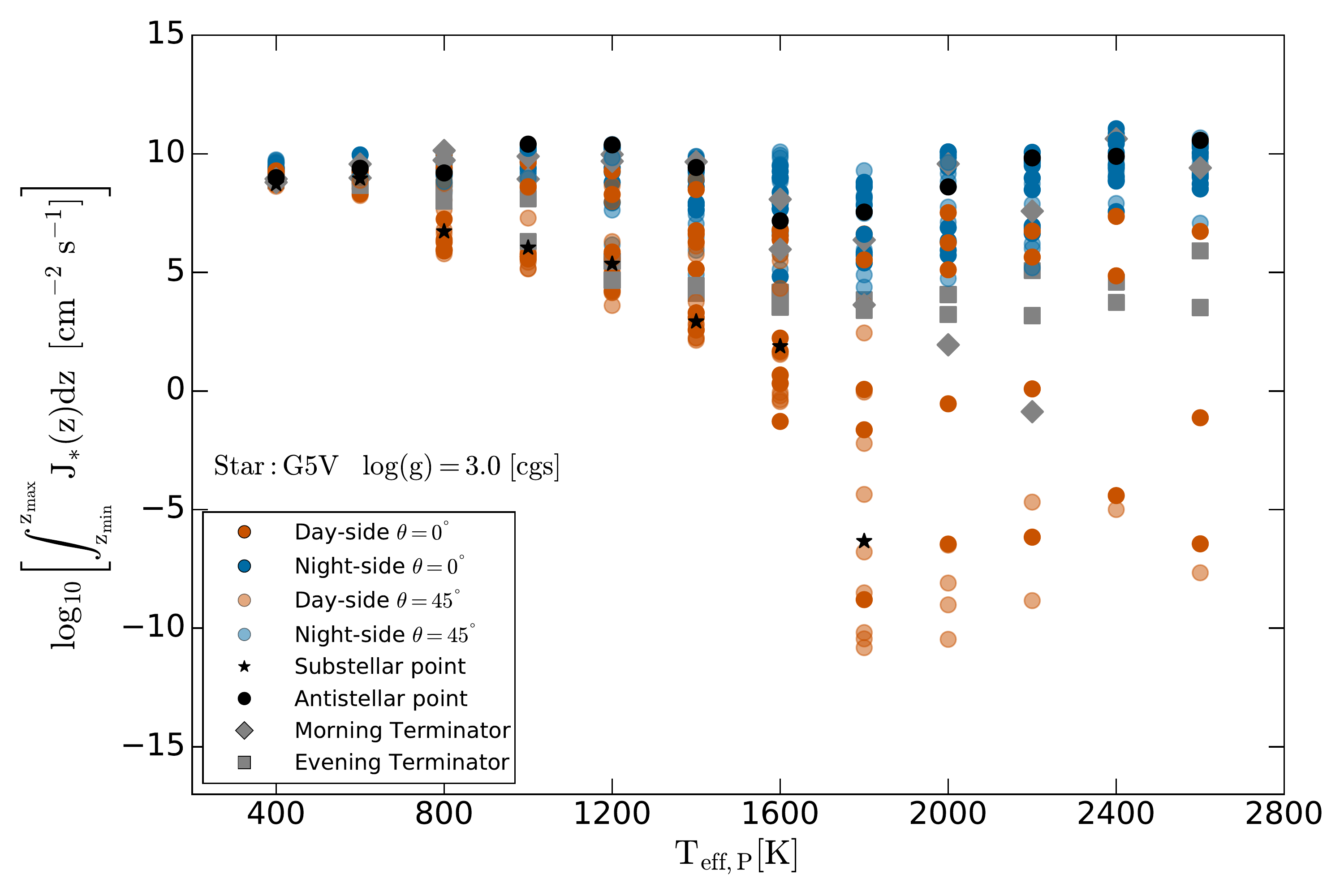}
    \includegraphics[width=20pc]{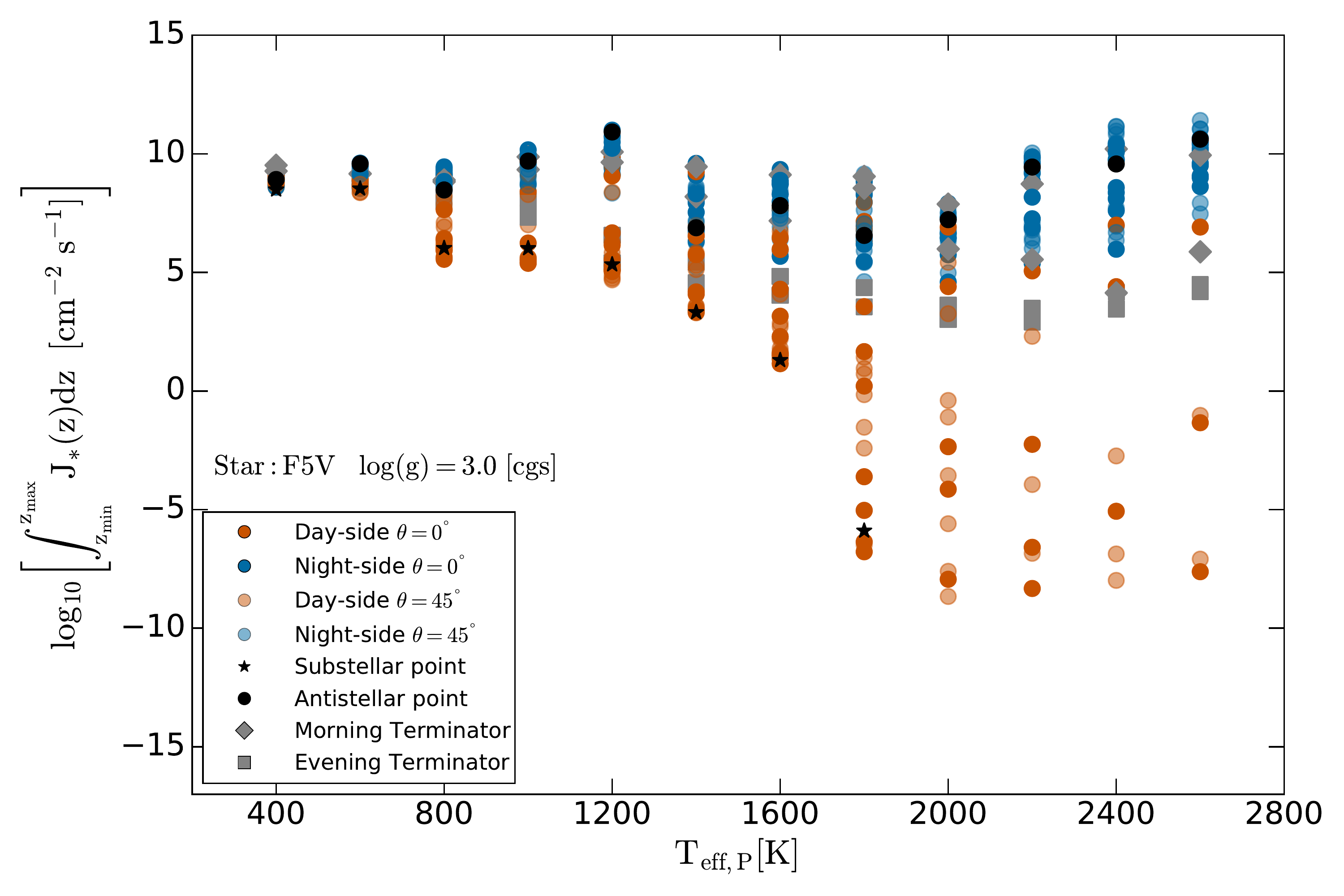}
    \caption{Column-integrated nucleation rates for the $\log(g) = 3 \rm{[cgs]}$ grid models and different stellar types. No one value suffices to describe the rate at which cloud particles form. At approximately 1600 K or 1800 K, the overall spread is much larger for all host stars. 
    }
    \label{fig:nuc_integrate_scatter}
\end{figure*}
\begin{figure*}
    \includegraphics[width=20pc]{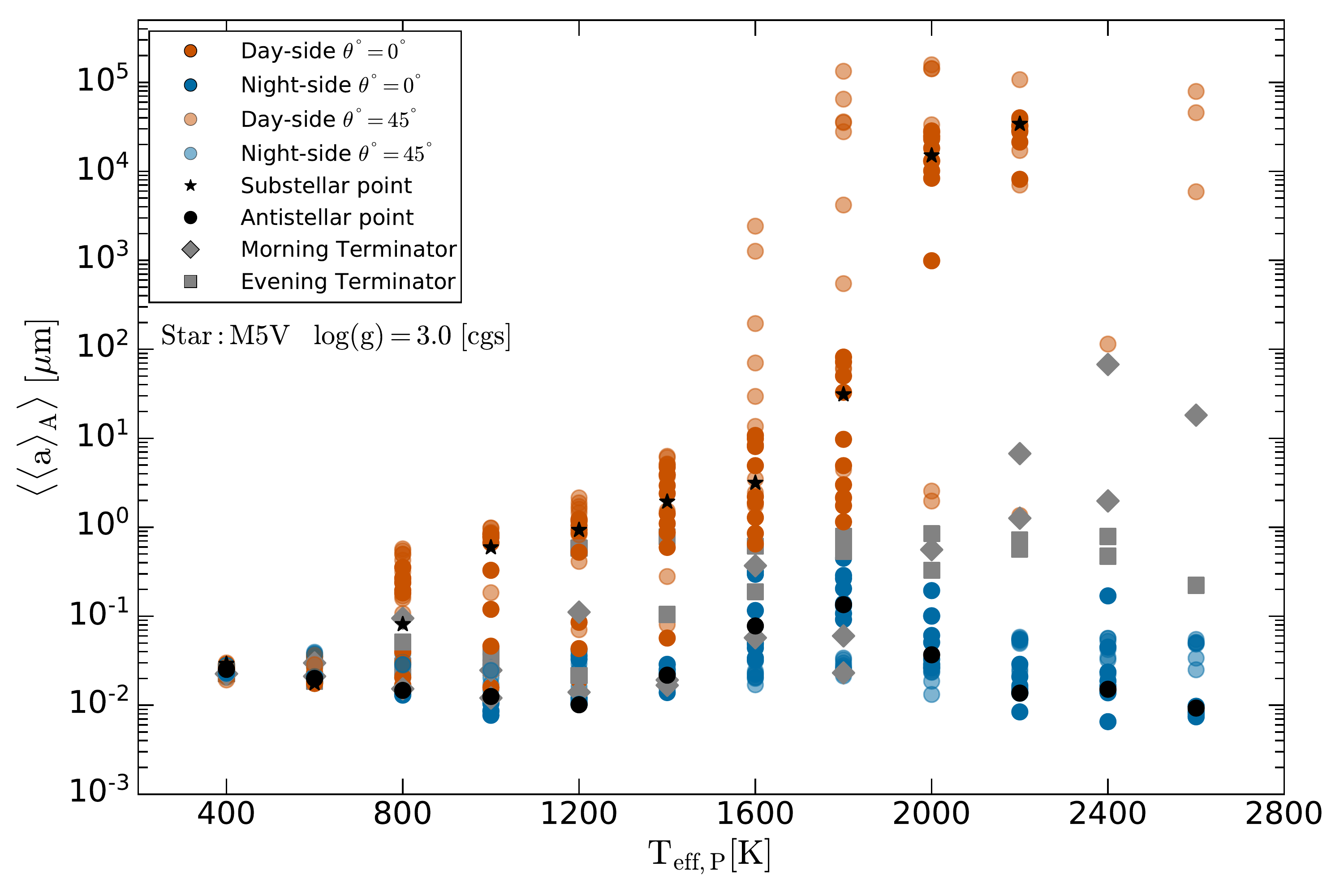}
    \includegraphics[width=20pc]{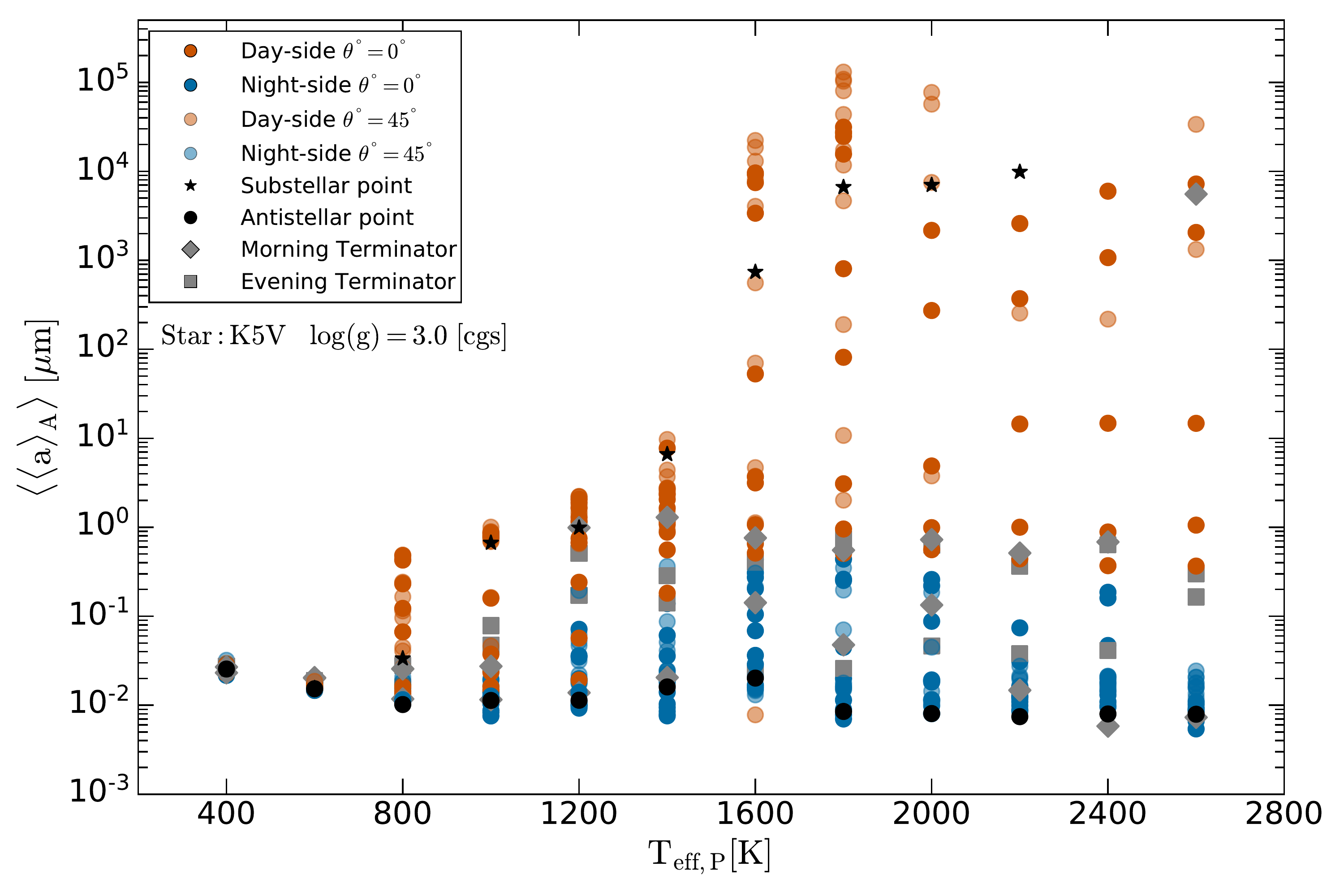}\\
    \includegraphics[width=20pc]{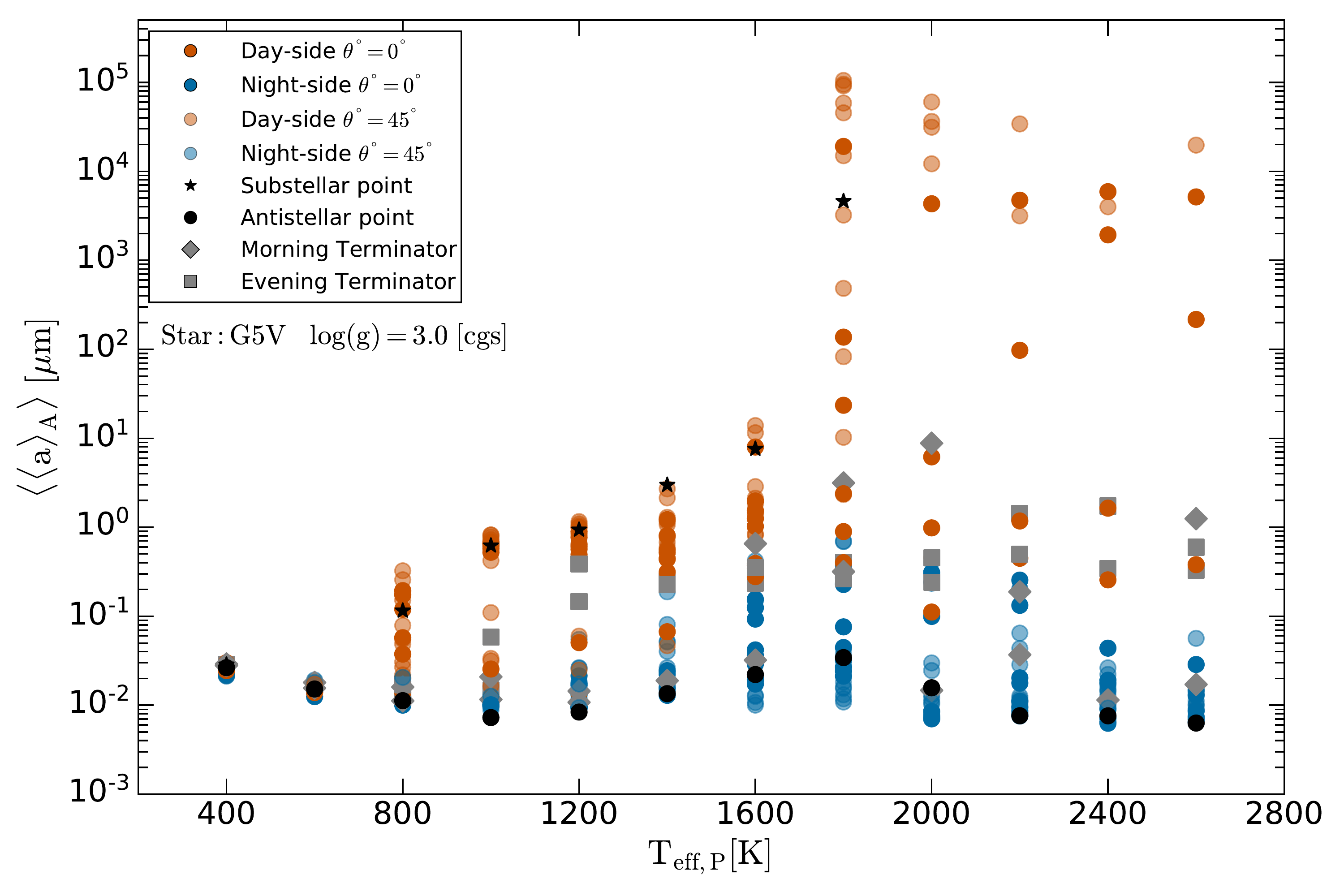}
    \includegraphics[width=20pc]{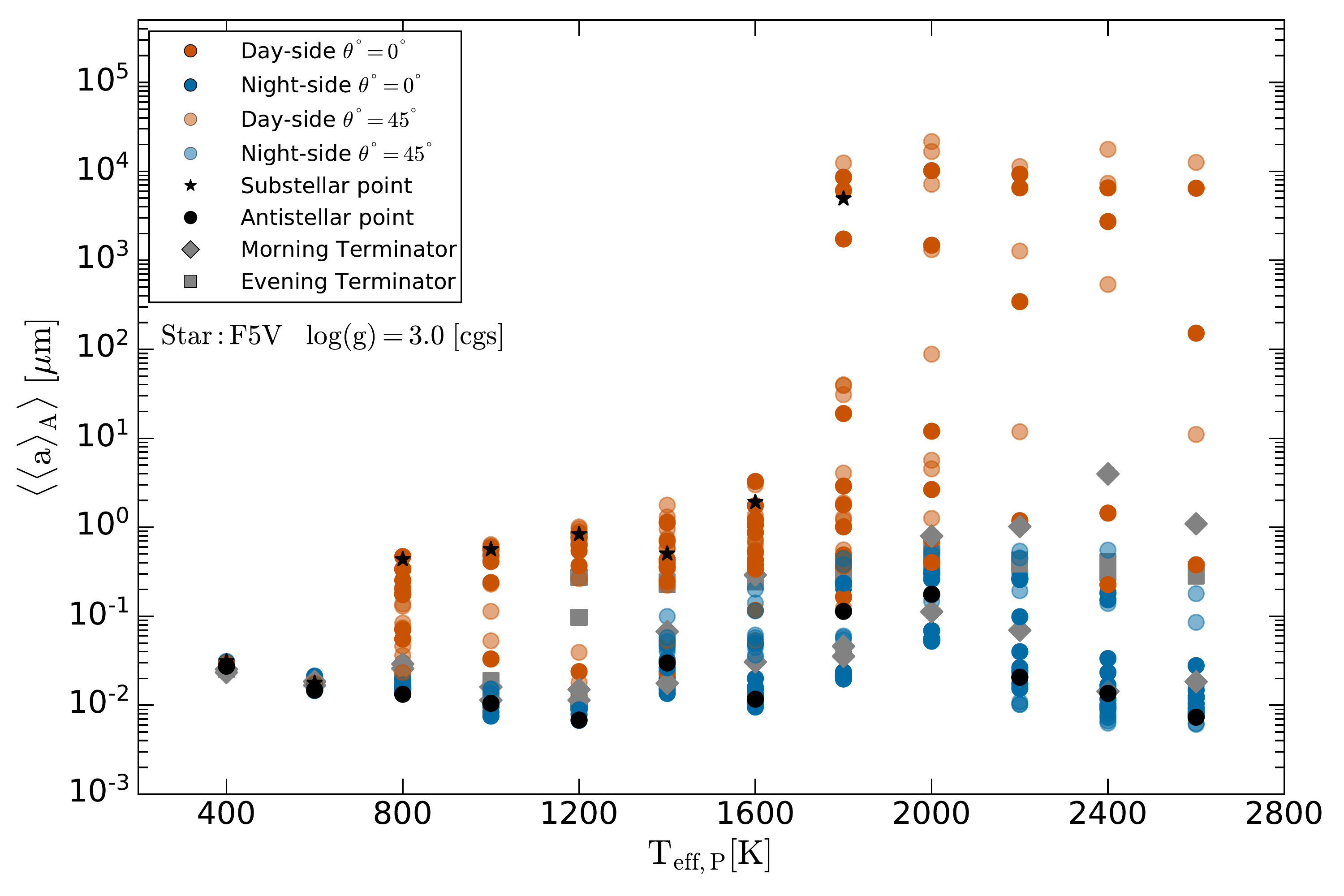}
    \caption{Column integrated, number density weighted surface averaged mean particle size, $\langle \langle a \rangle_{\rm A} \rangle$ [$\mu$m], for the $\log(g) = 3 \rm{[cgs]}$ grid models and different stellar types.}
    \label{fig:mean_particle_size_integrate_scatter}
\end{figure*}

\subsubsection{The formation of cloud condensation nuclei}\label{ss:nuc}

The total nucleation rate indicates the efficiency at which nucleation seeds form spontaneously from the gas phase, and is therefore used to identify the cloud formation regime. We consider here the nucleation of four species: TiO$_2$, SiO, KCl, NaCl. We calculate the total nucleation rate as $J_{\rm *,  tot}=\sum J_{\rm *, i}$ (i=TiO$_2$, SiO, KCl, NaCl). The equatorial distribution of the total nucleation rate is presented for all models with log(g)=3 [cgs] in Figs,~\ref{fig:global_slice_plots_logg3_nucleation_1},  \ref{fig:global_slice_plots_logg3_nucleation_2}.

There are crudely two nucleation regimes: one where nucleation occurs throughout the whole atmosphere across the whole globe (T$_{\rm eff, P}\leq 1200$ K), the  globally homogeneous nucleation regime, and one where nucleation occurs intermittently or asymmetrically distributed in the atmosphere (T$_{\rm eff, P} >1200$ K), the partial nucleation regime.

The lowest global  temperature corner of the globally homogeneous nucleation regime is characterised by an extremely efficient nucleation process.
In the upper atmosphere of the T$_{\rm eff, P } = 400$ K model the total nucleation has an initial value of approximately $10^{-3} ~{\rm cm^{-3} s^{-1}}$ which rises to a peak of $10^{3} ~{\rm cm^{-3} s^{-1}}$ by $p_{\rm gas} = 10^{-2}$ bar. This is due to the delayed onset of SiO nucleation, which begins at 10$^{-3}$ bar and dominates significantly over the TiO$_{2}$ nucleation between ~10$^{-2.5}$ bar and ~10$^{-0.5}$ bar. In the deep atmosphere at pressures greater than $p_{\rm gas} = 10^{-0.5}$ bar the efficiency of the SiO nucleation decreases such that TiO$_{2}$ becomes the dominant nucleation species. 

Nucleation occurs across almost the entire equatorial plane for the models T$_{\rm eff, P} \leq 1200$ K, with the dayside nucleation generally reduced compared to the nightside with a strong day-night asymmetry emerging with increasing planetary temperature. This asymmetry becomes most apparent in the models with T$_{\rm eff, P} \geq 1400$ K where regions of the dayside atmosphere do not exhibit nucleation. The extent of the dayside atmosphere where nucleation is not possible increases with the global planetary temperature T$_{\rm eff, P}$, starting east of the substellar point and varies in size between the four stellar types M, K, G and F. The size of the  decreased nucleation region on the dayside is larger for the faster rotating planets at a given temperature, i.e. largest for the M star and smallest for the F stars. The location of this dayside reduction in nucleation coincides with the temperature hot spot which is offset from the substellar point due to the equatorial jet. Nucleation still occurs on the dayside these hotter models, however, this only occurs where cool air has been carried across the morning terminator by the jet. In hotter planetary atmospheres, the radiative response becomes shorter and cold air is efficiently heated as it reaches the dayside. For T$_{\rm eff, P} \geq 2000$ K there is essentially no nucleation on the dayside of the M star orbiting planets, and the extension of the nucleation across the morning terminator is similar for the K, G and F stars. The nucleation rate will not only be affected by the global planetary temperature but also by the gravity which determines the density structure of the atmosphere. Higher gravity will shift the nucleation emergence towards higher temperature due to an increased thermal stability as result of a higher collision rate for increased gas densities.

To enable the comparison of the nucleation efficiency across the whole grid of global parameters, column integrated values are considered. We note that the integration column does vary for different planetary atmospheres due to the varying cloud extension as result of the local thermodynamic conditions (e.g. Fig.\ref{fig:extrapolated_clouds_2.0}).

Figure~\ref{fig:nuc_integrate_scatter} shows the column integrated total nucleation rate for each of the model planets for each stellar type to allow for a comparison of nucleation activity across the whole 3D GCM grid. 
The range of values for the column integrated nucleation rates is very narrow for both the T$_{\rm eff, P}$= 400 K ($\sim 10^{8}-10^{10} ~{\rm cm^{-2} ~s^{-1}}$ ) and T$_{\rm eff, P}$= 600 K ($\sim 10^{6}-10^{10} ~{\rm cm^{-2} ~s^{-1}}$ ) models for all stellar types. For warmer models the range in values widens and the divide between dayside and nightside becomes more apparent. The range of values is larger for the M and K stars ($\sim 10^{3}-10^{10} ~{\rm cm^{-2} ~s^{-1}}$) compared to the G and F stars ($\sim 10^{5}-10^{10} ~{\rm cm^{-2} ~s^{-1}}$) for planetary effective temperatures T$_{\rm eff, P} = 800 - 1200$ K. The small spread of values is representative for the globally homogeneous nucleation regime where the formation of cloud condensation nuclei is most efficient and possible across the globe.   For models T$_{\rm eff, P} = 1400, 1600, 1800$ K that range increases slightly and again separation between day and nightside becomes clearer and it is apparent that nucleation is less efficient for the higher $\theta = 45^{\circ}$ latitude compared to the equator for these models. Such a spread in values is representative for the partial nucleation regime, the largest spread will represent the atmospheres with the largest cloud formation asymmetry. The largest thermodynamic, and hence, nucleation asymmetry occurs for  T$_{\rm eff, P}\geq 2000$ K.

For all models the spread of values for the column integrated total nucleation rate is smaller for the nightside than the dayside. This is a reflection of the homogeneous local gas temperature of the nightside, whereas there is more variation in the temperatures on the dayside. The range of values on the dayside is reasonably consistent at $\sim 10^{-13}-10^{5} ~{\rm cm^{-2} ~s^{-1}}$ for the T$_{\rm eff, P} = 1400, 1600$ K models, and there is still significant overlap between the day and nightside values. A `cone' of diverging integrated nucleation rates emerges as function of the stellar effective temperature where the upper limit of integrated nucleation rate remains roughly similar ($10^{10}$cm$^{-2}$s$^{-1}$). A `bifurcation' occurs at the hotter end T$_{\rm eff, P} > 1400$K, most clear for the M5 host star, where there is a clear separation between the dayside and nightside nucleation efficiency.

We conclude that no one value of nucleation rate is sufficient to describe the first step for cloud formation in exoplanet atmospheres. Only for the coolest atmospheres one might describe the nucleation rate reasonably by one value. Here, the \cite{2001ApJ...556..872A} model may well be suited to speed up GCM efficiency.

\subsubsection{Mean particle size and dust-to-gas ratio}\label{amean}

Cloud particles sizes are essential to calculate the cloud opacity and are often seen as observationally accessible test of cloud properties and of cloud models, for example \cite{1998Sci...282.2063G,2002ApJ...568..335M,2006ApJ...648..614C, LeeHengIrwin2013,2016ApJ...830...96H,Benneke2019,Lacy2020}. We re-iterate previous results of  that exoplanet and brown dwarf clouds can not be characterised by one particle size only (\citealt{Helling2006,2017A&A...603A.123H,2019AREPS..47..583H}).  Here, however, the focus is on potential trends that might serve as input for automized retrieval efforts. We therefore chose to represent the mean cloud particle size in terms of surfaced averaged mean particle size,  $\langle a \rangle_{A}$.  The dust-to-gas ratio is a helpful property to locate the cloud mass load and to compare to other astrophysical objects where condensation processes take place (e.g. AGB stars, Wolf-Rayet stars, SNs).

The surfaced averaged mean particle size, $\langle a \rangle_{A}$ [cm], is 
\begin{equation}
    \centering
    \langle a\rangle_{\rm A} = \sqrt[3]{\frac{3}{4\pi}}\, \frac{L_3}{L_2},
    \label{eq:surf_size}
\end{equation}
with $L_{2}$ and $L_{3}$  the second and third dust moments \citep[Eq.A.1 in][]{helling2020mineral}. Further discussion of the mean particle size and the differing definitions can be found in Appendix A of \citet{helling2020mineral}. The column integrated, number density weighted, surface averaged mean particle size is 
\begin{equation}
\label{eq:aa}
\langle \langle a \rangle_{\rm A} \rangle = \frac{\int_{z_{min}}^{z_{max}} n_{d}(z)\langle a \rangle_{\rm A}(z) dz }
{ \int_{z_{min}}^{z_{max}} n_{d}(z) dz} \quad \mbox{with}\quad n_{\rm d}(z) = \frac{\rho(z) L_3(z)}{4\pi  \langle a(z)\rangle_{\rm A}^3/3}.
\end{equation}
The column-integrated properties are used to compare the cloud particle size  within the grid of 3D GCM model atmospheres.

All detailed results for  $\langle a \rangle_{A}$ as well as for the dust-to-gas ratio, $\rho_{\rm d}/\rho_{\rm gas}$,  are provided in the supplementary catalogue (\citealt{Lewis2022}).
A summary is given here as link for the understanding of the column integrated plots as well as for the comparison with the degree of thermal ionisation in Sect.~\ref{section:degree_of_ionisation}.

The  surfaced averaged mean particle size, $\langle a \rangle_{A}$, range from $10^{-2}\mu$m to $\approx 10^{4}\mu$m.
The largest particle sizes correlate with either low nucleation rates (in hot planetary atmospheres) or very high local densities (inside the planetary atmospheres). The local cloud particle distribution covers a larger volume of the planetary atmosphere than the nucleation rate due to transport processes like gravitational settling.

For all models with T$_{\rm eff, P} \geq 800$K, cloud particles on the dayside, where they exist, are on average larger than that of the nightside by 2-3 orders of magnitude. Cloud particles are generally not found on the dayside for models with T$_{\rm eff, P} \geq 2000$, except where the deep equatorial jet permits some cloud formation at 10$^{-2}$ bar.

Figure \ref{fig:mean_particle_size_integrate_scatter} shows the global distribution of the average particles sizes in terms of their column integrated, number density weighted values.
The results complement the nucleation rate results in Fig. \ref{fig:nuc_integrate_scatter}: In regions of low nucleation efficiency the average particle sizes are large. The range of integrated average particle sizes spans several orders of magnitude $\langle \langle a \rangle_{\rm A} \rangle \sim 10^{-3}-10^{5}\mu$m. The distribution follows the same `cone' like divergence structure of the average dayside particle size compared to the nightside which are consistently in the range $10^{-3}-10^{-1} {\rm \mu m}$.

For the colder models, the dust-to-gas mass ratio (Figs. 7 and 8 in te supplementary catalogue in \citealt{Lewis2022}) is lower across the dayside, especially near the morning terminator, hence, demonstrating the lower cloud formation efficiency in these atmospheric regions.  For models with T$_{\rm eff, P} \geq 1600$K, the dayside cloud formation is limited to the morning terminator regions on the dayside. This layer of cloud formation is more extended in the pressure scale for the slower rotators.
As the planetary effective temperature increases, this structure reduces in size until there is only limited dayside cloud formation near the evening terminator for the slower rotators and none whatsoever for the M star model. There are also inversions in the dust to gas mass ratio for the M star models with T$_{\rm eff, P}$=1600\,K-2000\,K

\begin{figure*}
    \centering
    \includegraphics[width=21pc]{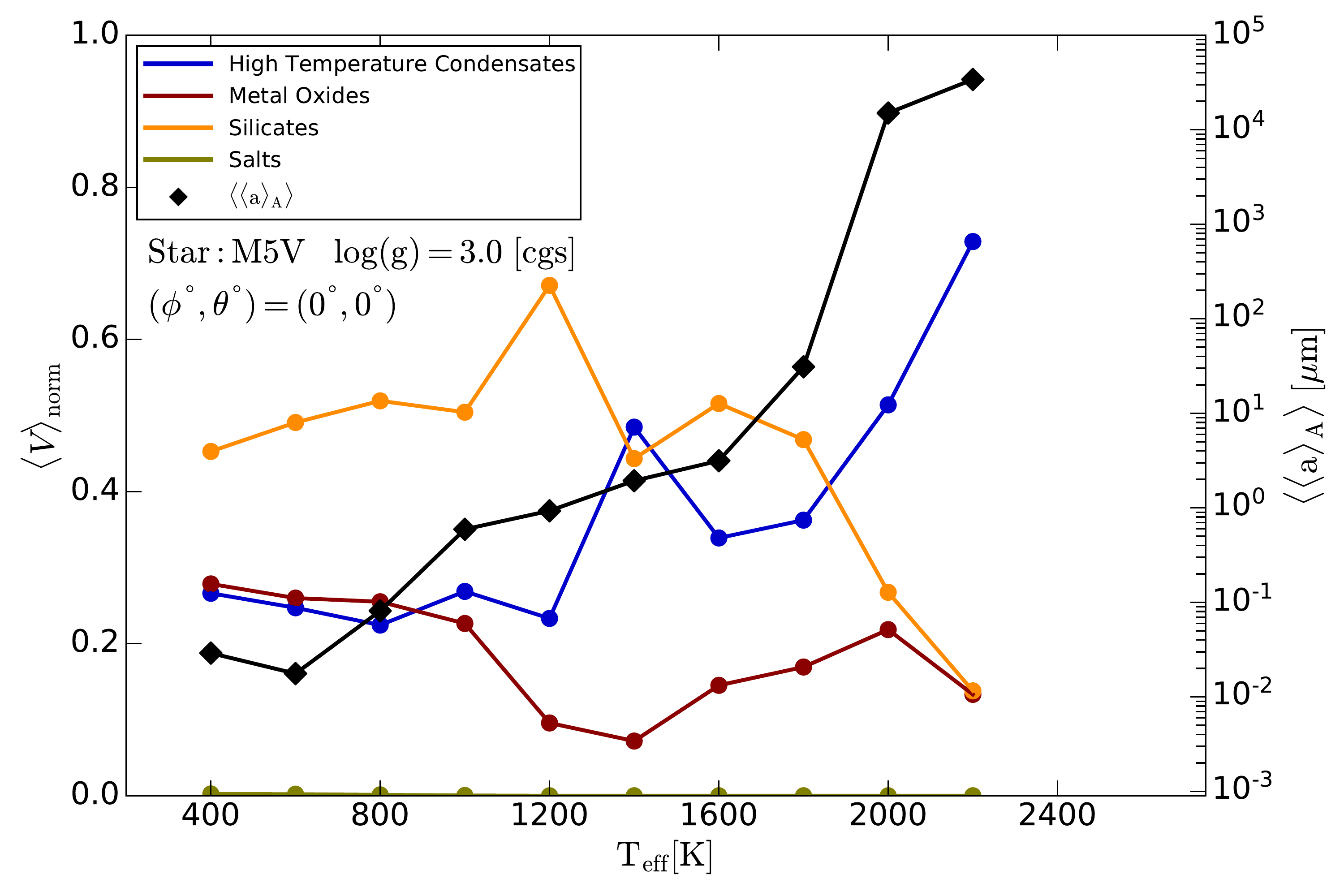}
    \includegraphics[width=21pc]{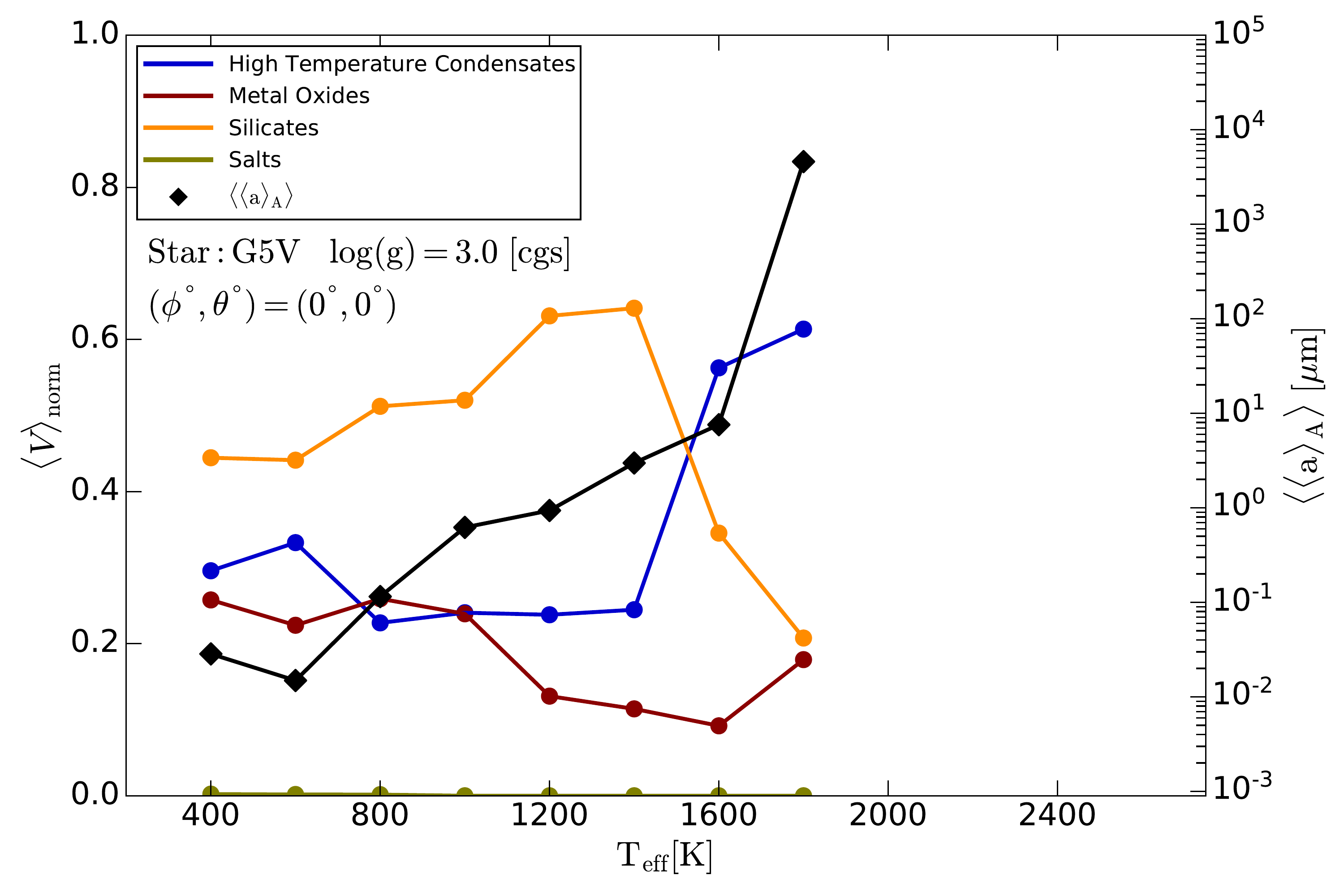}\\
    \includegraphics[width=21pc]{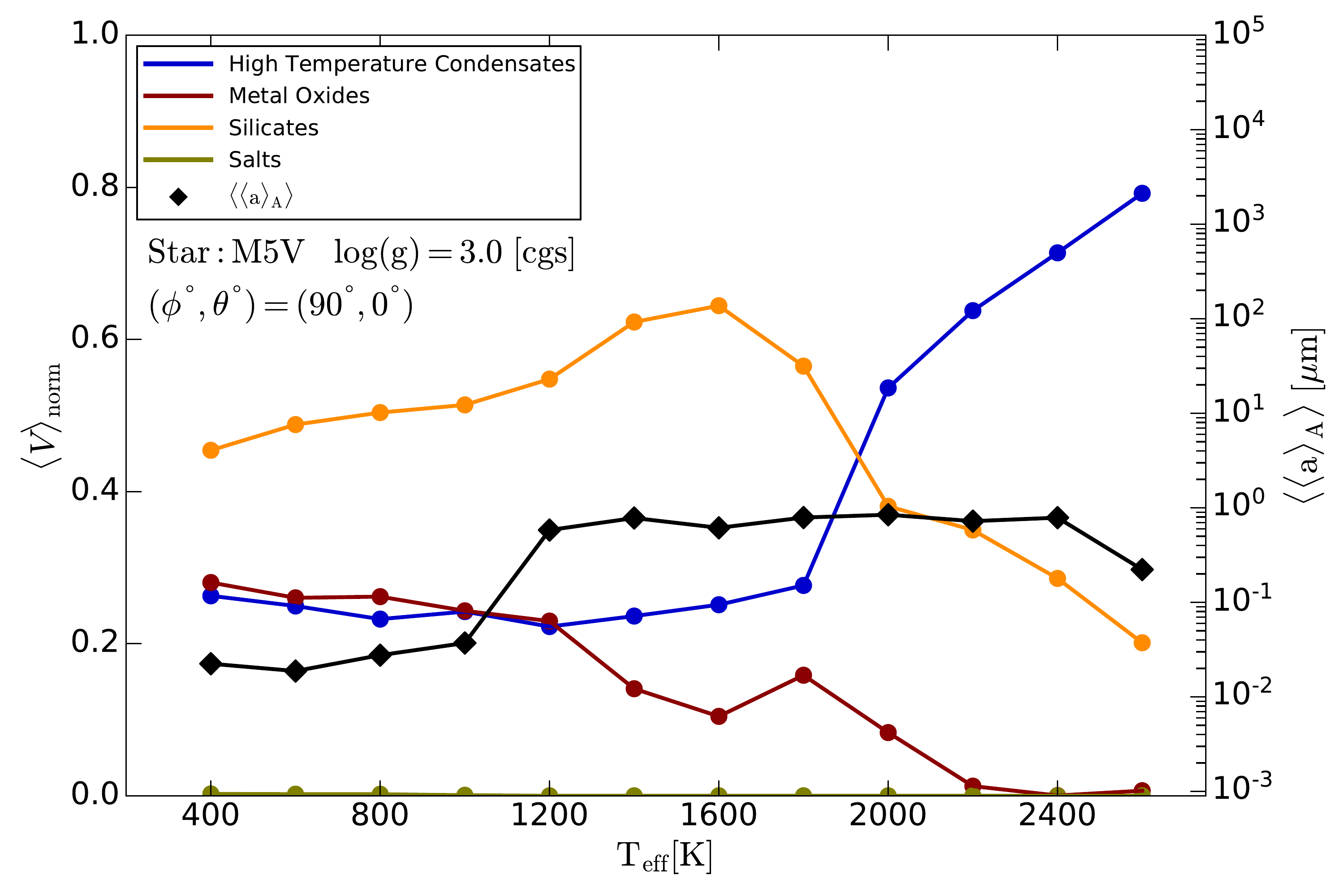}
    \includegraphics[width=21pc]{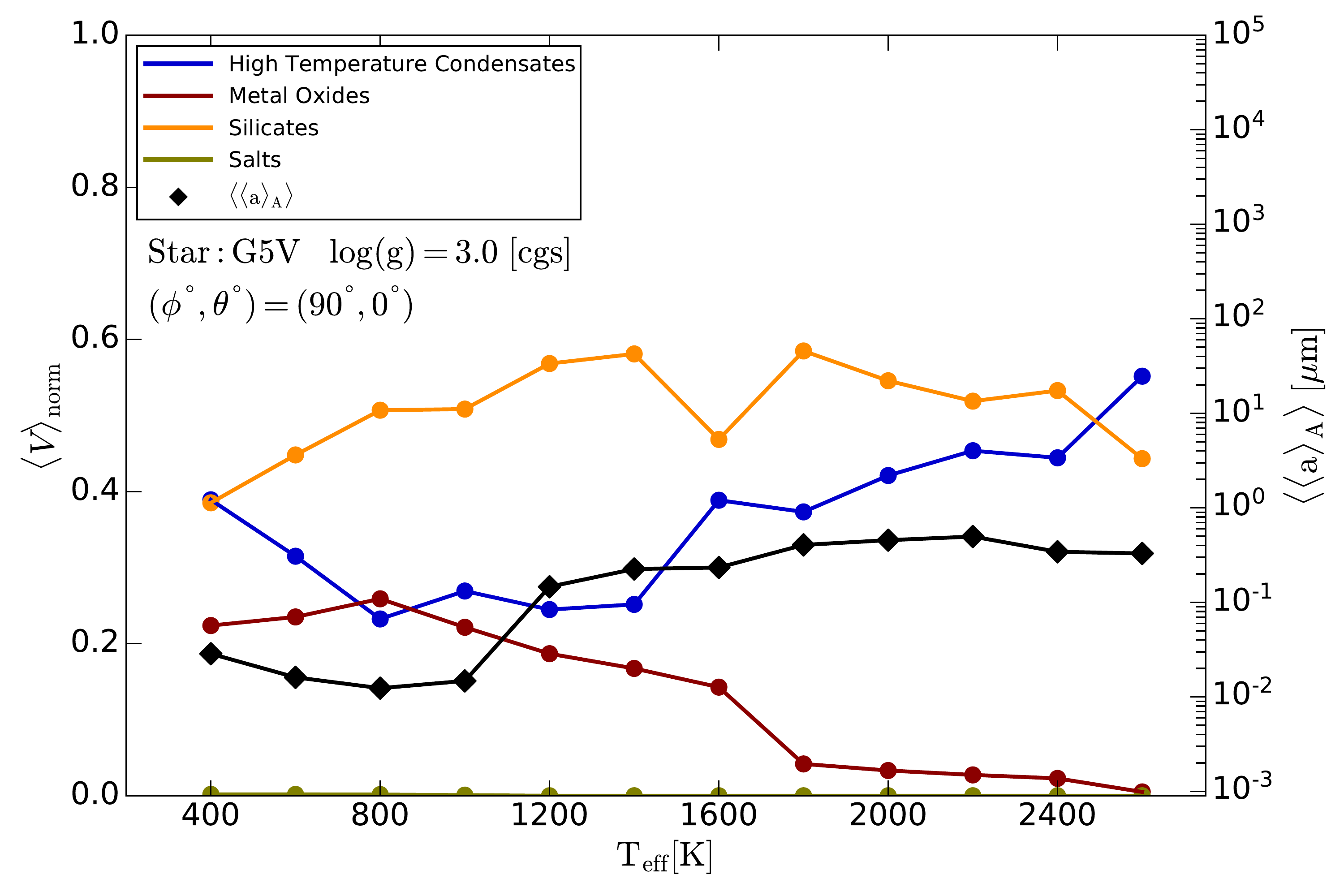}\\
    \includegraphics[width=21pc]{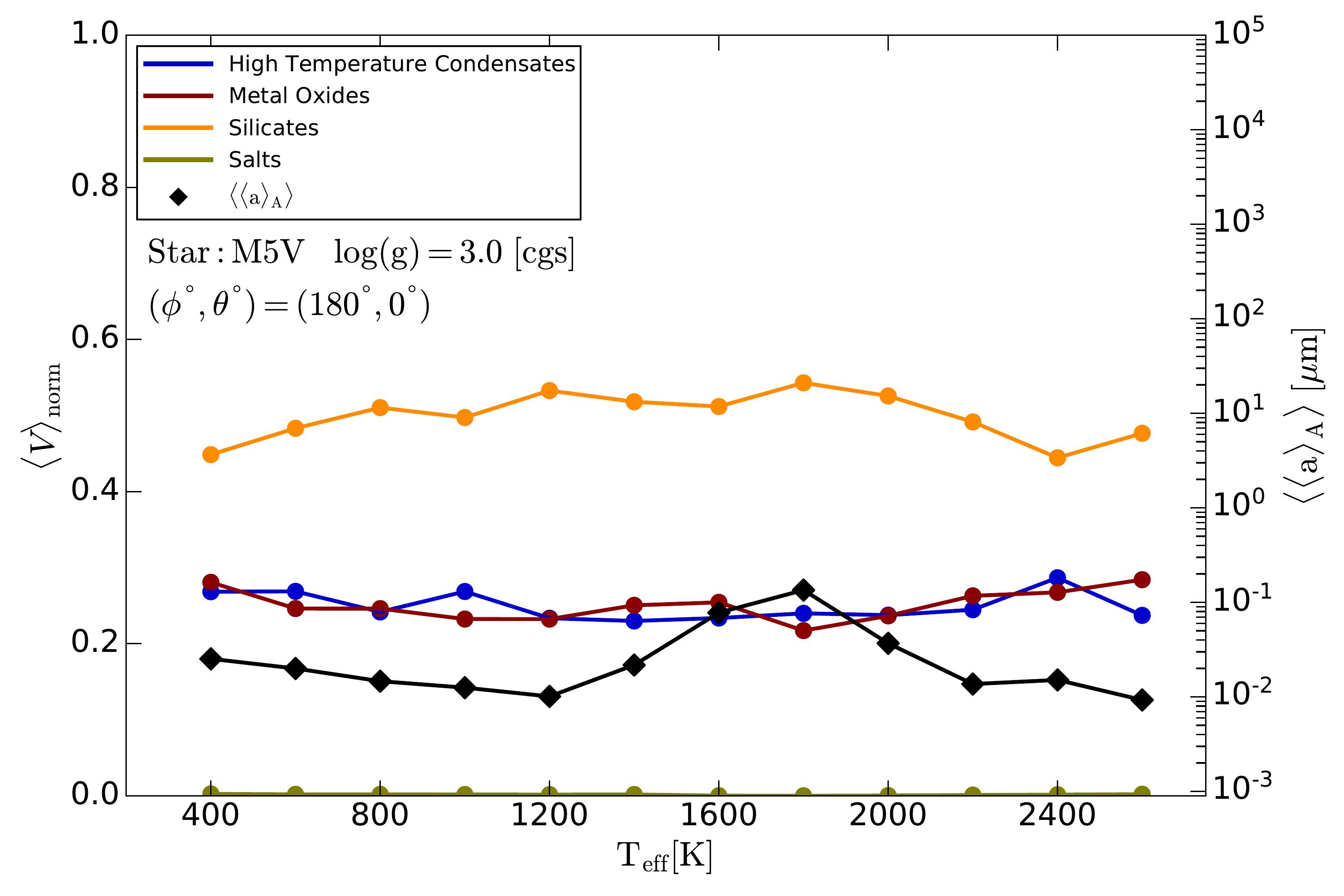}
    \includegraphics[width=21pc]{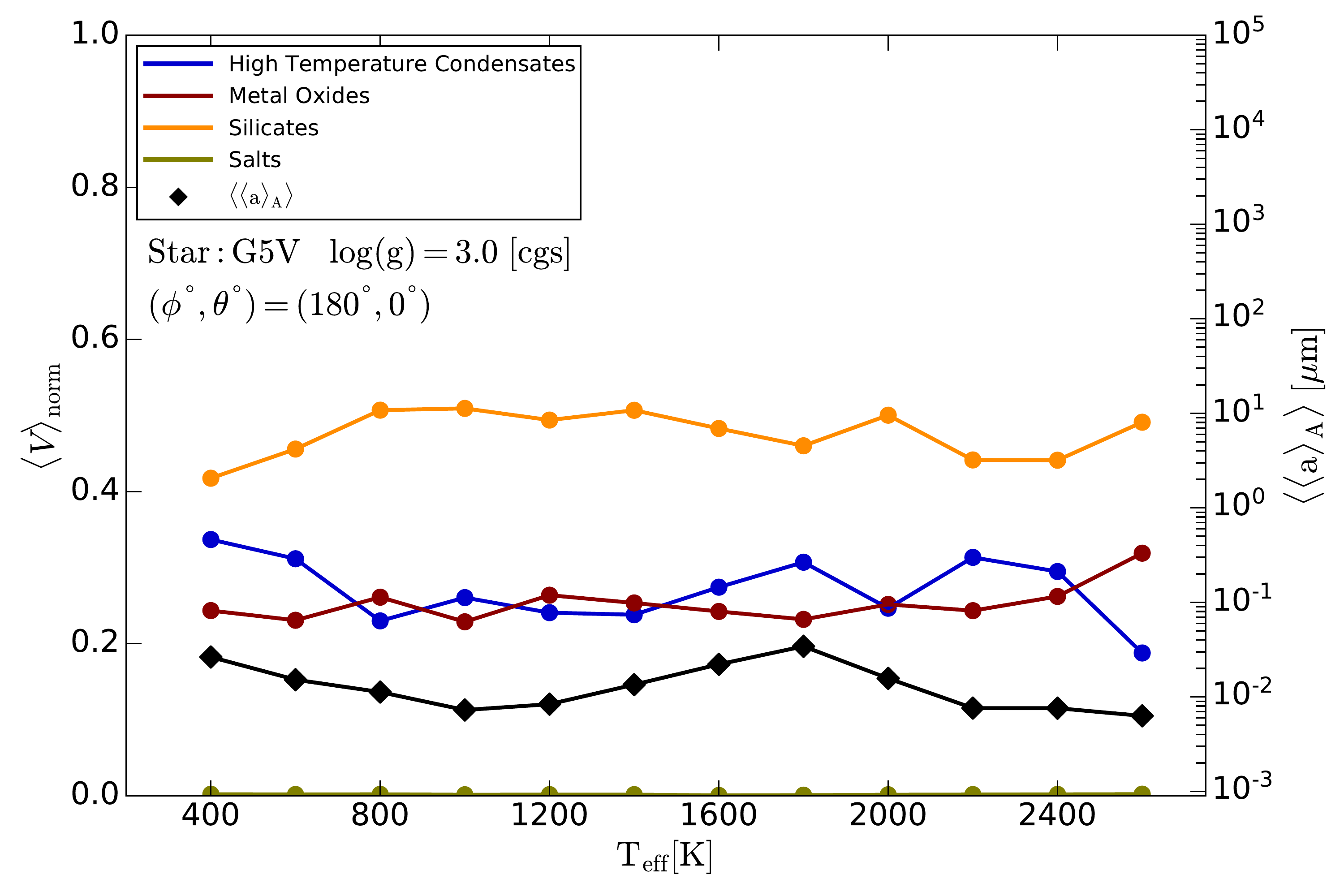}\\
    \includegraphics[width=21pc]{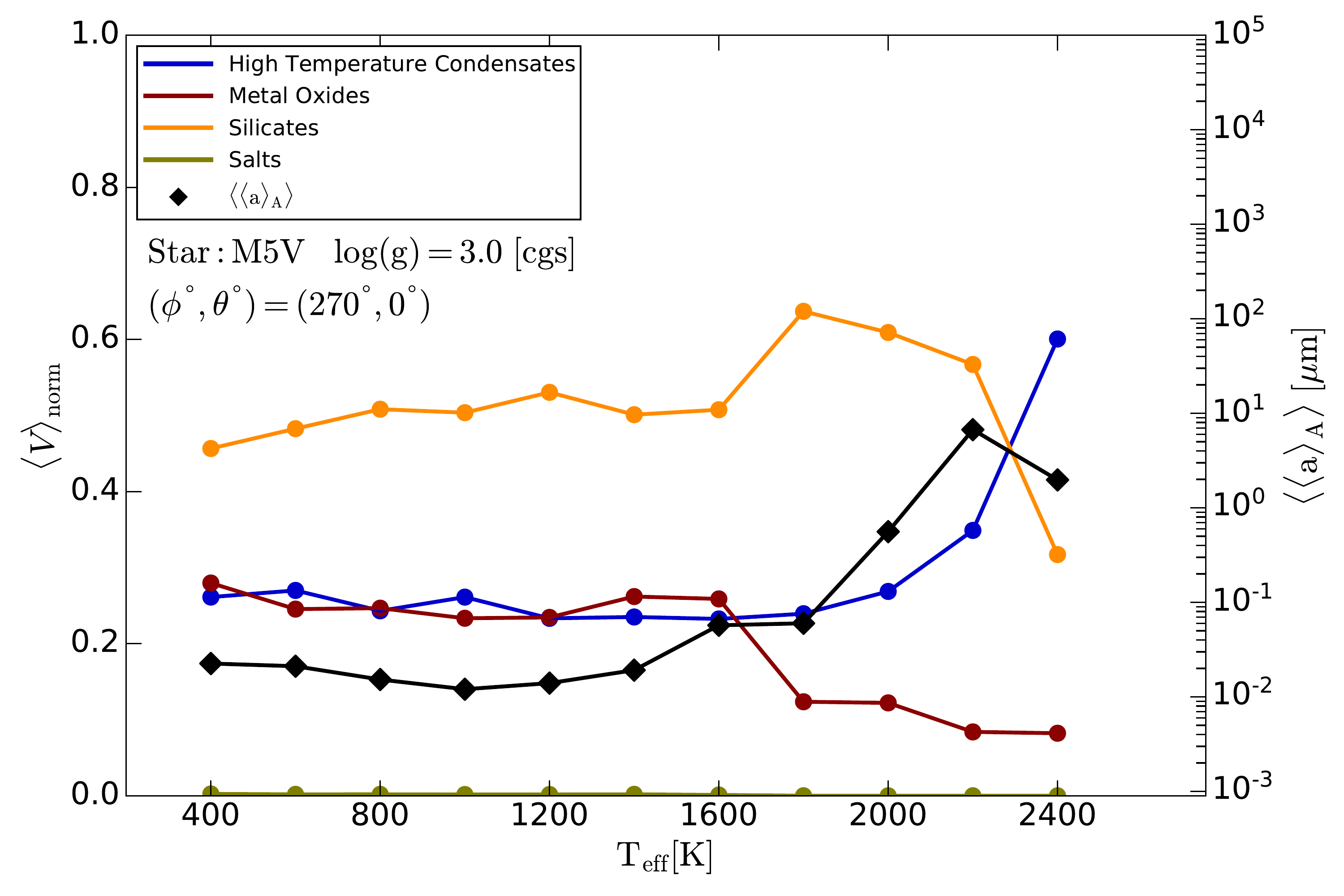}
    \includegraphics[width=21pc]{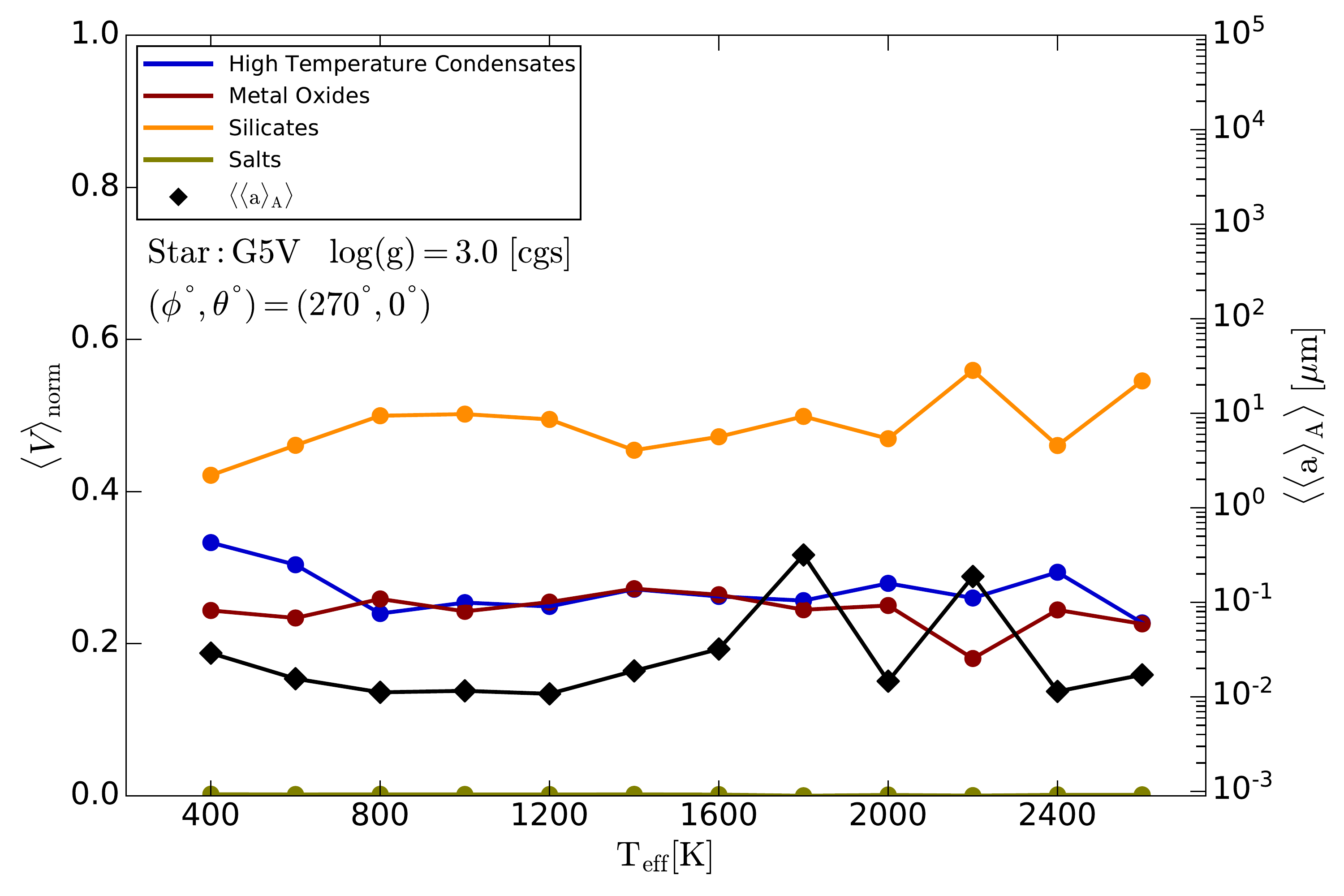}
   
    \caption{Normalised column integrated volume fractions and integrated number density weighted surface averaged mean particle size for the substellar point (\textbf{first row}), equatorial evening terminator (\textbf{second row}),  antistellar point (\textbf{third row}), and equatorial morning terminator (\textbf{fourth row}) for the M (\textbf{left}) and G (\textbf{right}) stellar types.}
    \label{fig:norm_col_vol_frac_w_grainsize_all_M_G}
\end{figure*}

\subsubsection{Material composition of cloud particles}
\label{section:cloud_material_compostion}

Another property used to characterise cloud  particles is the material composition of the cloud particles. It has been stated elsewhere that cloud particles change their material composition throughout their life time if they get transported into atmospheric regions of different thermodynamic conditions (for example, \citealt{Helling2006,2017A&A...603A.123H,2019AREPS..47..583H}). The global distribution of the individual materials for the whole 3D grid that we address here is included in the supplementary catalogue (\citealt{Lewis2022}).  Here  we explore the material composition of the cloud particles which form in the atmosphere in terms of material groups. The bulk growth of 16 condensate species is considered here, and for the purpose of extracting trends in types of material condensate, similarly to \cite{2021A&A...649A..44H}, we split the condensates into four groups: high temperature condensates,
metal oxides, silicates, and salts. The condensate species included and the group they are assigned to are shown in Table.~\ref{tab:vol_frac_type_table}. We choose to focus here on a comparison between the M and G stellar type models, with further discussion on the K and F stellar types and the equatorial distribution of material composition in \citet{Lewis2022}.

\begin{table}[h]
    \centering
    \begin{tabular}{p{3cm}|p{4cm}}
    \hline
         Condensate Group & Species Included\\
    \hline\hline
         Metal Oxides & SiO[s], SiO$_{2}$[s], MgO[s], FeO[s], Fe$_{2}$O$_{3}$[s] \\
         Silicates & MgSiO$_{3}$[s], Mg$_{2}$SiO$_{4}$[s], CaSiO$_{3}$[s], Fe$_{2}$SiO$_{4}$[s]\\
         High Temperature\newline Condensates & TiO$_{2}$[s], Fe[s], Al$_{2}$O$_{3}$[s], CaTiO$_{3}$[s], FeS[s]\\
         Salts & KCl[s], NaCl[s]\\
    \hline
    \end{tabular}
    \caption{The 16 bulk materials considered in our model are grouped in 4 categories. [s] indicates condensate materials.} 
    \label{tab:vol_frac_type_table}
\end{table}

Figure.~\ref{fig:norm_col_vol_frac_w_grainsize_all_M_G} shows the normalised column integrated volume fractions with s is a particular condensate species group,
\begin{equation}
 \langle V \rangle_{\rm norm}  = \frac{\int_{z_{min}}^{z_{max}} \frac{V_{s}(z)}{V_{tot}(z)} dz}{\sum_{i}  \int_{z_{min}}^{z_{max}} \frac{V_{i}(z)}{V_{tot}(z)} dz},
\label{eqn:V_norm}
\end{equation}
and $i$ runs through all the condensate species groups (Table \ref{tab:vol_frac_type_table}), with the integrated, number density weighted, surface averaged mean particle size (Eq.~\ref{eq:aa}) for the substellar, antistellar, equatorial morning and evening terminator of the models with  M and G stellar type host stars.

The antistellar points for both the M and G star are similar with small cloud particles ($\langle \langle a \rangle_{\rm A} \rangle \sim 10^{-2}-10^{-1} {\rm \mu m}$) which are dominated in composition by silicates ($\sim40-50\%$) forming for all model planets. The remaining volume is approximately equally distributed between the high temperature condensates and metal oxides (each $\sim20-25\%$). This is similarly the case for the morning terminator, excepting the highest temperature models, T$_{\rm eff} \geq 1800$ K, planets orbiting the M5V star where the silicates comprise $\sim55-60\%$ of the total volume, and the high temperature condensates dominate over the metal oxides by more than $10\%$. The highest temperature model where clouds form for the M star, T$_{\rm eff} = 2400$ K, has high temperature condensates as the largest contributor to the cloud particle load by volume at $\sim 60\%$, with the remaining volume comprised of silicates ($\sim 35\%$) and metal oxides ($\sim 5\%$). The average value for $\langle \langle a \rangle_{\rm A} \rangle$ increases gradually with increasing planet temperature for the M star planets, compared to an almost consistently small particle size for the G star.

At the substellar point the coolest planets exhibit a similar pattern of material composition as is found at the antistellar and morning terminator points, however, the range of values for the integrated number density weighted surface averaged mean particle size is larger here, $\langle \langle a \rangle_{\rm A} \rangle \sim 10^{-2}-10^{0} {\rm \mu m}$. For the hottest temperature planets (M: T$_{\rm eff} \geq 1800$ K, G: T$_{\rm eff} = 1800$ K) there are temperature inversions between $p_{\rm gas} = 10^{-1}...10^{0.5..1}$ bar which are sufficiently large to reduce the temperature such that TiO nucleation can occur and the supersaturation ratios of certain material condensates are large, $S \gg 1$. Of the species which can condense in these regions to form the bulk composition of the cloud particle Fe[s], Al$_{2}$O$_{3}$[s], CaTiO$_{3}$[s], Mg$_{2}$SiO$_{4}$[s], CaSiO$_{3}$[s] and SiO[s] are the most dominant. The nucleation is particularly inefficient for the daysides of the hottest models ($\log_{10}\left(J_{*,{\rm tot}}\right) \leq -13\,\ldots-15$ cm$^{-3}s^{-1}$) resulting in larger average particle sizes ($\langle\langle a \rangle_A\rangle \sim 10^{3.7}\ldots10^{5}\mu$m). Hence, clouds forming in the deeper atmosphere at the substellar point  are comprised of large particles  made of high temperature condensates and silicates. For models with T$_{\rm eff, P} > 1800$ K,  no cloud particles exist at the substellar point for the G star orbiting planets as the atmospheres are too warm to permit any cloud formation.

The planetary atmosphere clouds differ most significantly at the equatorial evening terminator. For the cooler planets the material composition is  similar to that of the other three points previously discussed, with silicates dominating and the high temperature condensates and metal oxides contributing equally to the remaining volume. With increasing planetary global temperature, T$_{\rm eff, P} \geq 1800$ K there is significant drop in the fraction of metal oxide condensates in favour of high temperature condensates. For the G star orbiting planets, silicates remain the dominant contributing species, excepting the hottest T$_{\rm eff, P} = 2600$ K model where high temperature condensates dominate. For the M star orbiting planets, however, the fraction of high temperature condensates increases and the fraction of silicates decreases with increasing model temperature. The average particle size  in the atmospheric clouds is similar between both stellar types.

The salts (KCl[s], NaCl[s]) do not contribute significantly to the average cloud particle volume composition at any point for any of the stellar types and planetary temperatures.

We note that carbonaceous materials have not been considered in the cloud formation model applied here because the undepleted element abundance was considered solar, hence, oxygen rich. The effect of element depletion/enrichment by cloud formation will be addressed in Sect.~\ref{section:chemical_regimes}. Carbon-rich materials in form of condensates (not hazes) have been discussed in the literature 
to study the effect of non-solar elemental abundances on the cloud structure and composition (\citealt{2017A&A...603A.123H,2017MNRAS.472..447M,2019AREPS..47..583H,2021arXiv211114144H}). The formation of hydrocarbon hazes (which are not condensates)  may be a potential explanation for radii that maybe 1.5-2$\times$ larger in the UV than in the optical (\citealt{2021AJ....162..287C}). This, however, requires to free the necessary carbon which is chemically blocked by CO and/or CH$_4$ in an oxygen-rich gas. While  photochemically triggered  hydrocarbon hazes can form in the upper atmosphere, their optical depth may not be large enough %
compared to mineral cloud particles  (\citealt{2020A&A...641A.178H}).

\medskip
Our systematic study has demonstrated that atmospheric thermodynamic structures lead to the formation of clouds that differ in particle sizes and numbers, and to a lesser extent, in their material composition. This finding is based on a cloud formation model in contrast to \cite{2021MNRAS.501...78P} who use particle sizes as parameters for their study of individual cloud materials. The suggestion of nightside clouds being similar for different planets promoted in \citet{Gao2021} can not be supported by our simulations, nor `the simple explanation that hot Jupiters all have the same species of nightside clouds' (\citealt{2019NatAs...3.1092K}).

\subsection{Changing  C/O regimes with changing global system parameters}
\label{section:chemical_regimes}

Cloud formation affects the local atmospheric chemistry in two ways. First, through the opacity feedback onto the temperature structure, and second, by element depletion of all elements that participate. Since cloud radiative feedback is not included in the GCMs that represent the input state for our cloud models, we focus here on the latter effect. Specifically,  we only refer to the effect on the oxygen abundance (which affects the carbon-to-oxygen ratio) and the mean molecular weight, and the major effects will be summarised below. The detailed spatial results can be found in the accompanying  catalog  (\citealt{Lewis2022}). Extensive studies regarding other elements have been presented in previous papers of our group.

\subsubsection{The carbon-to-oxygen ratio}
\label{section:C_to_O_ratio}

The carbon-to-oxygen ration (C/O) can be used as a fingerprint for the two nucleation regimes

of exoplanet atmospheres introduced in Sect.~\ref{ss:nuc}: the exoplanets that are covered in clouds homogeneously globally  (like HATS-6b) and those with a partial cloud coverage some of which feature strong day/night asymmetries (like WASP-18b).
These characteristic are:
\begin{itemize}
    \item the undepleted (e.g., solar)  C/O on cloud-free hot daysides,
    \item increased C/O ratio close to 0.75-0.8 
    where clouds form in the cool upper nightside atmosphere
    \item decreased C/O with increased pressure as cloud particles evaporate and replenish the gas phase with oxygen that was trapped in silicates and metal oxides. \end{itemize}
\noindent
The C/O threshold of  0.8 matches result of \citet{2021A&A...649A..44H} and  provides confirmation of findings by \citet{2020A&A...639A..36B} for hot and ultra-hot Jupiters to have C/O < 0.8.

{  C/O changes are a direct consequence of depletion of the oxygen by the nucleation and the surface growth. The carbon abundance is not affected as all planetary atmosphere considered here are assumed to be oxygen rich, hence, all carbon will be locked in CO (or CH$_4$). Mixing processes can only affect C/O (any local element abundance) if the respective processes are faster than the chemical processes involved in cloud formation.}

\begin{figure*}
{\ }\\*[-0.5cm]
   \includegraphics[width=0.97\textwidth]
    {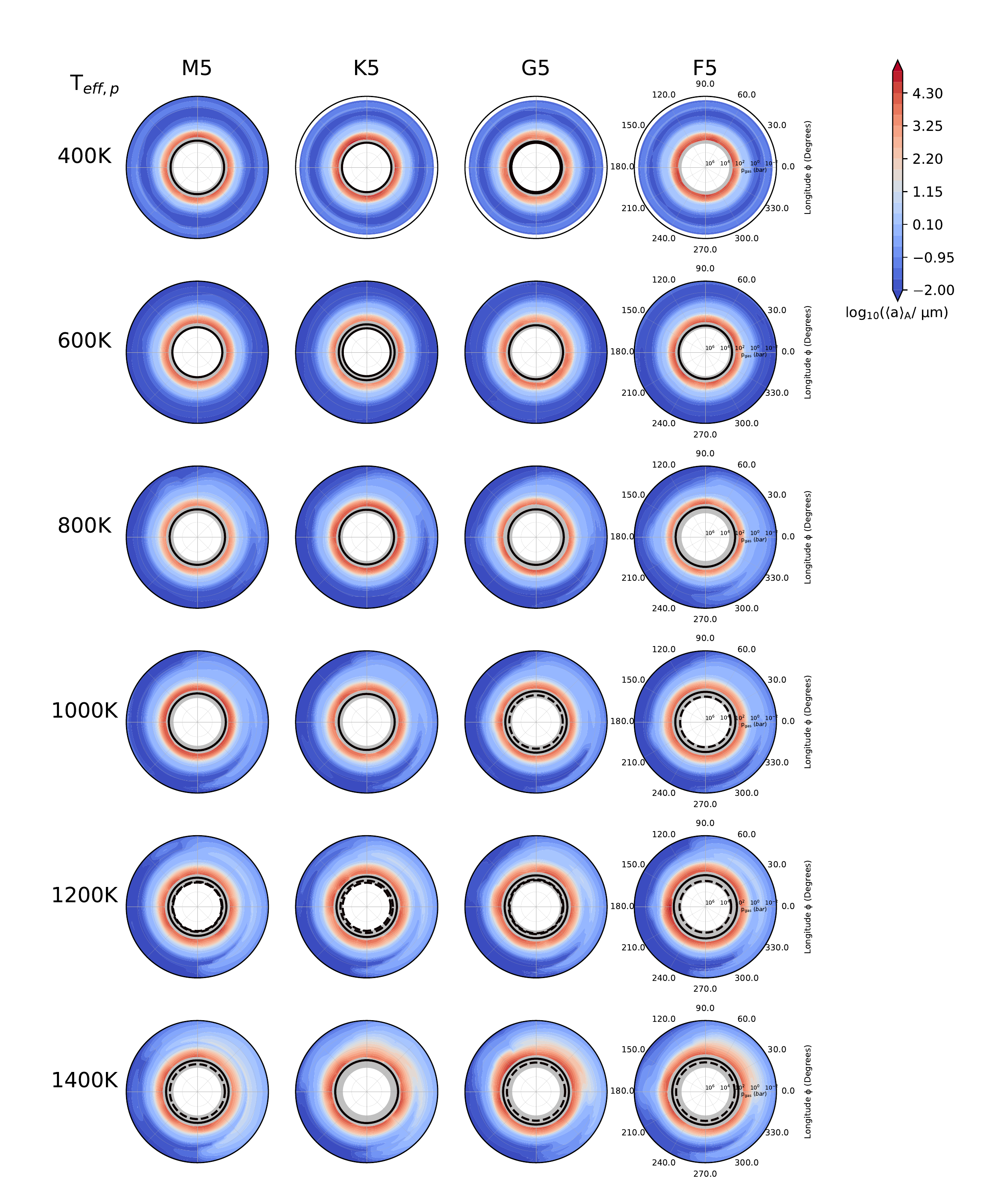}
    \caption{Surface averaged mean particles size, $\rm{log_{10}\left(\langle a \rangle_{A} /\ \mu m\right)}$,  in 2D  equatorial slices  ($\theta = 0\degree$) superimposed with degree of ionisation threshold of  f$_e$=10$^{-7}$ (solid; f$_e$=10$^{-6}$ is dashed line)  for the log$_{10}$g = 3 [cgs] 3D GCM models for.  T$_{\rm eff, P} = 400 \,\ldots\, 1400$ K for all stellar types. The outer limit for the pressure is fixed at 10$^{-3.5}$ bar for all models.}
   \label{fig:global_slice_plots_logg3_ionisation_rate_1}
\end{figure*}

\begin{figure*}
{\ }\\*[-0.5cm]   \includegraphics[width=0.97\textwidth]
    {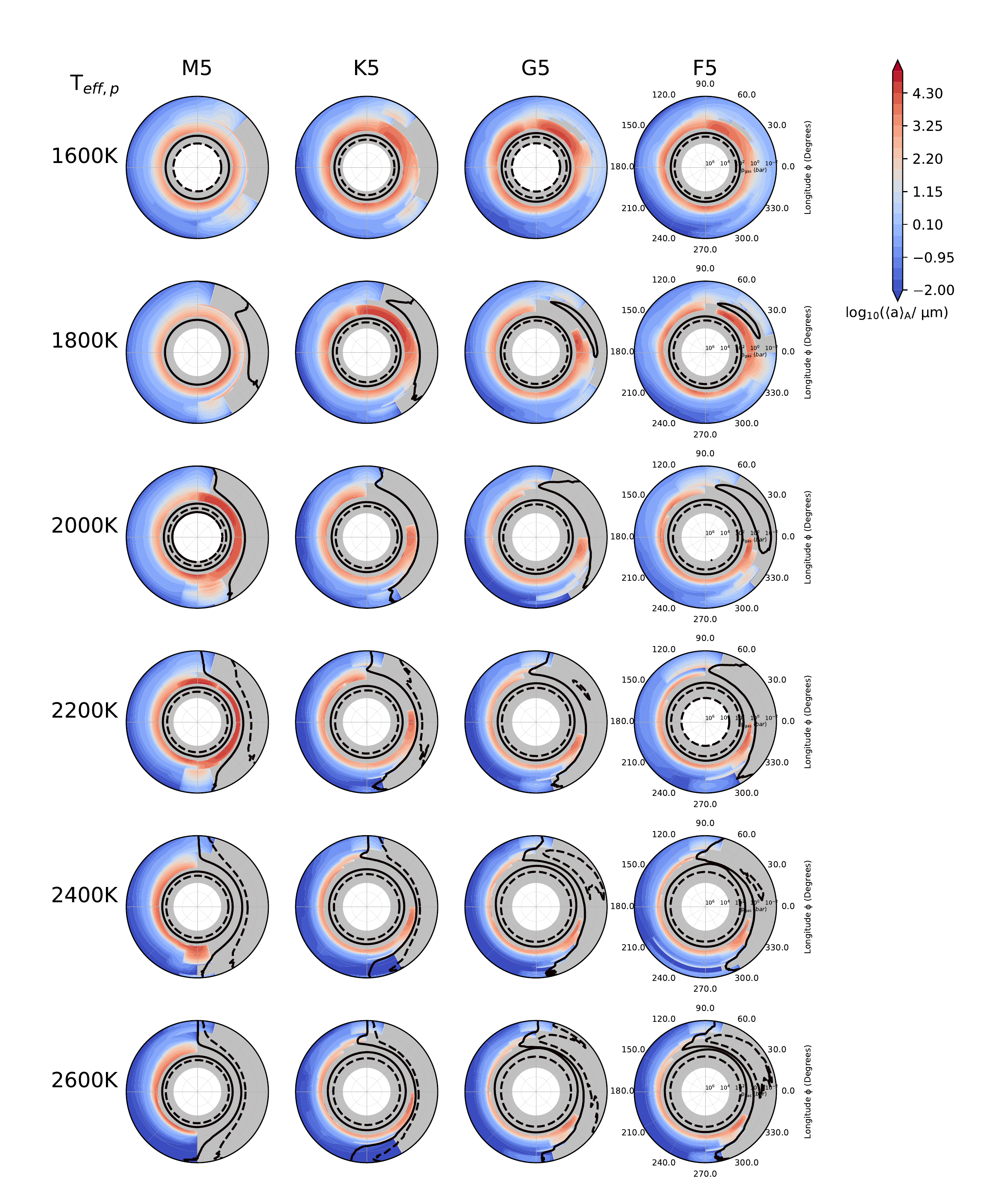}
    \caption{ Surface averaged mean particles size, $\rm{log_{10}\left(\langle a \rangle_{A} /\ \mu m\right)}$,  in 2D  equatorial slices  ($\theta = 0\degree$) superimposed with degree of ionisation threshold of  f$_e$=10$^{-7}$ (solid; f$_e$=10$^{-6}$ is dashed line)  for the log$_{10}$g = 3 [cgs] 3D GCM models for.  T$_{\rm eff, P} = 1600 \,\ldots\, 2600$ K for all stellar types. The outer limit for the pressure is fixed at 10$^{-3.5}$ bar for all models.}
   \label{fig:lobal_slice_plots_logg3_ionisation_rate_2}
\end{figure*}

\subsubsection{Mean molecular weight}
\label{section:mean_molecular_weight}

A constant mean molecular value,  $\mu$,  is typically assumed when running 3D GCMs to reduce the computational demands of the simulations \citep{Drummond2018a}. A constant mean molecular weight of 2.3 is likewise adopted in the GCMs underlying this study, in agreement with an atmospheric composition dominated by H$_2$ and He. However previous work has shown that this assumption may not be valid for all planets and generally for the whole atmosphere. We show in \citet{2021A&A...649A..44H} that in the case of ultra-hot Jupiters (e.g. HAT-P-7b, WASP-121b), where the day and nightside temperatures can differ by greater than 1500 K, there are substantial differences in the ionisation state of the atmospheric gas phase. The dayside is dominated  by atomic and ionised species compared to a nightside dominated by molecular species. 

Different classes of exoplanet atmosphere can therefore be expected {   be characterised by different}  mean molecular weight {   regimes}. In this paper, these classes are determined by the irradiation the planet receives and which affects the local thermal ionisation.

For the grid models with T$_{\rm eff, P} \leq 1600$ K the entire planet maintains a constant value of $\mu$ = 2.35 which is consistent with a molecular hydrogen dominated atmosphere. The nightside of all model planets has a constant $\mu = 2.35$. The hotter models (T$_{\rm eff, P}\geq1800$ K) show a decrease in $\mu$ only on the dayside initially, at the hottest region, offset from the substellar point. With increasing model temperature $\mu$ is decreased across more of the upper atmosphere on the dayside. The lowest value of mean molecular weight achieved is $\mu = 1.8$ and is centred around the substellar point in the upper atmosphere of the hottest T$_{\rm eff, P} = 2400, 2600$ K models which orbit the M and K stars. The maximum value of the decrease is $\mu = 1.8$ and is centred around the substellar point in the upper atmosphere of the hottest T$_{\rm eff, P} = 2400, 2600$ K models which orbit the M and K stars. For the G and F stars, $\mu \gtrsim 1.95$ across the dayside. 

\smallskip
 We conclude that it is reasonable to assume a constant mean molecular weight consistent with a H$_2$-dominated atmosphere for planets with T$_{\rm eff, P} \leq 1600$ K, such as warm Saturn or some hot Jupiter class planets, but not for hotter planets. The implication of the changing mean molecular weight due to the dissociation of H$_2$ is demonstrated in 
\cite{2021MNRAS.505.4515R}.

\subsection{Thermally driven ionospheres and the emergence of exoplanetary global electric circuits}
\label{section:degree_of_ionisation}

The degree of thermal ionisation ($f_{\rm e} = p_{\rm e}/\left(p_{\rm gas}+p_{\rm e}\right)$, where $p_e$ is the electron pressure) may be used to indicate plasma-like behaviour of the atmosphere, and by extension the potential for a thermally driven ionosphere to exist.  \citet{2015MNRAS.454.3977R} propose that values of f$_{e}\geq10^{-7}$ mark the transition between gas to plasma behaviour which is relevant for discussing the magnetic coupling of exoplanet atmospheres (e.g., \citealt{2022AJ....163...35B}). Here we are also interested in comparing the extension and location of the ionised part of the atmosphere to the location of the clouds. A net electron flux may be established that causes the  cloud particles to gain a net electrical charge in an ionised atmosphere, and more generally, a global electric circuit will establish if sufficient global background ionisation is available (\citealt{2019JPhCS1322a2028H}). 

Figures~\ref{fig:global_slice_plots_logg3_ionisation_rate_1} and \ref
{fig:lobal_slice_plots_logg3_ionisation_rate_2} present the thermal degree of ionisation in comparison to the global distribution  of the exoplanet clouds.  We utilize  the mean particle sizes in 2D equatorial slices for our comparison to regions of high ionisation based on thermal ionisation. The f$_{e}\geq10^{-7}$ threshold is shown as solid contour line and the location of f$_{e}\geq10^{-6}$ (dashed contour line) indicates the region where the thermal ionisation increases. 
The degree of ionisation resulting from thermal processes can be of the same magnitude as the degree of ionisation resulting from other processes like cosmic rays and UV radiation if the temperature is high enough (\citealt{Barth2021}).
The Lyman-continuum irradiation in star forming regions may be considerably higher than thermal effects in the very rarefied gases of the outer atmosphere of planets (\citealt{2018A&A...618A.107R}).

The degree of thermal ionisation never exceeds $f_{\rm e} = 10^{-7}$ in the upper atmosphere for any of the model planets where T$_{\rm eff, P} \leq 1600$ K. For T$_{\rm eff, P}=1800{\rm K}$, a dayside ionosphere emerges for all models, though it is more extended for the slower rotators. For the F, G and K star models, there is some overlap between the dayside cloud layers and the edge of the ionosphere, with the K star models having the most overlap. The most overlap for all stellar types occurs at T$_{\rm eff, P}$=2400\,K; beyond this planetary effective temperature the overlap decreases due to the reduction in the size of the dayside cloud layers.

\medskip

While the day/night asymmetry in $f_{\rm e}$ begins to appear at T$_{\rm eff, P}$=1000\,K, the thermal degree of ionisation does not reach ~10$^{-7}$ in the outer atmosphere until T$_{\rm eff, P}$=1400\,K, but only for the M and the F stars. Larger cloud particles  ($\langle a \rangle_{\rm A}\geq10^{2.2}\mu$m) form at pressures below this level of ionisation ($\leq10^{-2}$ bar).  Such cool  exoplanets are therefore unlikely to have a thermally driven ionospheres. 
At T$_{\rm eff, P}$=1600\,K and T$_{\rm eff, P}$=1800\,K, the ionisation level exceeds 10$^{-7}$ above 10$^{-1}$ bar across the dayside for the M and the K star exoplanets. However, there is no cloud formation in these regions in the M and the K star exoplanet atmospheres. The G and the F star exoplanets with  T$_{\rm eff, P}$=1600\,K form medium sized cloud particles in their atmospheres  ($\langle a \rangle_{A}\geq10^{0.8}\mu m$) in regions (near the evening terminator) where f$_e\geq10^{0.7}$. A electron flux induced cloud particle charging may occur in these atmospheres.
Above T$_{\rm eff, P}$=1800\,K, what little clouds remain form in regions where f$_{\rm e}$<10$^{-7}$.

It is not clear yet what effect the dayside ionization has on the wind flow and thus the 3D thermodynamics of utra-hot Jupiters. \citet{Tan2019} propose that reduced molecular mean weight at the dayside, which we neglect in the 3D GCM here, leads to larger wind speeds. Further, the thermal effect of hydrogen dissociation at the dayside and re-combination at the nightside reduces the horizontal temperature gradient. The first was proposed to lead to dayside cooling and the latter to nightside warming. However, \citet{Tan2019} did not consider cloud formation and their results seem to contradict \citet{2020A&A...639A..36B}. The latter study find a particularly large dayside emission, which the authors attribute to low dayside albedo and inefficient heat circulation, as we also re-produce in this study.

\citet{Beltz2022} suggest that coupling between the ionized wind and magnetic fields of 3 G can disrupt wind jets at the dayside completely, in particular for the upper atmosphere ($p<0.01$~bar), leading instead to equatorial-to-polar flow, greatly diminishing heat circulation between nightside and dayside. It is not clear yet how such a flow regime may affect vertical transport and cloud formation at the nightside.

In any case, it is quite clear that the daysides of ultra-hot Jupiters are fundamentally different from those of colder planets and that further work is needed here to study wind flow and cloud formation in the context of dayside ionization and magnetic field coupling. Further implications of atmospheric ionisation are discussed in Sect.~\ref{ss:asymion}.

\smallskip
We conclude that all gaseous exoplanets can be expected to have a thermally ionised inner atmosphere for pressure $\gtrsim 10$bar where, therefore, magnetic coupling of the atmosphere can occur. Exoplanet atmospheres with T$_{\rm eff, P}>2000{\rm K}$ can be expected to have deep thermally driven ionospheres on their dayside. This ionsophere reached into the terminator regions the hotter the planetary atmosphere is. We therefore conclude that these exoplanets have a) a geometrically more extended, cloud-free dayside compared to their nightside which b) can undergo magnetic coupling to a global planetary magnetic field should it exists, and c) that a global electric circuit {  may determine} the charge distribution within the atmosphere longitudinally, i.e. east-west-ward. Both, the global and local extension of the ionosphere and the global electric circuit will increase if additional ionisation processes occur. We further conclude that both regimes of exoplanet atmospheres, those with a globally homogeneous cloud coverage and those with an intermittent or asymmetric cloud coverage will undergo different degrees of magnetic coupling inside their atmospheres and, hence, may exhibit different magnetic field geometries.

\begin{figure*}
{\ }\\*[-0.5cm]
   \includegraphics[width=\textwidth]
    {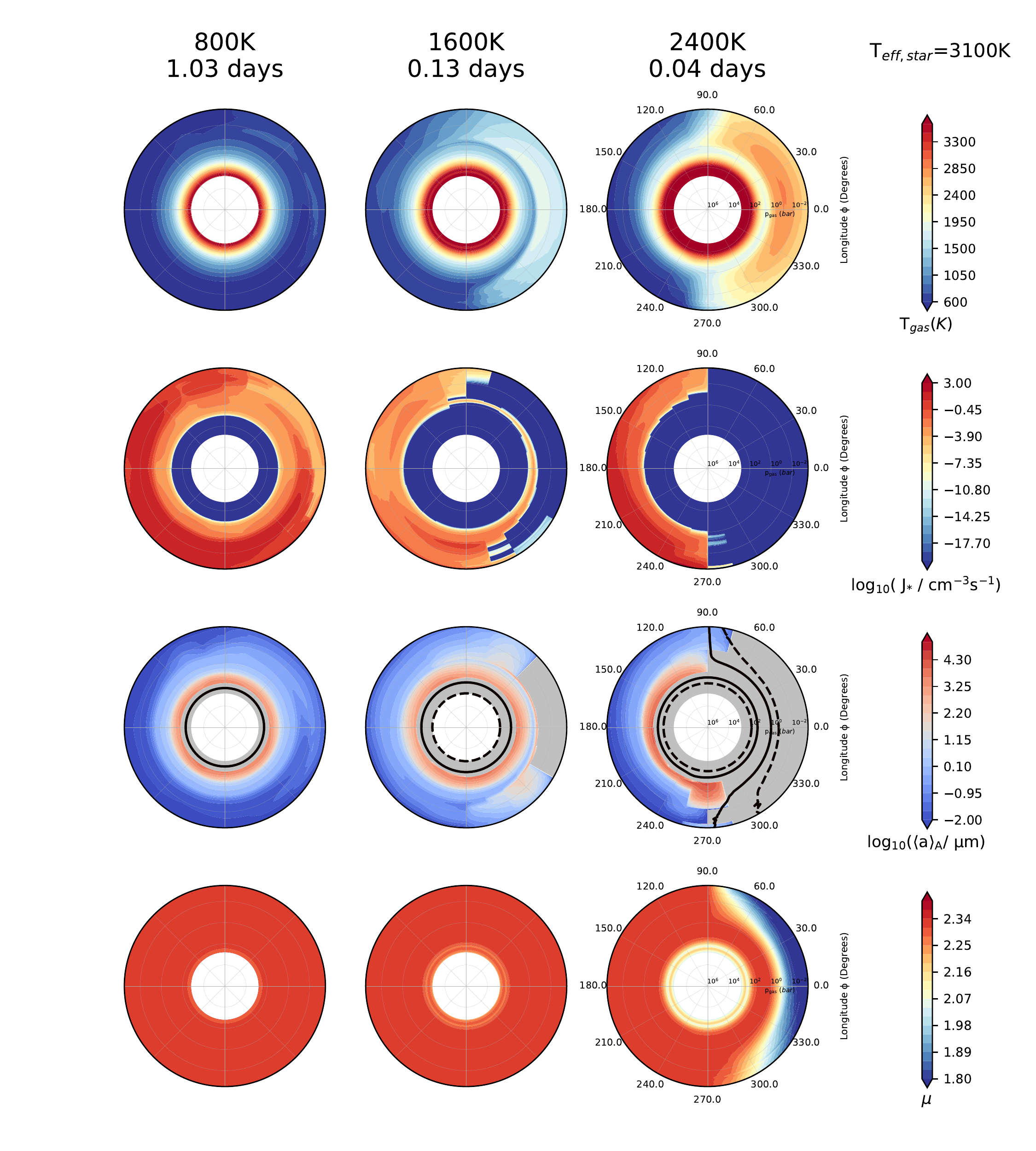}
    \caption{Cool (T$_{\rm eff, P}=800$K; e.g. HATS-6b, NGTS-1b), the transient (T$_{\rm eff, P}=1600$K; e.g. WASP-43b, NGTS-10b), the hot (T$_{\rm eff, P}=2400$K; brown dwarfs WD 0137 and EPIC 2122) exoplanet atmospheres (log$_{10}$g = 3 [cgs]) with an M-dwarf host star. 2D  equatorial plane slices ($\theta = 0\degree$) for the 1D profiles extracted from the 3D GCM models (\citealt{2021MNRAS.tmp.1277B}) show the main cloud formation properties:  local gas temperature [K] (first row), total nucleation rate [$\rm{log_{10}(\ J_{*}\ /\ cm^{-3} s^{-1})}$] (second row), surface averaged mean particle size [$\rm{log_{10}\left(\langle a \rangle_{A} /\ \mu m\right)}$] overlayed with f$_e$=10$^{-7}$ (solid line; f$_e$=10$^{-6}$ (dashed); third row),  mean molecular weight (fourth row).  The outer limit for the pressure is fixed at 10$^{-3.5}$ bar for all models.    }
   \label{fig:global_slice_plots_main_3100}
\end{figure*}

\begin{figure*}
{\ }\\*[-0.5cm]
   \includegraphics[width=\textwidth]
    {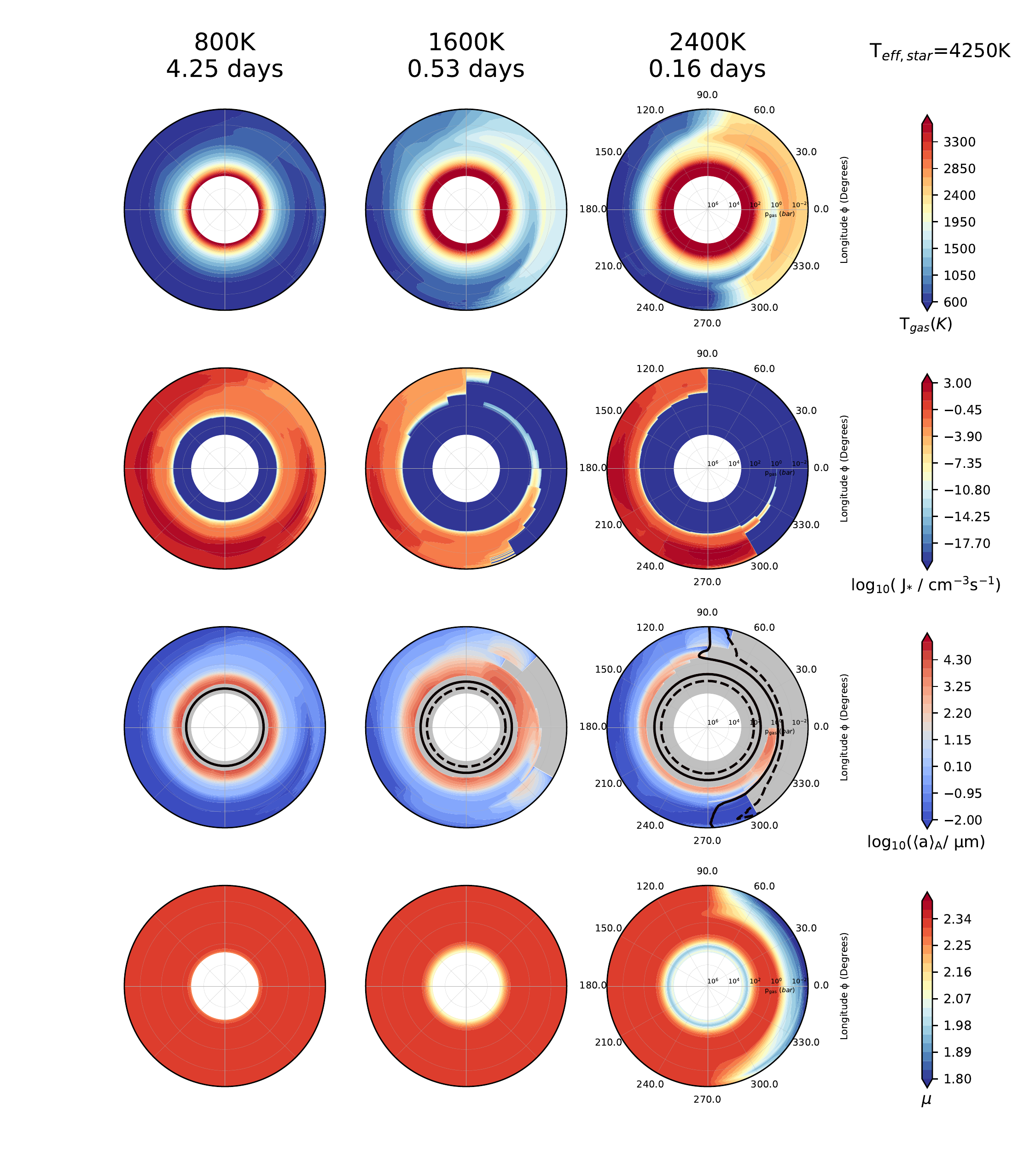}
    \caption{Cool (T$_{\rm eff, P}=800$K), the transient (T$_{\rm eff, P}=1600$K), the hot (T$_{\rm eff, P}=2400$K) exoplanet atmospheres (log$_{10}$g = 3 [cgs]) with an K-dwarf host star. 2D  equatorial plane slices ($\theta = 0\degree$) for the 1D profiles extracted from the 3D GCM models (\citealt{2021MNRAS.tmp.1277B}) show the main cloud formation properties:  local gas temperature [K] (first row), total nucleation rate [$\rm{log_{10}(\ J_{*}\ /\ cm^{-3} s^{-1})}$] (second row), surface averaged mean particle size [$\rm{log_{10}\left(\langle a \rangle_{A} /\ \mu m\right)}$] overlayed with f$_e$=10$^{-7}$ (solid line; f$_e$=10$^{-6}$ (dashed); third row),  mean molecular weight (fourth row).  The outer limit for the pressure is fixed at 10$^{-3.5}$ bar for all models.}
   \label{fig:global_slice_plots_main_4250}
\end{figure*}

\begin{figure*}
{\ }\\*[-0.5cm]
   \includegraphics[width=\textwidth]
    {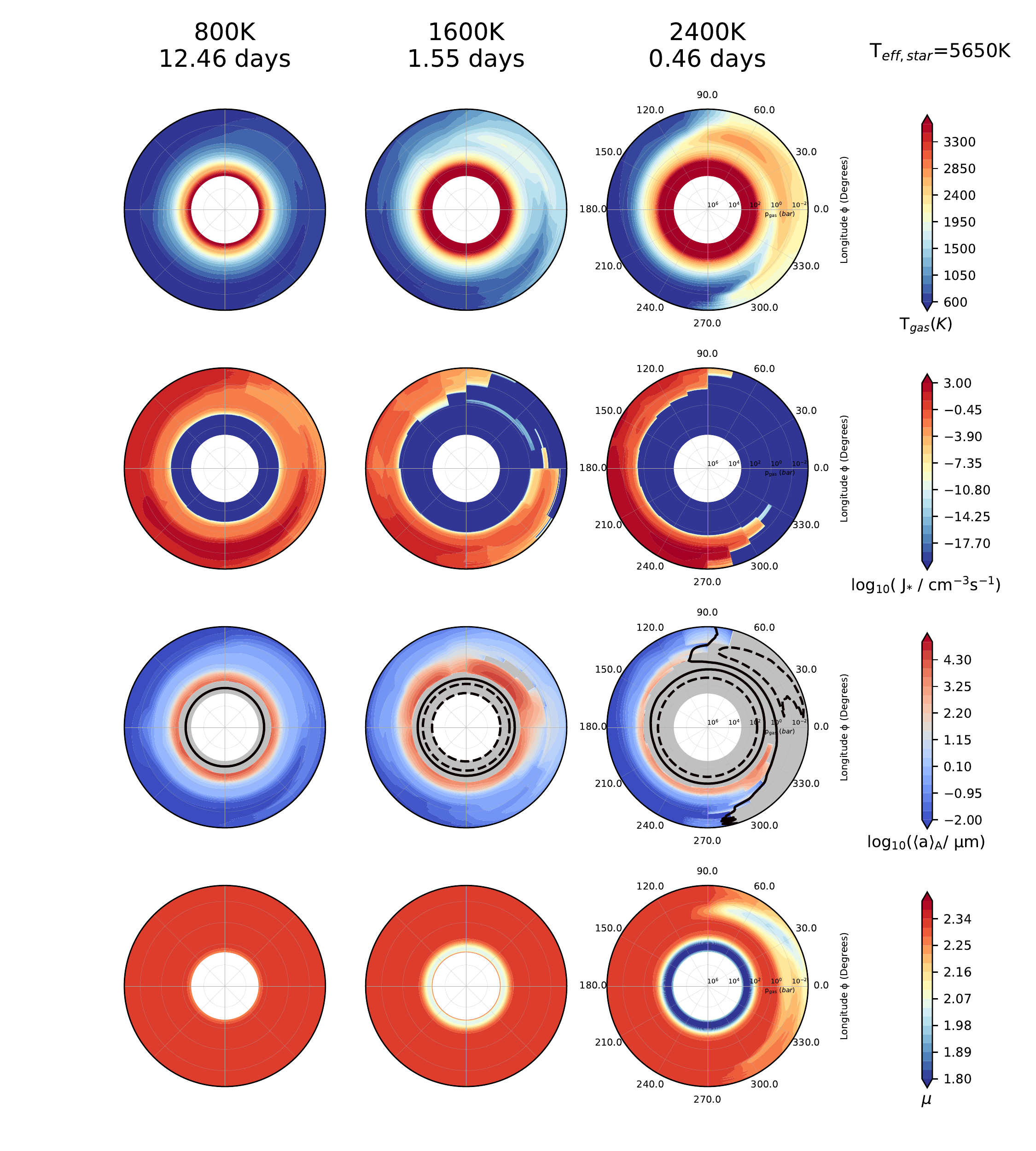}
    \caption{Cool (T$_{\rm eff, P}=800$K), the transient (T$_{\rm eff, P}=1600$K; e.g., HD\,209458b), the hot (T$_{\rm eff, P}=2400$K) exoplanet atmospheres (log$_{10}$g = 3 [cgs]) with a G-type host star. 2D  equatorial plane slices ($\theta = 0\degree$) for the 1D profiles extracted from the 3D GCM models (\citealt{2021MNRAS.tmp.1277B}) show the main cloud formation properties:  local gas temperature [K] (first row), total nucleation rate [$\rm{log_{10}(\ J_{*}\ /\ cm^{-3} s^{-1})}$] (second row), surface averaged mean particle size [$\rm{log_{10}\left(\langle a \rangle_{A} /\ \mu m\right)}$] overlayed with f$_e$=10$^{-7}$ (solid line; f$_e$=10$^{-6}$ (dashed); third row),  mean molecular weight (fourth row).  The outer limit for the pressure is fixed at 10$^{-3.5}$ bar for all models.
    }
   \label{fig:global_slice_plots_main_5650}
\end{figure*}

\begin{figure*}
{\ }\\*[-0.5cm]
   \includegraphics[width=\textwidth]
    {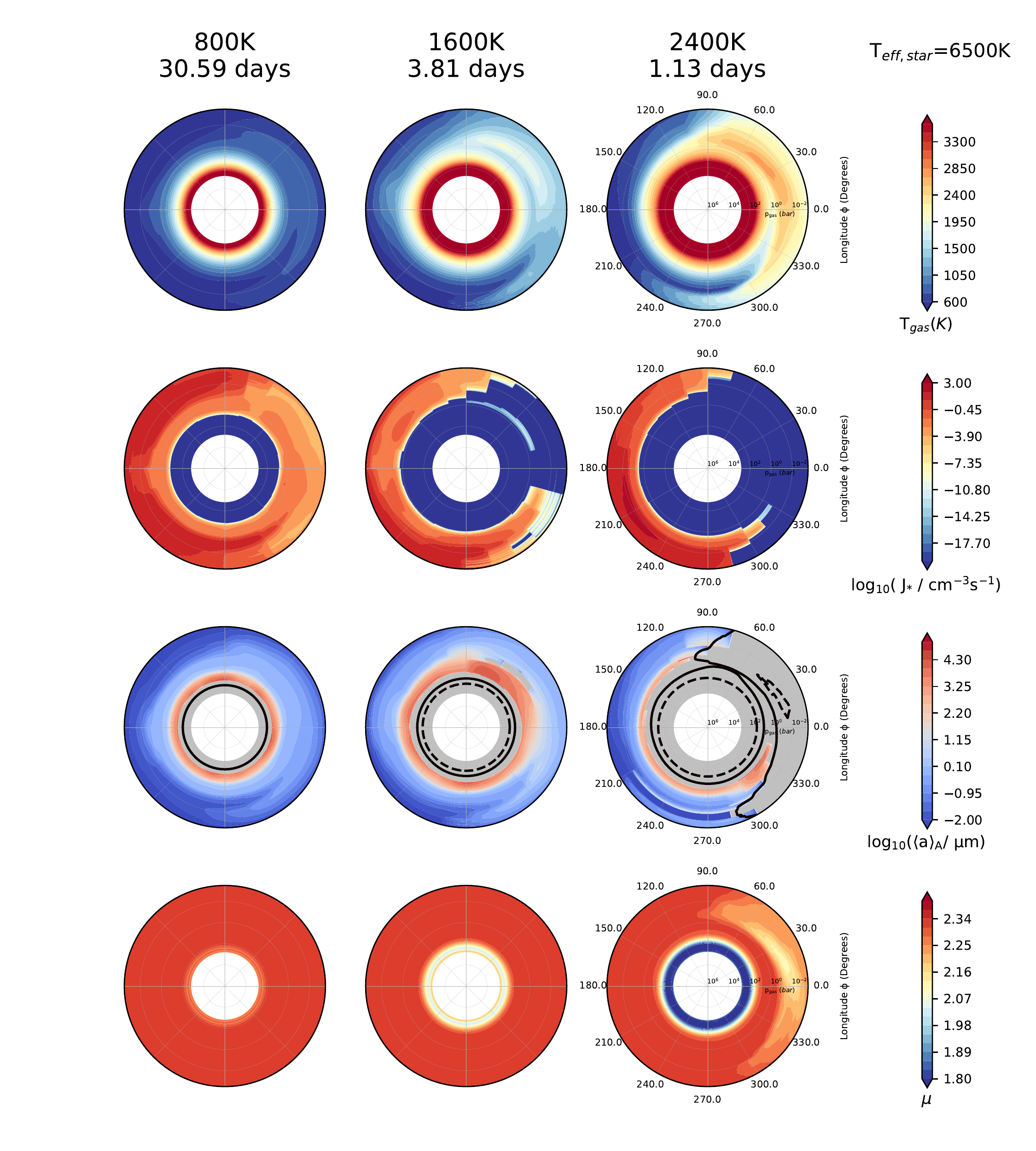}
    \caption{Cool (T$_{\rm eff, P}=800$K), the transient (T$_{\rm eff, P}=1600$K), the hot (T$_{\rm eff, P}=2400$K) exoplanet atmospheres (log$_{10}$g = 3 [cgs]) with an F type host star. 2D  equatorial plane slices ($\theta = 0\degree$) for the 1D profiles extracted from the 3D GCM models (\citealt{2021MNRAS.tmp.1277B}) show the main cloud formation properties:  local gas temperature [K] (first row), total nucleation rate [$\rm{log_{10}(\ J_{*}\ /\ cm^{-3} s^{-1})}$] (second row), surface averaged mean particle size [$\rm{log_{10}\left(\langle a \rangle_{A} /\ \mu m\right)}$] overlayed with f$_e$=10$^{-7}$ (solid line; f$_e$=10$^{-6}$ (dashed); third row),  mean molecular weight (fourth row).  The outer limit for the pressure is fixed at 10$^{-3.5}$ bar for all models.     }
   \label{fig:global_slice_plots_main_6500}
\end{figure*}

\section{Exoplanet weather/climate regimes: the cool, the transient and the hot exoplanet atmospheres} \label{sec:3cases}

The previous sections have presented a global picture of cloud formation in gaseous extrasolar planets that orbit M, K, G and F host stars at different distances. Most of the details for individual models, like the individual nucleation rates or material compositions of the cloud particles, have been left to a supplementary catalogue. Now we provide a comparison of three groups of exoplanets of the same planetary effective temperature that orbit different host stars at  different distances.

A closer inspection of the 3D GCM model grid shows that the planetary model atmospheres  fall into three cloud formation regimes, which we call classes, separated by planetary temperature: class i) \textit{the cool planets} (T$_{\rm eff, P} \leq 1200$ K), class ii) \textit{the transition planets}  (T$_{\rm eff, P} = 1400 - 1800$ K), and class iii) \textit{the hot planets} (T$_{\rm eff, P} \geq 2000$ K). The temperature thresholds should  be considered as approximate and may shift by $\pm200$K due to modelling uncertainties and additional parameter dependencies (e.g., log(g)). These cloud forming regimes are representative of weather regimes on short timescales but also  of climate regimes if understood for longer time scales. Example objects for class (i) include HATS-6b and  NGTS-1b, for class (ii) WASP\,43b, NGTS-10b and HD\,209458b, and for class (iii) WASP-18b, WASP-121b, WASP-103b and also brown dwarfs in close orbits around white dwarfs like WD\,0137b and EPIC\,2122B. {  We note here that, young objects excepted, brown dwarfs may have a stronger internal heating than giant gas planets at a similar age. The overall trend, however, can be expected to be very similar with respect to the large day/night difference in cloud presence, the far stronger ionisation on the day side, as well as the lower mean molecular weight in the night side. This hypothesis is supported by the finding that \cite{2021MNRAS.tmp.1277B}
shows that the mentioned BD-WD pairs are in general agreement in circulation regimes with \cite{2020MNRAS.496.4674L, 2021MNRAS.502.2198T}}.

To provide an overview of each of these three regimes we select a specific temperature planet model which is representative of the whole regime, these are T$_{\rm eff, P} = 800, 1600, 2400$ K. The transition and hot planet temperatures line up with the approximate temperature space occupied by the hot and ultra-hot Jupiter class planets \citep[Table. 1 in][]{2021A&A...649A..44H}. The cool planets represent the homogeneous regime from previous sections, while the transient and the hot planets represent the regime of intermittent cloud formation. In the following, we will confirm  that the intermittent regime is the best suitable to look for morning/evening terminator differences (\citealt{2021MNRAS.tmp.1277B}) which is greatly amplified by the cloud distribution

\subsection{Asymmetric cloud coverage}
The complex interplay between dynamics and irradiation is reflected by the asymmetric cloud coverage. Further, for a given global (equilibrium or effective)  temperature,  planets around K dwarfs are  substantially faster rotators than  G dwarf planets. Thus,  K dwarf planets tend to have more asymmetric and smaller dayside cloud coverage compared to G dwarf planets of the same global temperature. This is particularly apparent for $T_{\rm eff, P}= 1200\,\ldots\, 1600$ K (classes i) and ii)), of which several will be observed by JWST to investigate differences in evening/morning clouds (WASP-63b and HD 189733b).  We note that the cloud layers will be able to extend into deeper atmospheric layers for planets with higher masses where the increasing pressure increases the thermal stability of the cloud particles despite the increasing local temperature. This effect can be manipulated for low-mass planets by the choice of the inner boundary for the computational domain in 3D GCMs (see Sect. 6 in \citealt{2021A&A...649A..44H}).

\paragraph{Nightside cloud coverage:} In all three classes (i, ii, iii), clouds do form on the nightside. This is an essential result  because the greenhouse feedback of the nightside clouds will reduce the heat redistribution in the atmosphere. This affects the atmospheric temperature structure of the whole planet and the observed phase curves \citet{Parmentier2021}. We conclude that the nightside clouds play a particularly large role for exoplanets in the  intermediate temperature regime (class  ii).  In case of an increasing nightside temperature due to radiative transfer effects, for example \cite{2022arXiv220209183S}, the location of  thermally stable materials will change somewhat. Other effects need a more detailed consideration: A decrease of local density may not affect the cloud's optical depth due to an increased geometrical extension of the cloud, for example.

\paragraph{Dayside cloud coverage:}
The dayside  cloud coverage distinguishes the cool, the transition and the hot planets in their global weather and climate appearance. A nearly homogeneous cloud coverage emerges for the coolest planets (case i), a partial or transient cloud coverage for class (ii)  and a cloud-free dayside with considerable ionisation for the hot planets of class (iii). The details of these cloud layers have been discussed in Sects.~\ref{ss:nuc}--~\ref{section:cloud_material_compostion}. Associated with these changing cloud coverage are effects on the local chemistry which we presented in terms of the local C/O and the mean molecular weight (Sect.~\ref{section:chemical_regimes}). The dayside of the cool planetary regime will, hence, show a strongly depleted set of element abundances and a C/O that differs from the undepleted, pristine value. Planets in the transient regime will show such depleted elements in regions where cloud formation occurs which will be near the morning terminator. Hence, the dayside of these planets will be determined by a mix of depleted and undepleted areas. The dayside of planets in the hot regime  show a mainly undepleted set of element abundances and therefore a solar (or pristine) C/O in combination with a deceased mean molecular weight due to the thermal instability of H$_2$ in such hot atmospheres. Large  day/nightside temperature  differences further result into a different geometrical extension of  the dayside as well as the terminator regions. Hence, the cold planets in our grid (class i)  are likely to have no large geometrically asymmetries between the day- and the nightside, while the hot class (iii) planets may have a substantial geometrical asymmetry already within the atmosphere below p$_{\rm gas}<10^{-8}$ bar as footpoint of a planetary mass loss.

\smallskip
The dayside cloud coverage for planets of similar global temperature may well be different for different host stars. This is the case for the class (ii) planets,  WASP-43b/NGTS-10b (Fig.~\ref{fig:global_slice_plots_main_4250}) that orbit an K dwarf and HD\,209458b (Fig.~\ref{fig:global_slice_plots_main_5650}) that orbits a G star.  WASP-43b and NGTS-10b have a 10 times greater surface gravity compared to HD\,209458b  which, in addition,  leads to cloud extending deeper into the atmosphere. 

We note that recent WFC3/UVIS-HST observations in scattered light between 346-822 nm of WASP-43b are interpreted as seeing a (very dark) cloud-free dayside for pressure >1 bar \citep{Fraine2021}. This may be supported by the fast rotation models presented here.  Conversely, observations of HD\,209458b that canonically concluded it to be very cloudy would only pick up the very cloudy half of the dayside and  not the cloud-free region that occur towards the evening terminator as suggested in Fig.~\ref{fig:global_slice_plots_main_5650}. Extensive modelling studies are conducted for HD\,209458b (e.g., \citealt{2018A&A...615A..97L,2022arXiv220209183S}) which may enable a detailed modelling comparison.

Class (ii)  enables further to study the effects of rotation on the cloud patterns that characterise the weather and climate of these gaseous exoplanets. For example, planets with T$_{\rm eff, P}=1600$K that orbit a cooler star (M dwarf; Fig.~\ref{fig:global_slice_plots_main_3100}) need to orbit their host star at a smaller orbital separation of 0.13 days compared to a planet of the same effective temperature orbiting a G star orbiting with 1.55 days  (Fig.~\ref{fig:global_slice_plots_main_5650}). Our simulations therefore 
support our hypothesis that WASP-43b and NGTS-10b around K dwarfs may indeed show rotational deviations compared to G dwarf planets like HD~209458b of the same temperature but with much slower rotation. 

We note that the transition class (ii) of cloud climates arises because of a tight interplay between the thermal stability and the thermal background, which depends itself on the heat redistribution and wind field. Direct horizontal advection of cloud species is not included in our models, and could thus smear out the boundaries of  the partial cloud coverage. This smearing will be limitted by the thermal stability of the cloud particle in particular towards high-temperature regions like on the dayside. GCMs with 3D cloud-coupling \citep[e.g.][]{2018MNRAS.481..194L} may be used to refine the boundaries of the classes defined in this work.

\subsection{The asymmetry of the thermal ionosphere}\label{ss:asymion}
How does the deep ionosphere affect vertical transport of elements observed with high resolution observations? Can we readily compare observed abundances in the upper atmosphere of (partly) ionized planets with those of un-ionized planets?  While these questions are outwith the scope of this paper, we shortly review them with respect to the  introduced cloud formation regimes that are characteristic for potentially distinct  weather and climate scenarios. 
Weather and climate scenarios are determined by the thermo- and hydrodynamics behaviour of the atmosphere which does determine the global and local gas phase and cloud characteristics, but also secondary processes like ionisation and the emergence of global electric circuits (GEC) (\citealt{2016SGeo...37..705H}). The conditions for the emergence of an exoplanet GEC (eGEC) include a sufficient ionisation of the atmosphere and the presence of clouds than may produce lightning (\citealt{2019JPhCS1322a2028H}). Little to no eGEC  effects are expected on the atmosphere structure for the solar system planets (\citealt{2020SSRv..216...26A}), a conclusion that most likely can be extrapolated to extrasolar planets. The ionisation processes that drive the eGEC, however, do affect the local chemistry and may support the formation of cloud condensation nuclei in particular in the photo-dominated uppermost atmosphere layers in analogy to processes on Earth (\citealt{2017NatCo...8.2199S,2018ACP....18.5921T}). We therefore seek  to build our understanding of where in the atmosphere which ionisation processes act and how this may help to understand the global weather and climate on exoplanets. Based on the modelling framework of this paper, we concentrate on the thermal ionisation in what follows.
Figure~\ref{fig:global_properties} contextualises the model grid that we explore here with respect to a selection of possible candidates (dark golden dost with error bars; candidate data from Tables~\ref{t:UV1}, ~\ref{t:UV2})  for future UV missions, for example ARAGO \cite{2019BAAS...51c.219N}, PolStar \cite{scowen2021polstar} or 
POLLUX on LUVOIR \cite{18BoNeGo}.

It was suggested by \citet{Tan2019}, \cite{2021A&A...648A..80H} and \citet{Beltz2022} that the degree of ionization at the dayside of ultra-hot Jupiter would promote very efficient coupling between the ionized wind flow and the planetary magnetic field if it is of the order of a few Gauss. For example,  \cite{2018A&A...618A.107R} demonstrate that the magnetic flux threshold value for where the cyclotron frequency exceeds the local gas collisional frequencies  decreases to well below 1G  in the upper, low-density atmospheric layers. For the high-density atmosphere at $>1$bar, a local magnetic flux of > 100G may be required. Consequently, the  efficient magnetic field coupling of the atmospheric gas could lead to a very sharp transition from efficient day-to-nightside heat transport and very inefficient day-to-nightside heat transport as soon as a sufficient dayside ionization occurs. Thus, the transition between the intermediate and ultra-hot temperature regime would be affected by its degree of ionization. We find, however, that while the ionization can penetrate deep into the planet's atmosphere, the degree of ionization may not be sufficient to allow for efficient coupling between ionized gas and magnetic fields. In that case, the transition between intermediate and ultra-hot regime could occur for different temperature depending on rotational period, where faster rotators would exhibit a transition at cooler global temperature than slower rotators. That is, planets orbiting an K dwarf would transit at lower global temperatures than planet around F dwarf stars. 

{Observationally, phase curves of exoplanets around the 1800-2000 K temperature transition between case (ii) and (iii) with different orbital periods could be compared to determine if they exhibit differences in heat circulation and cloud distribution. Transiting planets orbiting bright stars with global temperatures around the case (ii) and (iii)  transition between intermediate and ultra-hot Jupiters are e.g. K2-31b ($P=1.26$~days, G type, \citet{Grziwa2016}\footnote{Its grazing transit will make this planet, however, difficult to characterize.}), WASP-14b ($P=2.2$~days, F type, \citet{Raetz2015,Southwood2012,Wong2015} and for $T_{\rm P}=2000$~K WASP-19b ($P=0.78$~days, G type, see e.g. \citet{Hebb2010,Maxted2013}).} Reduced hot spot shift with rotation period for both intermediate and hot Jupiters have been verified very recently by a systematic study of Spitzer data \citep{May2022}.

\subsection{Concluding discussion}

The grid study presented here supports and complements the ensemble of observational study of gas planets across different temperatures and rotation periods  in  preparation of the PLATO and the Ariel space missions.  Such complex models are needed to move on from single-case models for specific planets which does require a considerable adjustment in various physical and numerical parameters to fit the observational data (e.g., inner and outer boundary of computational domain, viscose damping, ...).

\citet{Roman2021} have also performed a grid study with a 3D GCM and clouds.  These authors, however, started with 1400~K, thus missing HD~189733b (with 1200 K). Both,  \citet{Roman2021}  and \citet{Parmentier2021} utilise simplified cloud models in contrast to our kinetic, multi-process approach, and they focus on only one rotational period.  These authors conclude, that when radiative feedback of clouds is included, the dayside-to-nightside temperature differences increase and the eastward hot spot offset decreases.%

For $T_{\rm gas}$=2000 - 3500 K, both authors see an apparent westward shift for phase curves in the optical wavelength range ($<1$~micron). This apparent westward shift in the optical phase curves 
is associated with a pile-up of clouds on the morning terminator 
and due to enhanced reflection of stellar light over these regions. This "westward shift" is thus mainly a radiative effect and only visible in optical wavelength ranges. \citet{Roman2021} excluded planetary rotation as a modifying factor in cloud feedback for phase curves on the basis that irradation on the planet is not modified by planetary rotation. \citet{Parmentier2021} did not investigate the impact of planetary rotation.

However, \citet{May2022} have shown that planetary rotation period definitely plays a role in moderating eastward shift and even allows the on-set of westward shift as observed in the IR Spitzer data. Thermal westward phase curve shifts, in contrast to optical westward phase shifts, can only be brought about by dynamical effects, that is, changes in the wind jet from eastward to westward flow. \citet{Carone2020} predicted that westward flow tendency on an intermediate Jupiter could appear for $P_{orb}<1.5$~days in agreement with the observations reported by \citet{May2022}. For hotter planets, the westward shift already appears for $P_{\rm orb}<2$~days according to \citet{May2022}. Thus, radiative effects alone are not sufficient when discussing cloud feedback and their effects on planetary phase curves. In this study, westward flow at the dayside is also included in the GCMs used here for intermediate to hot planets around M and K dwarf stars \citep[see][Fig.2]{2021MNRAS.tmp.1277B} but here we have not yet included the full radiative feedback.

Similar to \cite{Roman2021} and \citet{Parmentier2021}, our result suggest a  morning cloud pile-up for ultra-hot Jupiters for G, F and K planets, but not for M dwarf planets. Thus, while the cloud pile-up effect found is not sufficient to explain all aspects of phase curves for intermediate to Jupiters, it is probably amplifying the underlying westward flow tendency due to dynamical effect on the optical phase curves.

The closest analogues to ultra-hot Jupiters around M dwarfs in our simulations will also be important to consider to understand cloudy exoplanet phase curves. While ultra-hot Jupiters around M dwarfs do not exist,  ultra-hot brown dwarfs in close orbit around a white dwarf star were detected: WD 0137B (2400 K, P= 0.0803 days, \citealt{2020MNRAS.496.4674L}) and EPIC 2122B (4000 K, P=0.047~days). Observations by \citet{Zhou2022} indicate no asymmetries across the limbs. In addition, the 3D GCM of \citet{2020MNRAS.496.4674L} is very close to our ultra-hot Jupiters around M dwarf simulations in its strong day/night temperature dichotomy.

Interestingly, for WD 0137B water absorption was observed on the nightside. This may indicate that the nightside clouds lie deeper in the atmosphere in these two highly irradiated brown dwarfs due to the higher surface gravity of brown dwarfs compared to Jupiter mass planets as demonstrated, {  for example by the 1D {\sc Drift-Phoenix} models that solve cloud formation consistently as part of the whole atmosphere simulation (Fig. 2 in \citealt{2009A&A...506.1367W}).}
Recent results show that cloud particles and either magnetic drag or modified dynamics are needed to explain the phase curves of these objects \citep{Lee2022}. Large modelling studies like this work and those of \citet{Roman2021,Parmentier2021} that cover similar global parameters but consider different physical effects are thus highly timely and vital to interpret detailed observational studies with JWST and Ariel. {   Fully consistent radiation transfer solutions are required and progress is being made (e.g., \citealt{2022arXiv220209183S,2022ApJ...929..180L}).  }Only then can we identify which factors shape the phase curves of intermediate to hot Jupiters: radiative, dynamical, and magnetic effects and how clouds modify these.

\section{Clouds beyond:\\ The formation of mineral hazes }
\label{section:extrapolation_gas_and_cloud_results}

The gas pressure domain over which a 3D GCMs are simulated may differ for different authors. For example, \cite{Parmentier18} simulate the gas pressure ranging from  200 bar to 2 $\mu$bar, and \citet{2021A&A...649A..44H}  from 0.1 mbar to 700 bar. We began to explore the impact of inner pressure boundary of the GCM on the formation of clouds for the specific case of hot Jupiter WASP-43 b (\citealt{2021A&A...649A..44H}), showing that the increased thermal stability associated with higher pressures permits cloud formation to occur deeper in the atmosphere towards the hotter inner boundary. Here, we address the upper, low pressure boundary of the GCM simulations.

\begin{figure*}
{\ }\\*[-1cm]
\centering
    \includegraphics[width=0.85\textwidth]{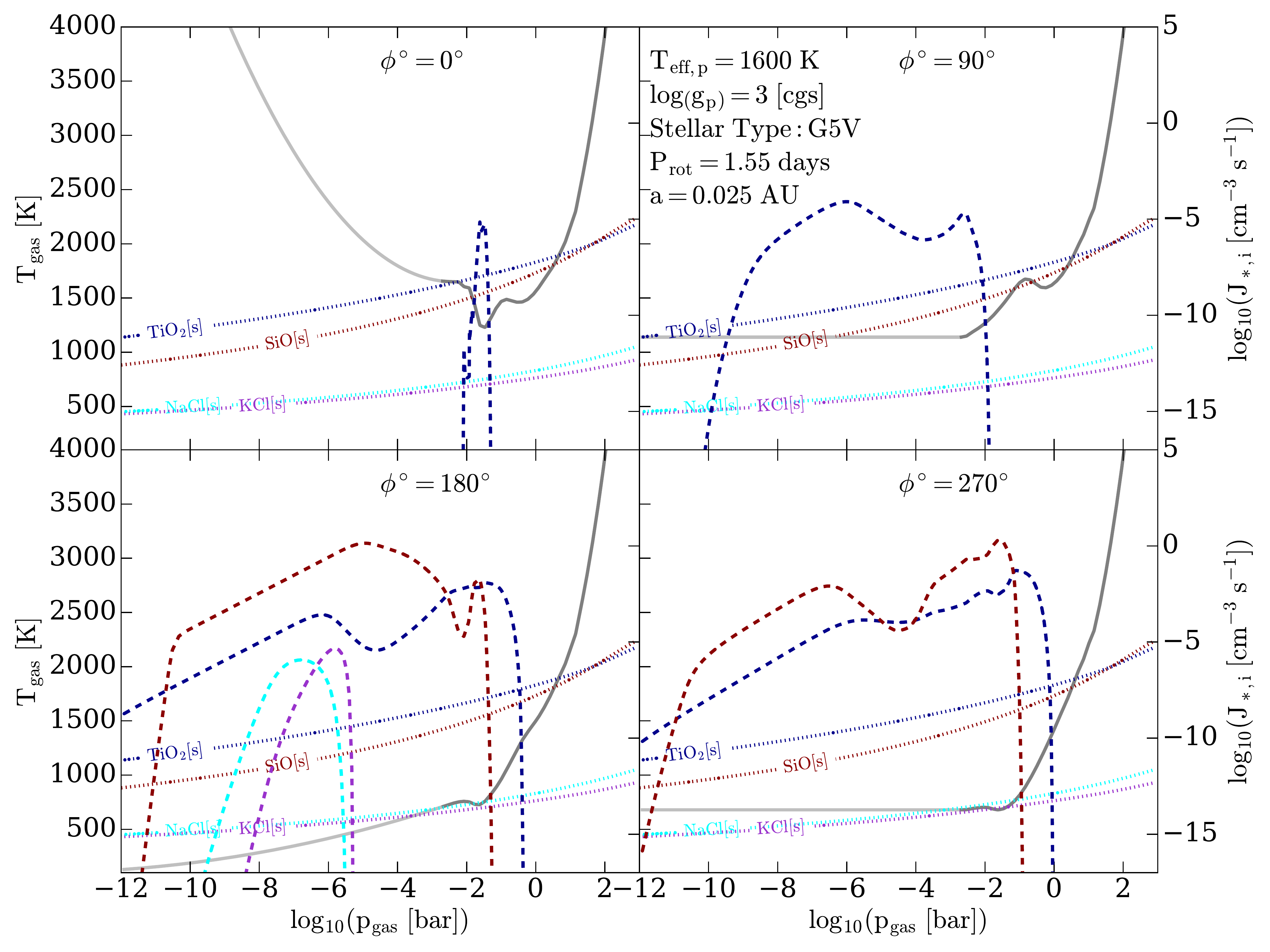}
    \caption{Nucleation rates, $J_{*,i}$ [cm$^{-3}$s$^{-1}$] (dashed lines), in the extended, low-pressure extrapolated substellar, antistellar, and equatorial morning and evening terminator (T$_{\rm gas}$, p$_{\rm gas}$)-profiles (solid dark grey lines) for the G star planet  atmosphere (T$_{\rm eff, P} = 1600$ K, $\log(g) = 3$ [cgs]).  The nucleation rates for TiO$_2$ (dark blue), SiO (brown), KCl (magenta), NaCl (cyan)  and the thermal stability curves (supersaturation ratio $S = 1$ for solar element abundances; dotted lines) for their respective solid condensates are shown. The $S_{\rm i} = 1$ curves do not represent our full kinetic model approach and are provided here for the purpose of visualisation only.}
    \label{fig:SatCurves}
\end{figure*}

We extrapolate a log-equidistant pressure grid and calculate corresponding temperatures using a parameterisation based on \citet{2009ApJ...707...24M}. We chose to extrapolate four profiles (substellar and antistellar point, equatorial morning and evening terminators) for on selected exoplanet atmosphere configuration (host star: G5V, T$_{\rm eff,~p} = 1600$ K, $\rm \log_{10}(g)$ = 3 [cgs]) to form the basis of the discussion on the potential for cloud formation outside of the commonly used computational domain.  The final temperature approached by the temperature parameterisation, T$_{\rm gas, outer}$, of the substellar extrapolated profiles is 10000\,K {  following works by \cite{2007P&SS...55.1426G,2004Icar..170..167Y}}. The antistellar extrapolated profile has a fixed T$_{\rm gas, outer}$ = 100 K. {  We note that the exact value of the outermost temperatures do not affect the result presented here since cloud formation stops at lower pressures in both cases because of too high temperatures or too low collision rates.} The terminator points are assumed to have an isothermal temperature structure for p$_{\rm gas} \gtrsim 10^{-3}$ bar where the temperature is fixed to the final value of the original 1D profile. The lowest pressure considered in this extrapolation is $10^{-12}$ bar.
The gas can safely be considered as a neutral hydrodynamic fluid to gas pressures as low as $10^{-8}$ bar (Appendix~\ref{section:hydrodynamics_validity}). If the gas can be assumed to be sufficiently ionised (see Sect.~\ref{section:degree_of_ionisation}) such that the collisional cross section increases accordingly, the validity shift to lower pressures of $10^{-15}$ bar. Such low pressures are still not without challenge since they imply very low particle growth rates as well as very little frictional interaction between the cloud particle and the gas. The transition from the original 3D GCM ($T_{\rm gas}, p_{\rm gas}$) domain into the extrapolated pressure domain is depicted by the transition from dark grey to light grey solid line in 
Fig.~\ref{fig:SatCurves} and occurs at $p_{\rm gas}\approx 10^{-3}$bar.

The extended 1D profiles are shown in Figure~\ref{fig:SatCurves}, and 
 despite the possible crudity of our first-order extrapolation, it becomes clear that the trends of the inner atmosphere {   with respect to their cloudiness} will continue into the upper atmosphere. 
 Figure~\ref{fig:SatCurves} demonstrates that the atmospheric range were cloud formation is triggered by the formation of condensations seeds (TiO$_2$, SiO, KCl, NaCl), extends into the very low pressure range for the terminators and the nightside profile. TiO$_2$ (dark blue dashed) and SiO (brown dashed) remain the dominating nucleation species.  Figure ~\ref{fig:SatCurves} further re-emphasizes that nucleation only occurs if the local gas temperature drops below the temperature were thermal stability (i.e. supersaturation ratio S=1) occurs. This explains the difference in cloud extension between the two terminators with the evening terminator ($\phi=90^o$) being somewhat hotter than the morning terminator. Furthermore, the  nightside ($\phi=180^o$) and the morning terminator ($\phi=270^o$) profile probed here could have mineral clouds extending even into higher atmospheric regions where $p_{\rm gas}<10^{-12}$ bar.

\begin{figure*}
{\ }\\*[-1cm]
\centering
    \includegraphics[width=\textwidth]{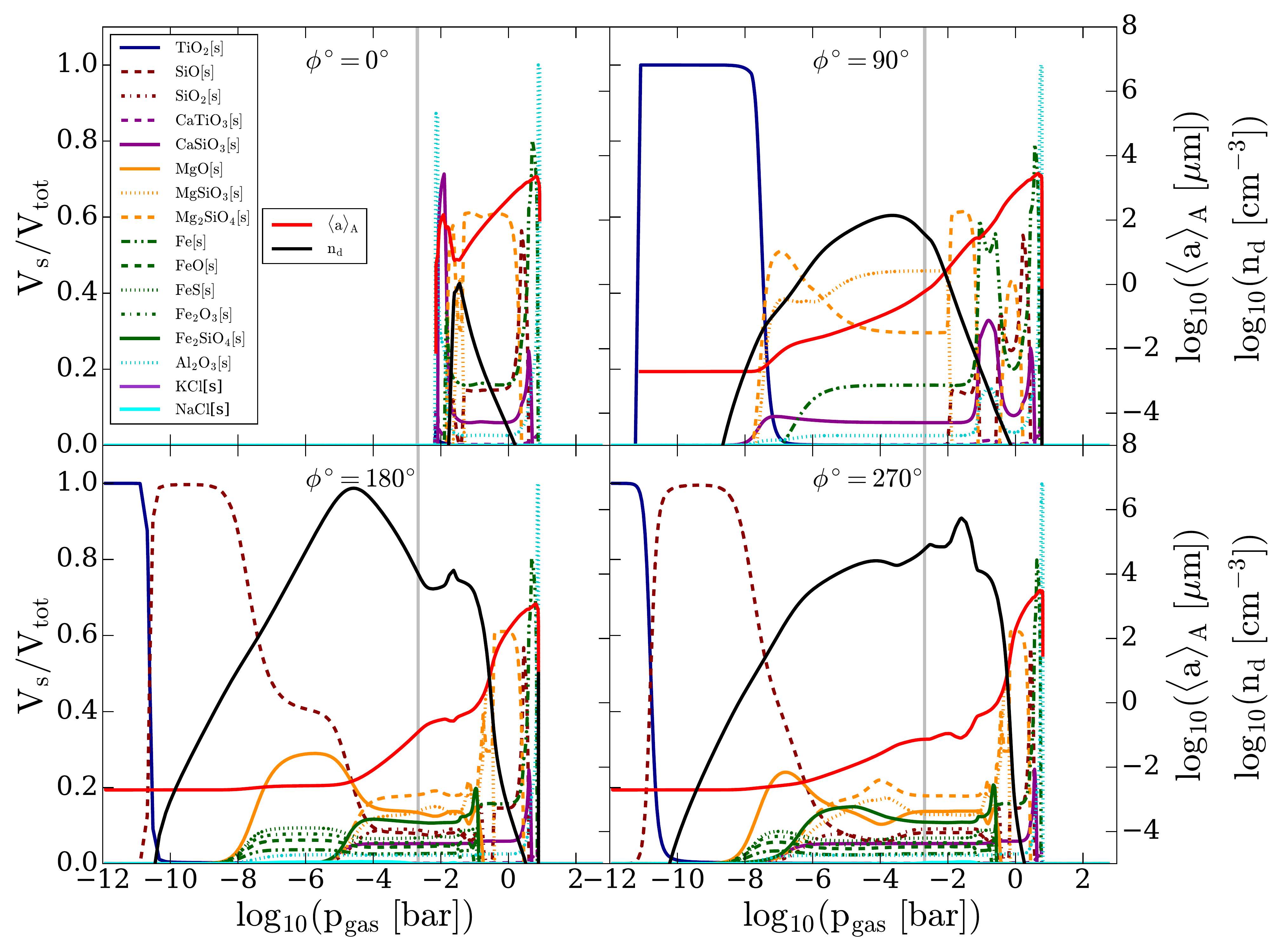}
    \caption{The material volume fractions ($V_{\rm S}/V_{\rm tot}$, color coded), surface averaged mean particle size ($\langle a\rangle_{\rm A}$ [$\mu$m], solid red), and cloud particle number density ($n_{\rm d}$ [cm$^{-3}$], solid black) at the low-pressure extrapolated substellar, antistellar, and equatorial morning and evening terminator points for the G star planet ( $\log(g) = 3$ [cgs], T$_{\rm eff} = 1600$ K) atmosphere. Both $\langle a\rangle_{\rm A}$  and $n_{\rm d}$ use the right-hand axis.}
    \label{fig:extrapolated_clouds_2.0}
\end{figure*}

Figure~\ref{fig:extrapolated_clouds_2.0}
presents the combined information about the clouds that form beyond the 3D GCM computational domain.  The significant increase in the extension of clouds into the low pressure, upper atmosphere ($p_{\rm gas}< 10^{-4}$ bar) is apparent for the two terminators and antistellar point.
 Figure~\ref{fig:SatCurves} shows that the efficient nucleation in the low-pressure, extrapolated atmosphere enables the formation of a layer of mineral hazes {  in the form of metal oxide clusters and cloud condensation nuclei}. The correspondingly low local densities do not enable efficient bulk growth until $p_{\rm gas}\approx 10^{-8}$ bar (Fig.~\ref{fig:extrapolated_clouds_2.0}). This suggest that the observation of these very upper atmospheric regions with $p_{\rm gas}< 10^{-8}$ bar, which are accessible through high resolution transmission spectroscopy,  might allow to study the nucleation process in more detail by searching for the spectroscopic signatures of (TiO$_2$)$_{\rm N}$, (SiO)$_{\rm N}$ clusters as proposed in \cite{2021A&A...654A.120K}. This mineral haze layer would be expected to be extant to $p_{\rm gas}\approx 10^{-8}$ bar, below which first mixed-material particles occur. A clear difference in cloud particle sizes may occur between evening and morning terminator due to different nucleation efficiencies. The nightside temperature is the lowest such that the nucleation efficiency is the highest, hence, the cloud particles remain the smallest in these extrapolated, low-pressure atmospheres.

\subsection{Limitations}
The computational efforts of 3D GCM requires sensible approximations and the assessment of those. Within the hierarchical approach that this paper and \cite{2021A&A...649A..44H} have followed, it can be concluded that the extension of the derived cloud layers is affected by the location of the inner and the outer boundary of the computational domain. An inner boundary at higher pressures will stabilise the cloud particles such that the cloud's backwarming can stronger affect the temperature in atmosphere-core transition region.

An extended upper boundary allows the mineral cloud formation to contribute in the domain that so far was understood as photochemically driven to form hydrocarbon hazes. This conclusion, however, is based on simulations that did not include the formation of metal-oxide clusters so far. A first comparison of their different efficiency was presented in \cite{2020A&A...641A.178H}. 

The limitation of our approach is that the simulations are not consistent, instead, the radiation hydrodynamics does not include the cloud formation effects nor the kinetic gas-phase effects. These are clearly topics for future work.
It is, however, reasonable to expect that photochemical effects will not only occur for the C/N/O/H/S chemistry but also for the metal-oxide chemistry. \cite{2021Univ....7..243G} demonstrate, however, that the ionisation energies of (TiO$_2$)$_{\rm N}$ clusters (N being the number of TiO$_2$ monomers forming the cluster) exceed the atomic ionisation energy of Ti. (SiO)$_{\rm N}$ is suggested as a better candidate for cluster ionisation but one can only reasonably expect the UV part of the stellar radiation field to affect the ionisation state of these clusters which contribute to the formation of cloud condensation nuclei. We further note that the stability of these clusters may be affect by their fall speed within these low-pressure regimes. The gravitational settling speed is rather high as little frictional interaction occurs. Hence, once such an interaction does occur, the kinetic energy of such interactions will be high.

\begin{table}[]
\centering
\caption{J band magnitudes for stellar types M5V, K5V, G5V and F5V. The absolute magnitudes are taken from \cite{PecautandMamajek2013} and the apparent magnitudes are calculated ($m_{\rm J} = 5\log_{10}\left( d/10 pc \right) + M_{\rm J}$) for three distances d = 50, 100, 200 pc. }\begin{tabular}{c||ccc}
    \hline\hline
    Stellar & Absolute & Distance & Apparent \\
    Type &  Magnitude &  & Magnitude\\
        &  ($M_{\rm J}$) &  (pc)   & ($m_{\rm J}$)\\
    \hline
    \multirow{3}{*}{M5V} & \multirow{3}{*}{9.09} & 50  & 12.58 \\
                         &                       & 100 & 14.09 \\
                         &                       & 200 & 15.6  \\ [1ex]
    \multirow{3}{*}{K5V} & \multirow{3}{*}{5.10} & 50  & 8.59  \\
                         &                       & 100 & 10.1  \\
                         &                       & 200 & 11.61 \\ [1ex]
    \multirow{3}{*}{G5V} & \multirow{3}{*}{3.73} & 50  & 7.22  \\
                         &                       & 100 & 8.73  \\
                         &                       & 200 & 10.24 \\ [1ex]
    \multirow{3}{*}{F5V} & \multirow{3}{*}{2.52} & 50  & 6.01  \\
                         &                       & 100 & 7.52  \\
                         &                       & 200 & 9.03  \\ [1ex]
    \hline
\end{tabular}

\label{tab:TSM_J_magnitudes}
\end{table}

\section{Observational implications}
\label{section:observational_implications}

\subsection{Transmission Spectroscopy Metric}
\label{section:transmission_spectroscopy_metric}

\medskip

We calculate a Transmission Spectroscopy Metric (TSM) \citep{2018PASP..130k4401K} for each of the grid model planets to give an indication on how amenable the planets would be to transmission spectroscopy observations. The TSM is calculated as

\begin{equation}
    {\rm TSM} = {\rm SF} \cdot \frac{R_{\rm P}^{3} T_{\rm eq}}{M_{\rm P} R_{*}^2} \cdot 10^{-\frac{\rm m_{J}}{5}}
\end{equation}
where $R_{\rm P}$ and $M_{\rm P}$ are the radius and mass of the planet in Earth units, T$_{\rm eq}$ is the planetary equilibrium temperature in Kelvin, $R_{*}$ is the radius of the host star, $m_{J}$ is the apparent magnitude of the host star in the J-band, and SF is a scaling factor. Whilst the radius of the model planets (1.35 $R_{\rm J}$) falls outside of the radius bins of the planets analysed by \cite{2018PASP..130k4401K}, we opt to use the scale factor SF $= 1.15$ calculated for the $4.0 < R_{\rm P} < 10 R_{\rm E}$ range. A J-band (central wavelength $\sim$1.2 ${\rm \mu m}$) apparent magnitude is used as the wavelength is closest to the centre of the NIRISS bandpass. We calculate the J-band apparent magnitudes of each of the host stars using the absolute magnitudes of \cite{PecautandMamajek2013} for three distances of d $ = 50, 100, 200$ pc. Both the absolute magnitudes of \cite{PecautandMamajek2013} and the calculated apparent magnitudes are listed in Table~\ref{tab:TSM_J_magnitudes}.

\begin{figure}[!ht]
    \centering
    \includegraphics[width=21pc]{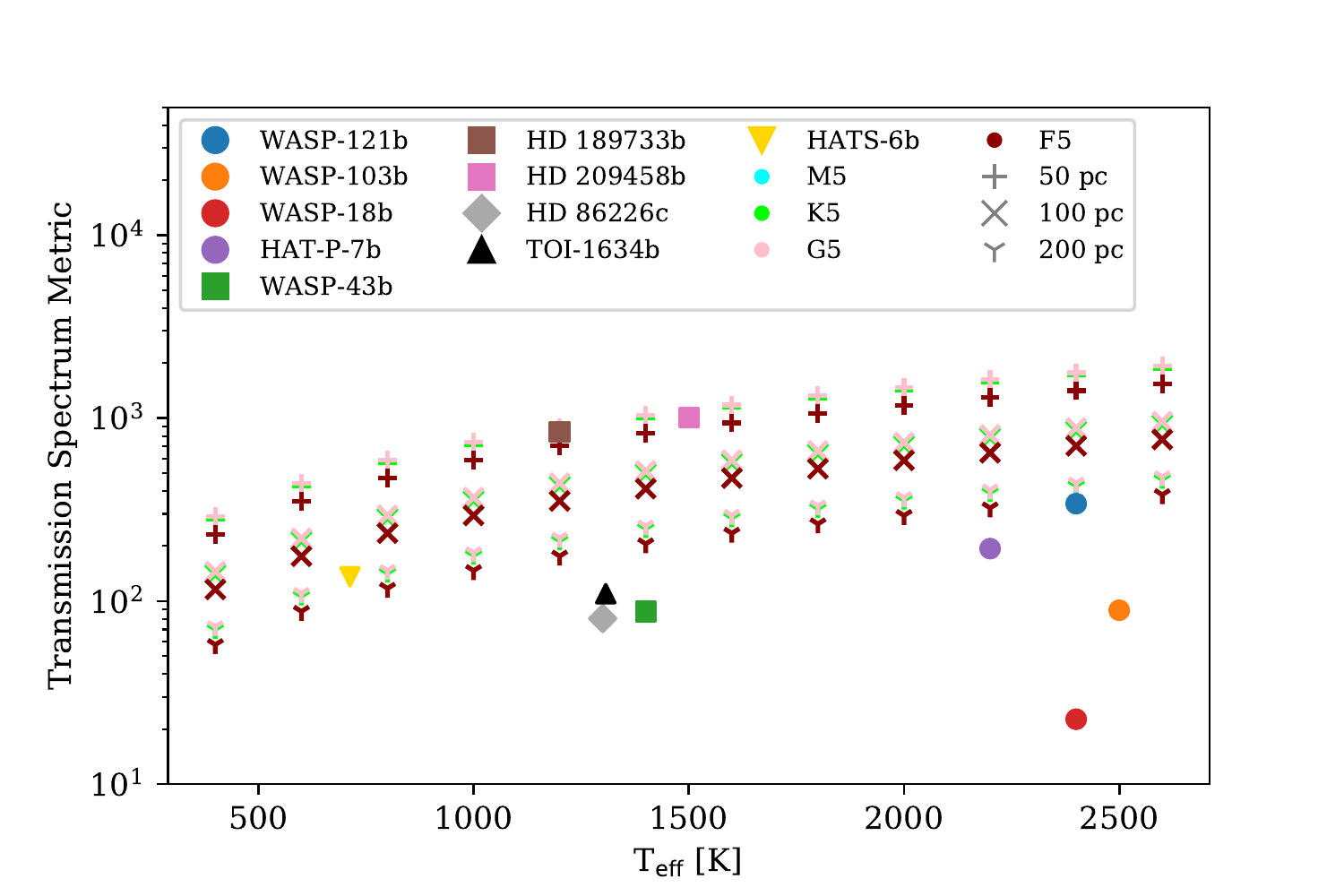}
    \caption{Transmission spectrum metric (TSM) \citep{2018PASP..130k4401K} for selected planets and $\log g = 3$ [cgs] grid models. The TSMs for the grid models are calculated using the J-band apparent magnitudes listed in Table.~\ref{tab:TSM_J_magnitudes}}
    \label{fig:transmission_metric_plot}
\end{figure}

Figure \ref{fig:transmission_metric_plot} compares the 3D GCM grid TSM with those for various gas-giants and ultra-hot Jupiters. Since the transmission spectroscopy metric $\propto H_{\rm p} \propto \frac{T}{g}$,  planets with high T$_{\rm eq}$ and low log(g) like HD~189733b, HD~209458b but maybe also HATS-6b, WASP-121b and HAT-P-7b are deemed to be easier observable through transmission spectroscopy.

\begin{figure*}
    \centering
{\ }\\*[-1cm]
    \includegraphics[width=0.85\textwidth]{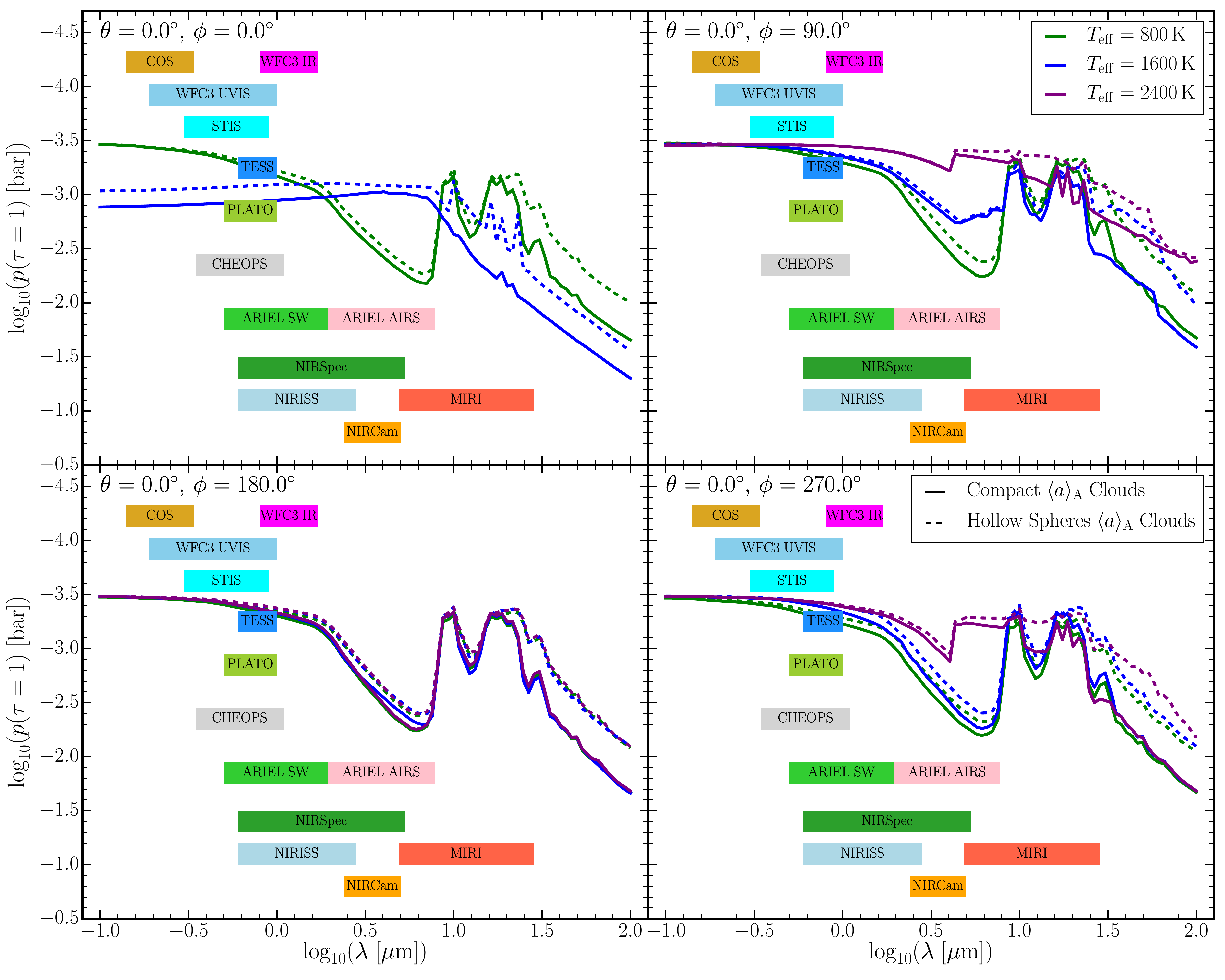}
    \caption{Wavelength-dependent pressure levels,  $p_{\rm gas} = p(\tau_{\lambda} = 1)$,  at which the optical depth due to cloud particles becomes unity, $\tau=1 $, for a grid model planet with an F-type host star, at T$_{\rm eff, P}=800, 1600$ and 2400 K, $\log(g) = 3$ [cgs].}
    \label{fig:Opt_Depth_tstar6500_test}
\end{figure*}

\begin{figure*}
\centering
    \includegraphics[width=0.85\textwidth]{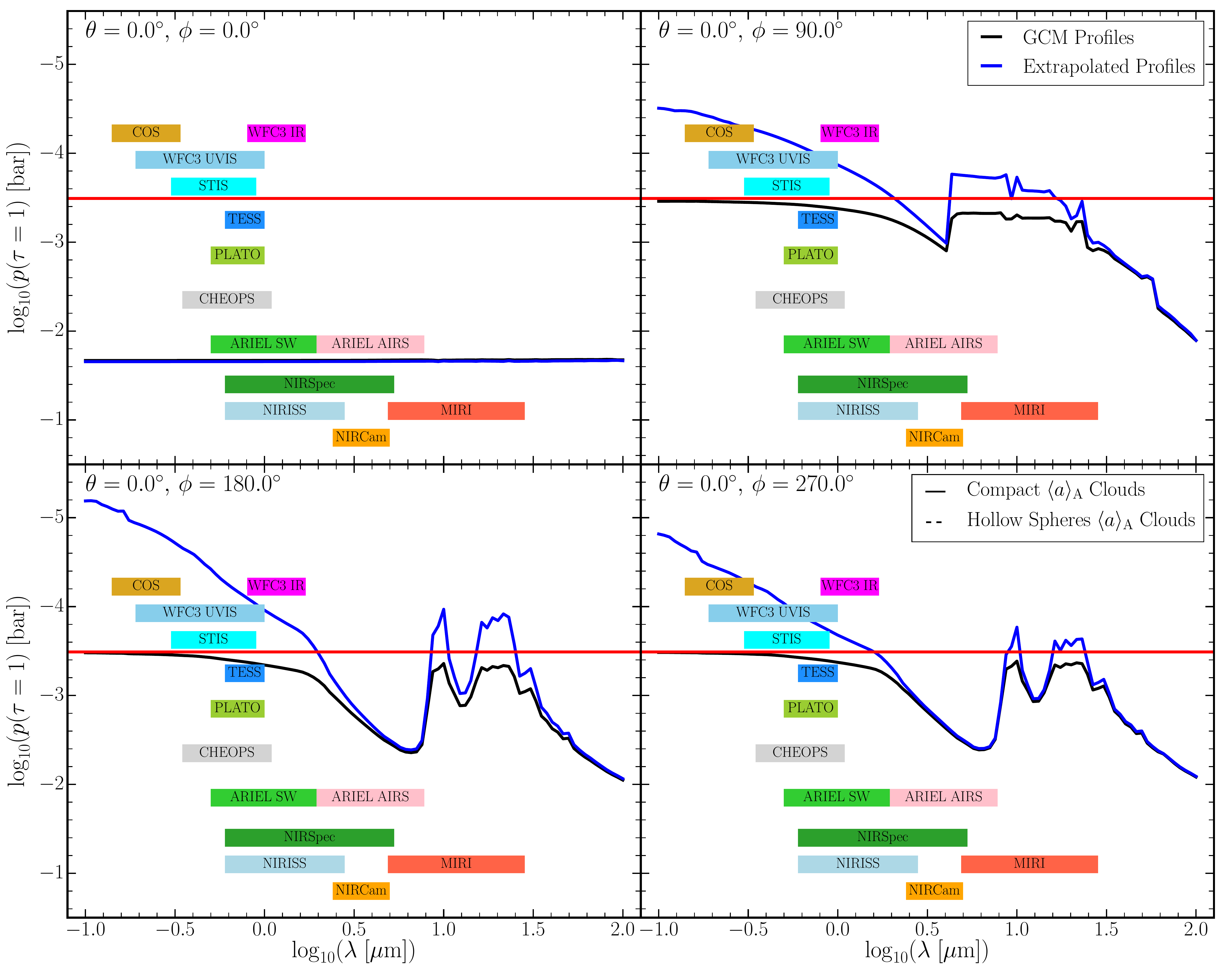}
    \caption{   Wavelength-dependent pressure level,  $p_{\rm gas} = p({  \tau_{\lambda} = 1})$, at which the optical depth due to cloud particles becomes unity, $\tau = 1$, for the grid model planet T$_{\rm eff, p}=1600$ K, $\log(g) = 3$ [cgs], with a G-type host star. The most suitable wavelength region for cloud investigation is $\lambda > 1\mu$m.  The solid black lines show the results based on the original 1D profile extracted from the \citet{2021MNRAS.tmp.1277B} 3D GCM grid, and the blue lines show the results based on the extrapolated profiles. Only the hollow-spheres opacity  are shown to demonstrate the maximum effects of the cloud opacity. The horizontal thin, red line indicates the upper boundary in pressure space of the 3D GCM computational domain. The atmosphere becomes optically thick already at the cloud top at shorter wavelength. The horizontal blue line indicates that the cloud is optically thick at $p>10^{-2}$bar at the substellar point.}
    \label{fig:GCM_start_impact_ptau1}
\end{figure*}

\subsection{Where the clouds gets optically thick: the p($\tau=1 $) levels}
\label{section:opacity_levels}

Figure ~\ref{fig:Opt_Depth_tstar6500_test} shows the atmospheric gas pressure levels where the optical depth reaches one for $0.1\,\ldots\,100\mu$m, $p(\tau(\lambda=1))$ [bar]. The atmosphere will be blocked by the clouds for higher pressures being equivalent to lower altitudes. For the calculation, we follow the same approach as in  \cite{2021A&A...649A..44H} (Sect.~7). Here, we focus on how the mineral absorption features change for planets within the cool (800\,K), transition (1600\,K) and the hot (2400\,K) exoplanet atmosphere regime for F-type host stars in Figure~\ref{fig:Opt_Depth_tstar6500_test} which represent the cases shown in Fig.~\ref{fig:global_slice_plots_main_6500}. The results for the M, G and K stars are presented in the Appendix (Fig.~\ref{fig:Opt_Depth_for_multiple_temps}) for the those models shown in Figs.~\ref{fig:global_slice_plots_main_4250} - ~\ref{fig:global_slice_plots_main_6500}. The comparison of the $p(\tau(\lambda=1))$ for different host stars is shown in Figs. 33 in the supplementary catalogue (\citealt{Lewis2022}).
All figures depict the sub-stellar (dayside: $\phi=0.0^o$), the anti-stellar (nightside: $\phi=180.0^o$) and the two terminator profiles (evening: $\phi=90.0^o$; morning $\phi=270.0^o$) at the equator ($\theta=0.0^o$).

The most suitable wavelength region for cloud investigation is $\lambda > 1\mu$m. In this region, the silicate features appear and the differentiation between compact and more agglomerate-like cloud particles can be made. The atmosphere becomes optically thick already at the cloud top at shorter wavelength such that the cloud provides a grey background opacity in the optical spectral region. 
Figure ~\ref{fig:Opt_Depth_tstar6500_test} suggest that the nightside would be almost indistinguishable for all three exoplanet regimes, but the considerable differences in the spectral range of the mineral features emerge on the dayside as well as in the terminator regions. The evening terminator, where the hot dayside gas flows towards the nightside, appears particular amenable to distinguish the three regimes.


For the terminators in the mid-infrared the cloud optical depth for all planets is dominated by the silicate spectral features and therefore the profiles are largely indistinguishable for different planetary effective temperatures. However, for the hot exoplanet regime ($T_{\rm eff,p}=2400\,{\rm K}$) irregularly shaped particles, modelled through a Distribution of Hollow Spheres (DHS, see \citealt{Min2005}, \citealt{Samra20}) appears to produce a flat, higher optically thick cloud pressure level, compared with the compact case. Including a DHS has the effect of increasing the optical depth of clouds in general, although for all other wavelengths and planetary effective temperatures the difference is relatively minor.

It is a different story in the near-infrared, the region covered by JWST NIRSpec and the Ariel infra-red spectrograph (AIRS). In the near-IR  are substantial (at greatest an order of magnitude in pressure) differences in the optically thick pressure level of the clouds for different planetary effective temperatures. Phase curve observations were used to infer cloud properties (e.g. \citep{2016NatAs...1E...4A,Oreschenko16_phase-curve,2017AJ....153...68S,2021ApJ...915...45C}). This provides a good incentive to investigate near-IR phase curves for a wide variety of planetary effective temperatures, as such observations could tease out details of cloud formation as affected by stellar installation and wind flow. Such a survey of phase curves is  proposed for Ariel for $\sim 50$ exoplanets (\citealt{CharnayARIEL2021}).
Furthermore the differences in the optical depth between the near-IR region and the optical may also allow for better constraints on the pressure-temperature structure of these planets, especially for hot gas-giants and transition temperature gas-giants. At such wavelengths, deeper pressure levels are observable, potentially providing information about the local gas temperatures which is unavailable to visible and UV observations. The near-infrared provides a `window' through the clouds to the deeper atmosphere below the observable cloud deck at other wavelengths.

The cloud's optical depth at the substellar points ($\phi=0^o$) differs dramatically for planet of different global temperatures because hot gas-giants have no clouds at $\phi=0^o$ (case i and iii)  and cool gas-giants exhibiting a cloud deck (case i). This strongly affects the dayside albedo.  Reflected light observations of WASP-43b have suggested a dark dayside \citep{Fraine2021}, and  reflected light in visible wavelengths for Kepler-7b have been found to also potentially discriminate between material composition of the clouds \citep{Webber2015}.

Finally, the effect of the stellar spectral type does also impact these conclusions. For the evening terminator ($\phi=90^o$) of M-type host star planets (Fig.~\ref{fig:Opt_Depth_for_multiple_temps}), the cloud optically thick pressure level is  dramatically different for the three effective temperatures in the mid-IR. The silicate features are absent for planets in the hot planet regime (case iii).

Figure ~\ref{fig:GCM_start_impact_ptau1} tests in how far the limit of the computational domain may affect our conclusions regarding the mineral spectra features: $p(\tau(\lambda=1))$ is compared for the original profile (black and similar to Fig.  ~\ref{fig:Opt_Depth_tstar6500_test}) to results for the extend atmosphere profile (blue) as discussed in Sect.~\ref{section:extrapolation_gas_and_cloud_results}. The horizontal thin, red line indicates the upper boundary in pressure space of the 3D GCM computational domain. It can be concluded that the $p(\tau(\lambda=1))$-level moves higher into the atmosphere due to cloud particles being able to form higher in the atmosphere and that the feature depth increases around $\lambda\approx 8\mu$m. 
The cloud's optical depth is also affected  in the optical and UV by the shifted upper boundary of the computational domain.  The $p(\tau(\lambda=1))$ increases steeper with increasing wavelength until $\lambda\approx 0.6\mu$m  for the terminators and antistellar points  due to the extended cloud decks to higher altitudes (lower pressures) but also due to the smaller average size of the mineral haze at these pressures.

\section{Conclusions}
\label{section:conclusions}

We propose to characterise the  weather and  climate on exoplanets by three classes that exhibit characteristic cloud and  gas-phase chemistry with clear implications for atmospheric asymmetries:
\begin{itemize}
\item class i) \textit{the cool planets} (T$_{\rm eff, P} \leq 1200$ K; e.g., HATS-6b, NGTS-1b) are characterised by:\\
-- globally homogeneous nucleation, hence, a globally homogeneous cloud coverage,\\
-- globally depleted element abundances, hence, increasing C/O in cloud forming layers,\\
-- homogeneous mean molecular weight,\\
-- globally low thermal ionisation,\\
-- metal-oxide clusters form a homogeneous haze layer.\\

\item class ii) \textit{the transition planets}  (T$_{\rm eff, P} = 1400 - 1800$ K; e.g., ASP\,43b, NGTS-10b, HD\,209458b)
are characterised by:\\
-- intermittent nucleation, hence, intermittent cloud coverage\\
-- intermittent element depletion and, hence, intermittent C/O across observable planet disk\\
-- cloud and gas chemistry emphasise day/night terminator difference\\
-- homogeneous mean molecular weight,\\
-- intermittent increases in thermal ionisation, \\
-- metal-oxide clusters may form mineral hazes on the nightside and on the morning terminator.\\

\item class iii) \textit{the hot planets} (T$_{\rm eff, P} \geq 2000$ K; e.g., WASP-18b, WASP-121b, WASP-103b,  brown dwarfs like WD\,0137b and EPIC\,2122B) are characterise by:\\
-- nightside confined nucleation,\\
-- cloud-free dayside with undepleted element abundances,\\
-- differences in day/night mean molecular weight implies a larger, geometrical extension of the dayside atmosphere, hence, a strong geometrical day/night asymmetry,\\
-- dayside exhibits an ionosphere that extends into the high-pressure, inner atmosphere suggesting a highly asymmetric magnetic coupling of these atmospheres,\\
-- metal-oxide clusters form mineral hazes on the nightside.
\end{itemize}

We, hence, conclude that for the cool planets (case i), 1D simulations suffice for the atmosphere up to 10$^{-5}$ bar. 
The homogeneity of the cloud cover suggest that the inferences of the C/O ratio based on observations of molecules over one planetary location is representative for the whole atmosphere.
Combined with the evidence that atmospheric mixing processes homogenize the chemical composition of cool planets \citep{2021MNRAS.tmp.1277B},  this further means that the non-detection of methane in cool gas planets as inferred from observations of WASP-107b \citep{Kreidberg2018b} (800 K, G type), WASP-117b (800 K, F type) \citep{Carone2021} and HD 102195 b (800 K, K type) \citep{Gandhi2020} indeed represent the atmosphere composition and are indicative of methane quenching.  Consequently, the presence of multiple carbon and nitrogen bearing species as inferred for HD~209458b \citep{Giacobbe2021}, which lies in the intermediate regime (1400\,K, G type), should not be interpreted in the 1D framework to represent the whole planet due to the complex interplay between 3D dynamics, chemistry and cloud formation. 

The intermediate temperature regime, not just hot extrasolar planets, may need substantial efforts to treat particularly the cloud distribution in three dimensions, or at the very least, with two profiles for asymmetric terminators for  transmission retrieval efforts.

\begin{acknowledgements}
    Ch.H. and P.W. acknowledge funding from the European Union H2020-MSCA-ITN-2019 under Grant Agreement no. 860470 (CHAMELEON). D.L.  and G.H. acknowledge the School of Physics \& Astronomy at the university of St Andrews for financial support of the summer project, R.C. acknowledges the Laidlaw Foundation. D.S. acknowledges financial support from the Science and Technology Facilities Council (STFC), UK. for his PhD studentship (project reference 2093954), O.H. acknowledges PhD funding from the St Andrews Center for Exoplanet Science.  D.S. and O.H. acknowledge financial support from the \"Osterreichische Akademie der Wissenschaften. R.B.~acknowledges support from the KU Leuven IDN/19/028 grant ESCHER. L.C. acknowledges the Royal Society  University Fellowship URF R1 211718 hosted by the University of St Andrews. K.L.C. acknowledges STFC funding under project number  ST/V000861/1.
   
\end{acknowledgements}
\bibliographystyle{aa}
\bibliography{reference.bib}

\begin{appendix}

\section{Testing validity of hydrodynamics regime}\label{section:hydrodynamics_validity}

Here we determine over which pressure range the utilized model atmospheres and their extrapolations are collision dominated, i.e. the fluid assumption is valid.

The validity of the hydrodynamic assumption is assessed via the Knudsen number $Kn=\lambda/L$, where $\lambda$ is the mean free path and $L$ is the characteristic length scale. For the hydrodynamic assumption to be valid, $Kn<1$. .

For this model, the characteristic length scale is taken as the scale height. The mean free path can be calculated via

\begin{equation}
    \lambda = \frac{1}{\sqrt{2}n\pi d^2}
\end{equation}
where $d$ is the covalent radius of hydrogen: $d_{H}=2.25\times10^{-12}$m (\cite{Slater1964}) and $n$ is the number density of the local gas phase [cm$^{-3}$]. The scale height is derived from the assumption of hydrostatic equilibrium.

\begin{equation}
    H_{S} = \frac{k_{B}T}{\mu m_{H}g}
\end{equation}
where $T$ is the temperature of the local gas phase and $\mu$=2.35 [amu].

\citet{Debrecht2020} point out that the upper atmospheres will be affected by ionisation such that the mean free paths of the gas phase does change. Irradiation by the planet's host star is the most likely cause, but the interstellar radiation may already suffice to ionise the uppermost atmospheric layers (\citealt{2018A&A...618A.107R}). 
 
\begin{figure*}
\centering
    \includegraphics[width=18pc]{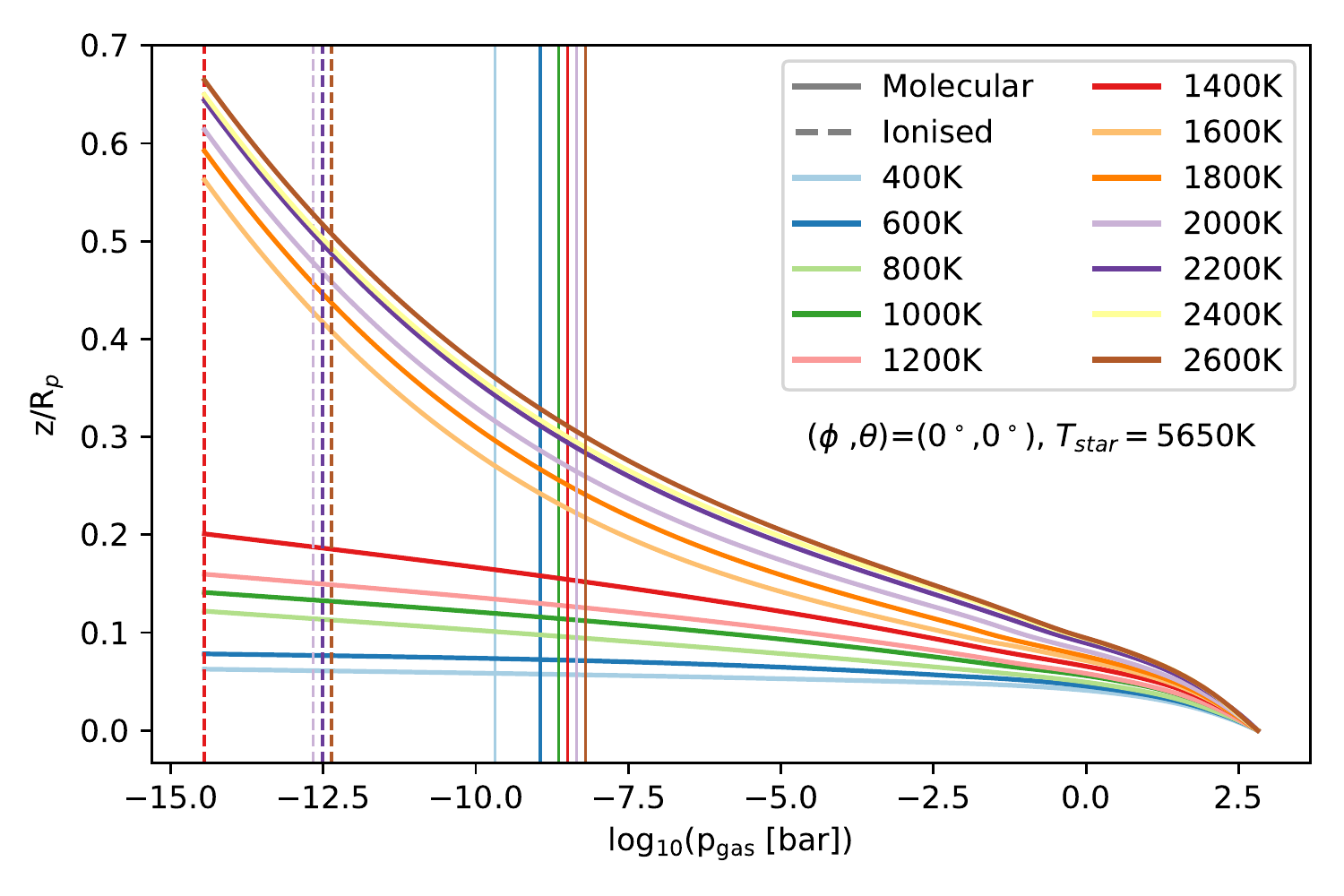}
    \includegraphics[width=18pc]{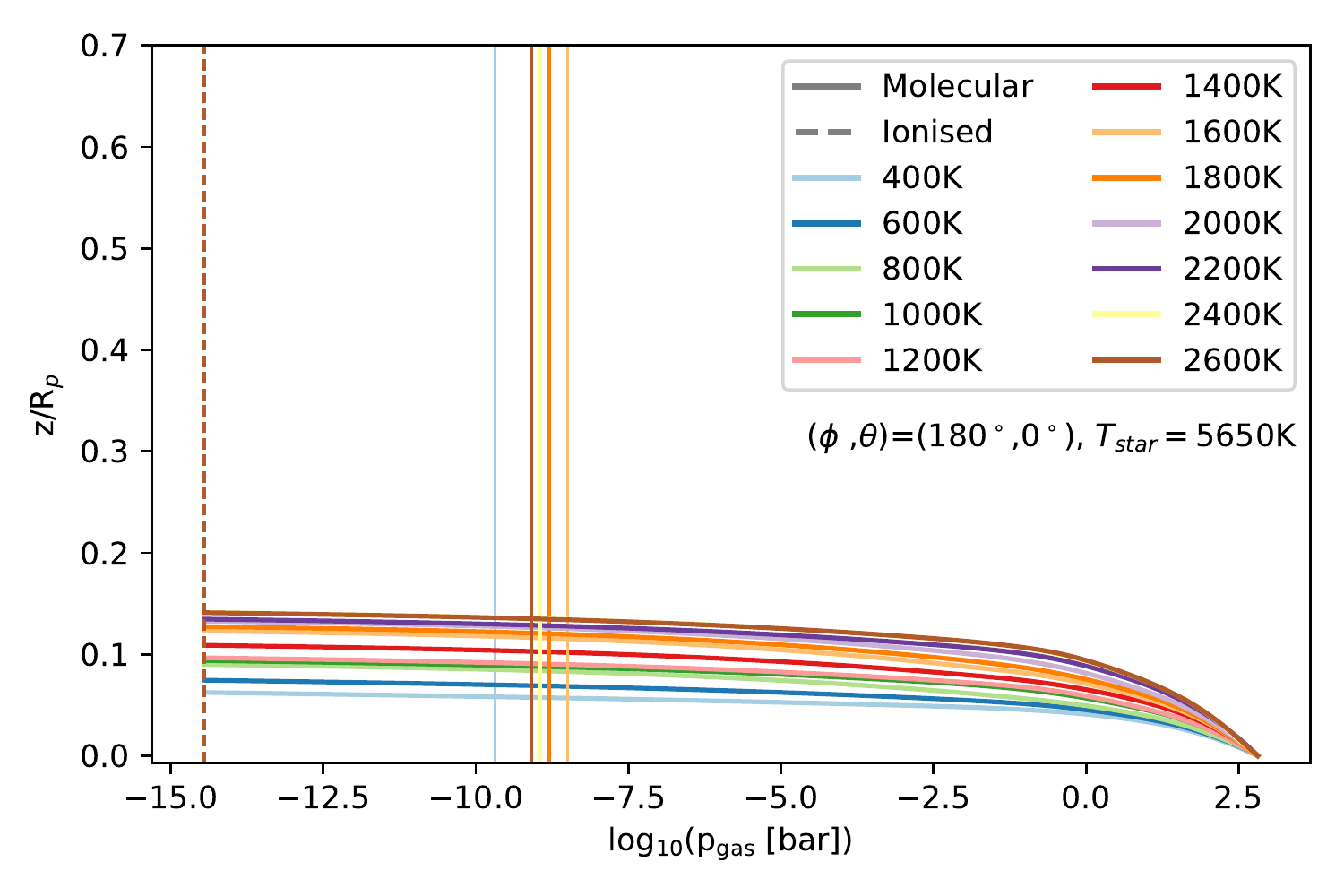}\\
    \includegraphics[width=18pc]{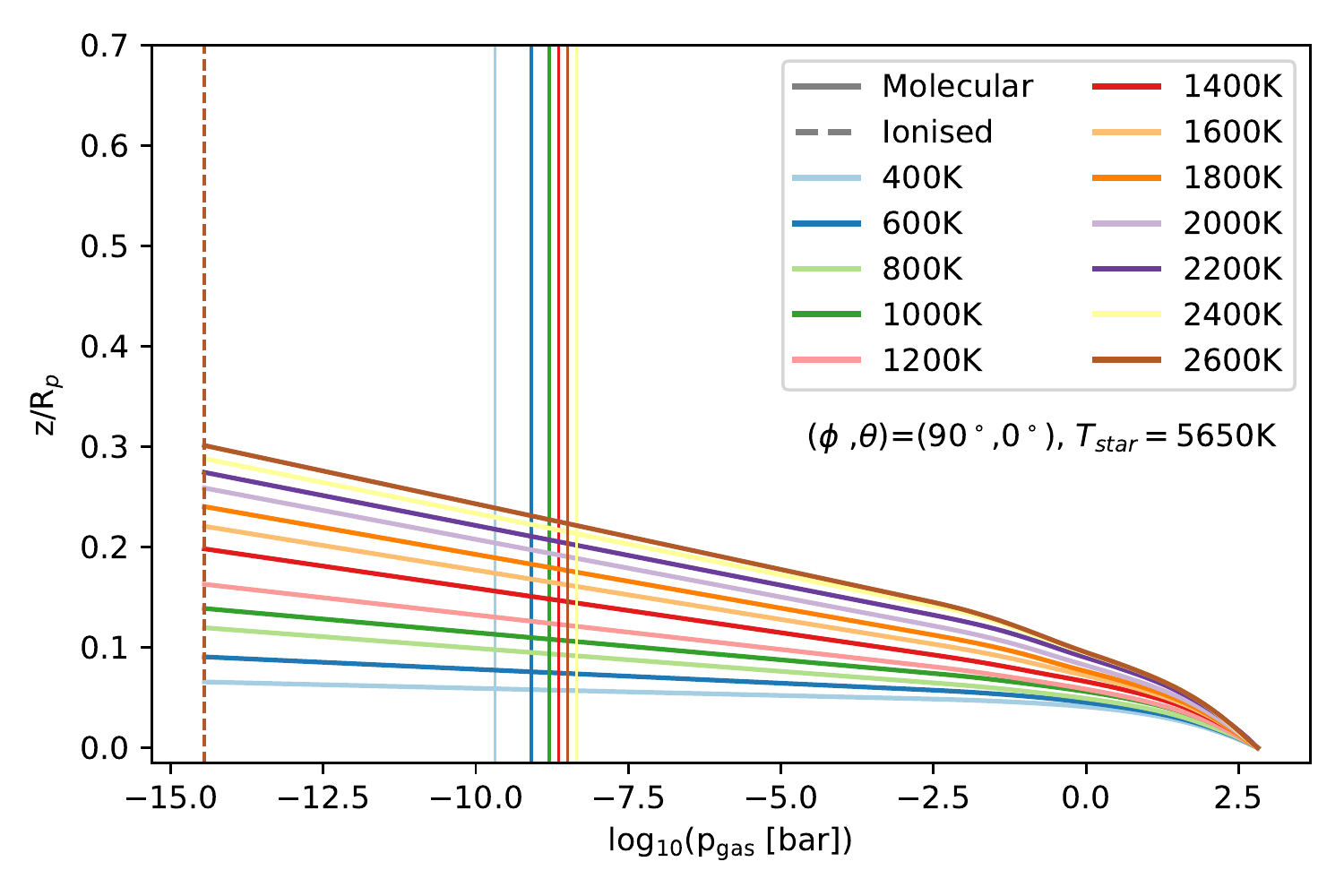}
    \includegraphics[width=18pc]{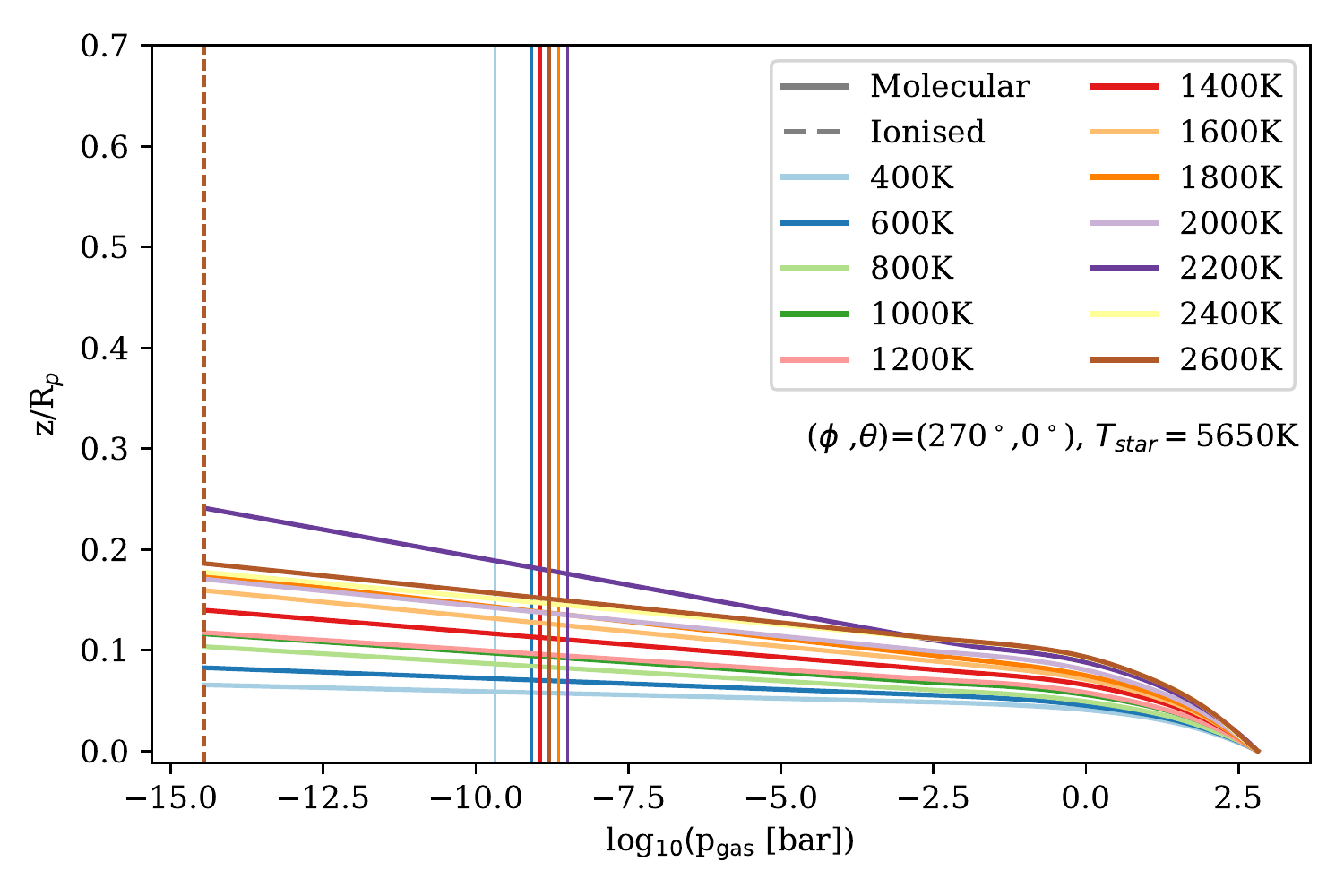}
    
    \caption{Height of the atmosphere as a function of pressure, with vertical lines denoting the pressure at which the Knudsen number exceeds 1. The solid vertical lines show the pressure limit for the molecular gas, and the dashed lines show the pressure limit for the fully ionised gas. The height profiles are shown for all planetary effective temperatures for the sub-stellar, anti-stellar, morning terminator and evening terminator points. All the models have log(g)=3 [cgs] and orbit G5 stars.     }
   \label{fig:Height_vs_Pressure_Knudsen}
\end{figure*}

To determine the Knudsen number limit for the ionised atmosphere, the mean free path is calculated as 

\begin{equation}
    \lambda = \frac{1}{\sigma n}
\end{equation}

where n is the number density of the local gas phase [m$^{-3}$] and $\sigma$ is the cross sectional area [m$^2$] where $\sigma=10^{-11}/T^2$ (\citet{Debrecht2020}).
Figure \ref{fig:Height_vs_Pressure_Knudsen} shows the height of the atmosphere as a function of pressure, with vertical lines denoting the pressure at which the Knudsen number exceeds 1 for the molecular and ionised gas.

We also plot the minimum cross sectional area of the gas particles for which the gas is collision dominated throughout the entire pressure regime (ie. the upper pressure limit at which the gas is collision dominated is 3.58$\times$10$^{-15}$) bar) in figure \ref{fig:cross_sectional_area_plots}. 

\begin{figure*}
\centering
    \includegraphics[width=18pc]{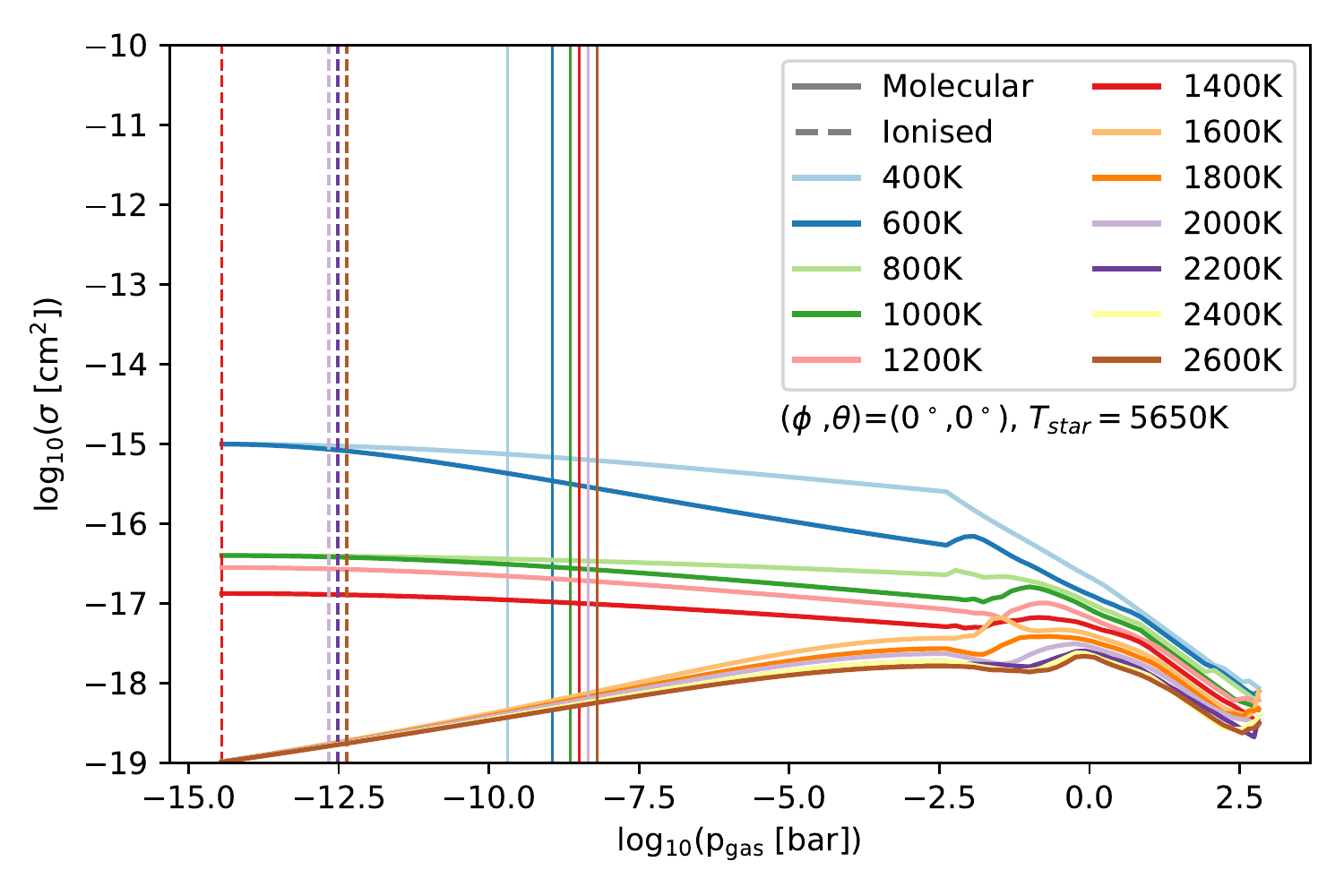}
    \includegraphics[width=18pc]{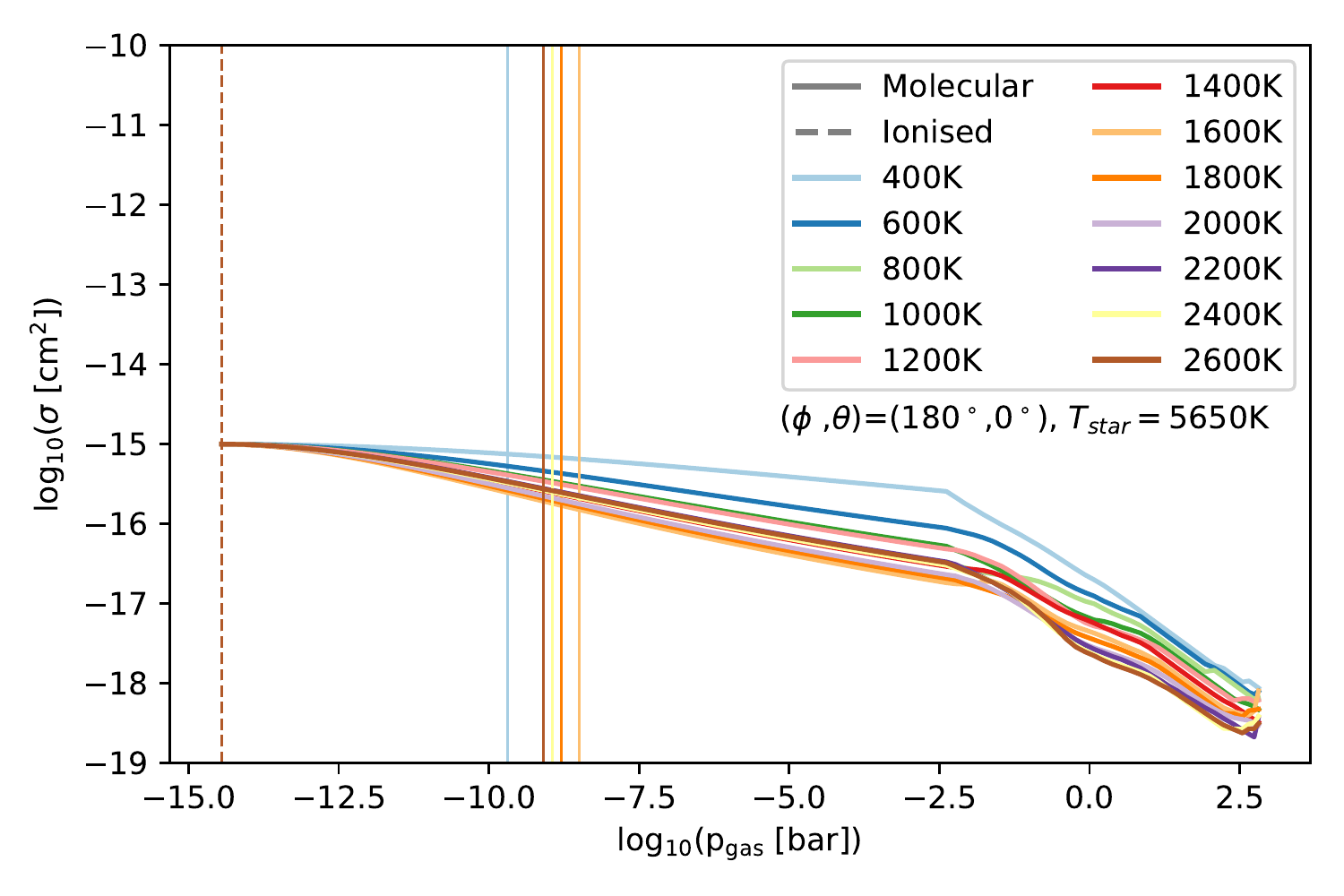}\\
    \includegraphics[width=18pc]{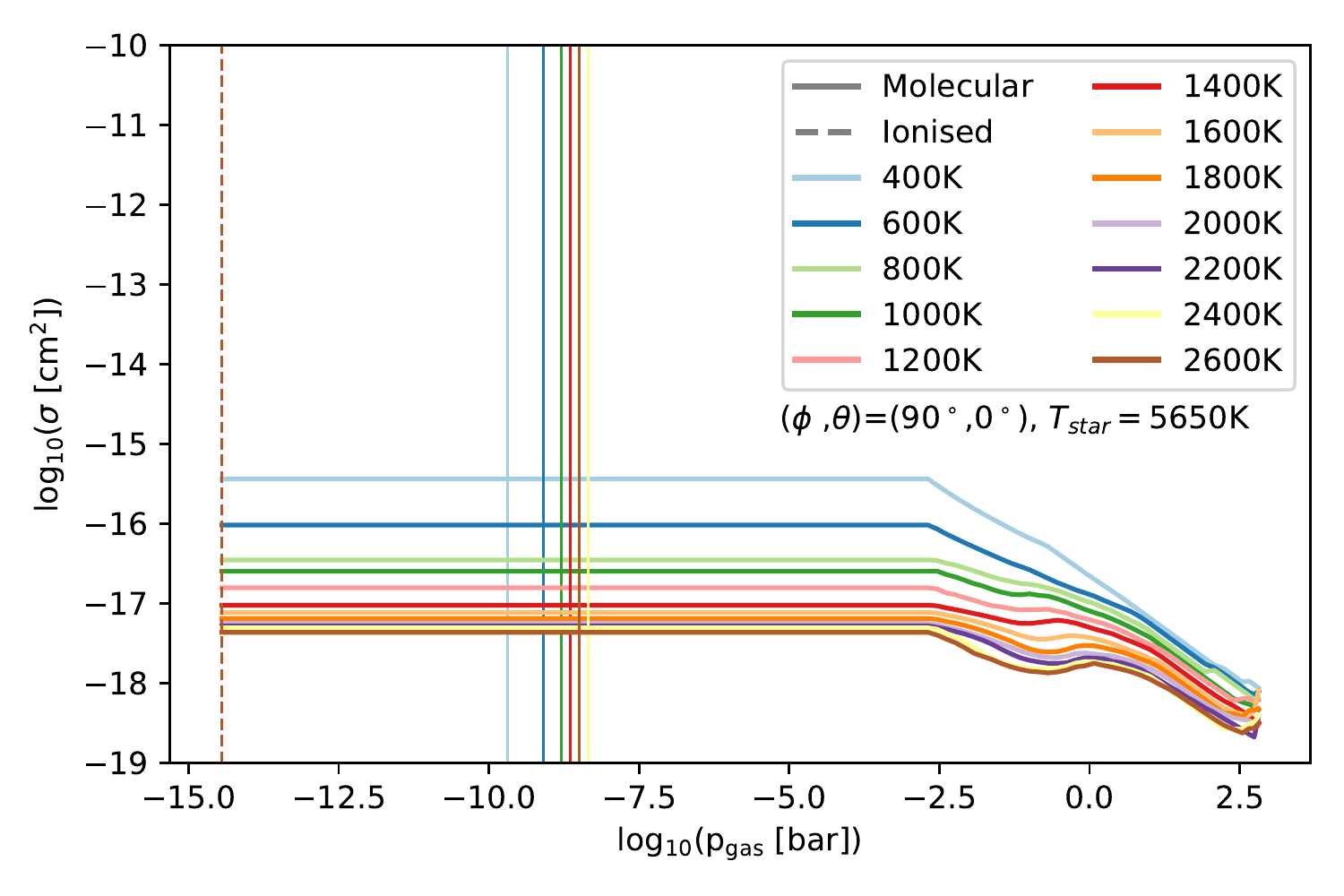}
    \includegraphics[width=18pc]{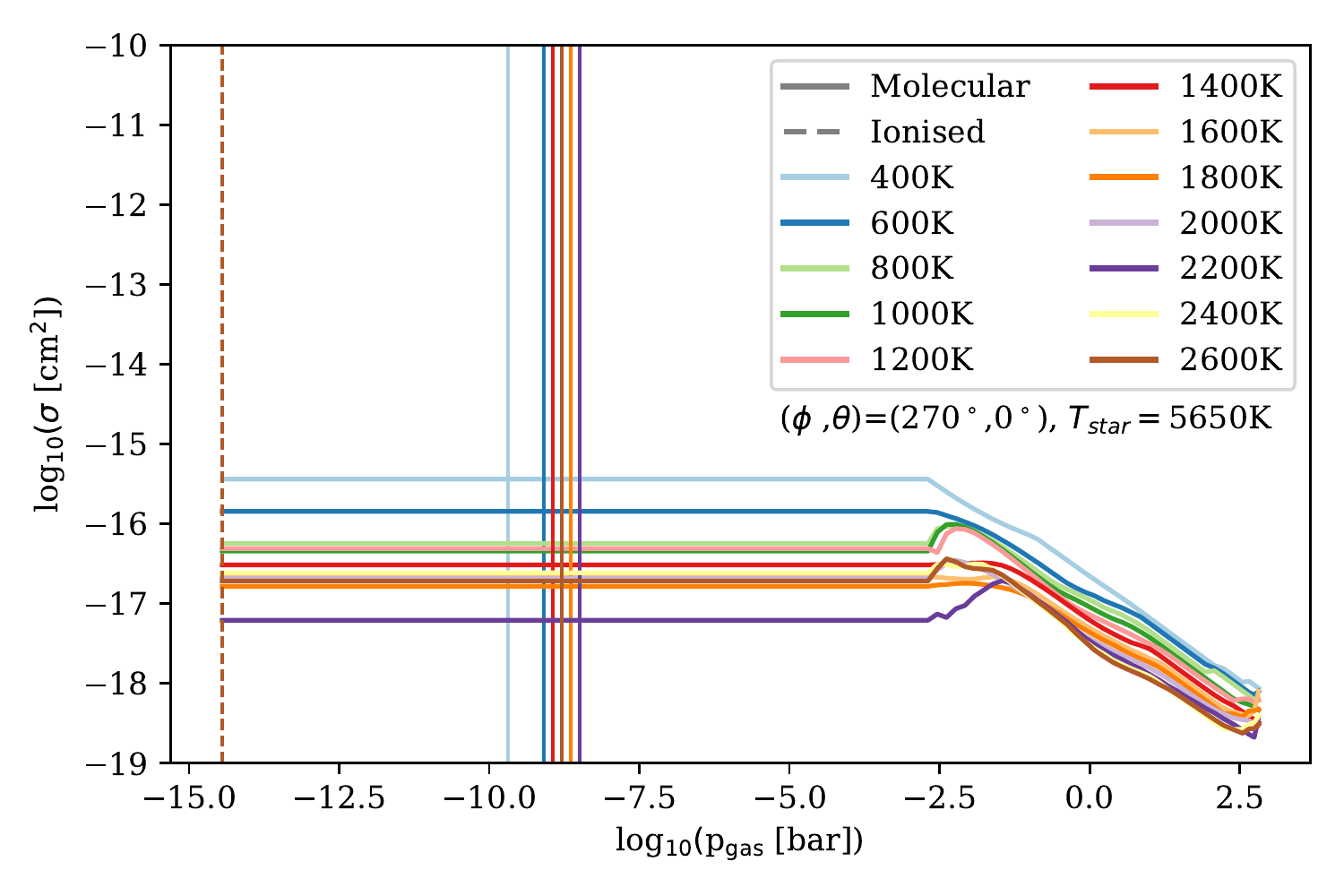}
    
    \caption{Cross sectional area of the gas particles as a function of pressure, with vertical lines denoting the pressure at which the Knudsen number exceeds 1. The solid vertical lines show the pressure limit for the molecular gas, and the dashed lines show the pressure limit for the fully ionised gas. The profiles are shown for all planetary effective temperatures for the sub-stellar, anti-stellar, morning terminator and evening terminator points. All the models have log(g)=3 [cgs] and orbit G5 stars.}
    
   \label{fig:cross_sectional_vs_pressure_plots}
\end{figure*}

\begin{figure*}
\centering
    \includegraphics[width=20pc]{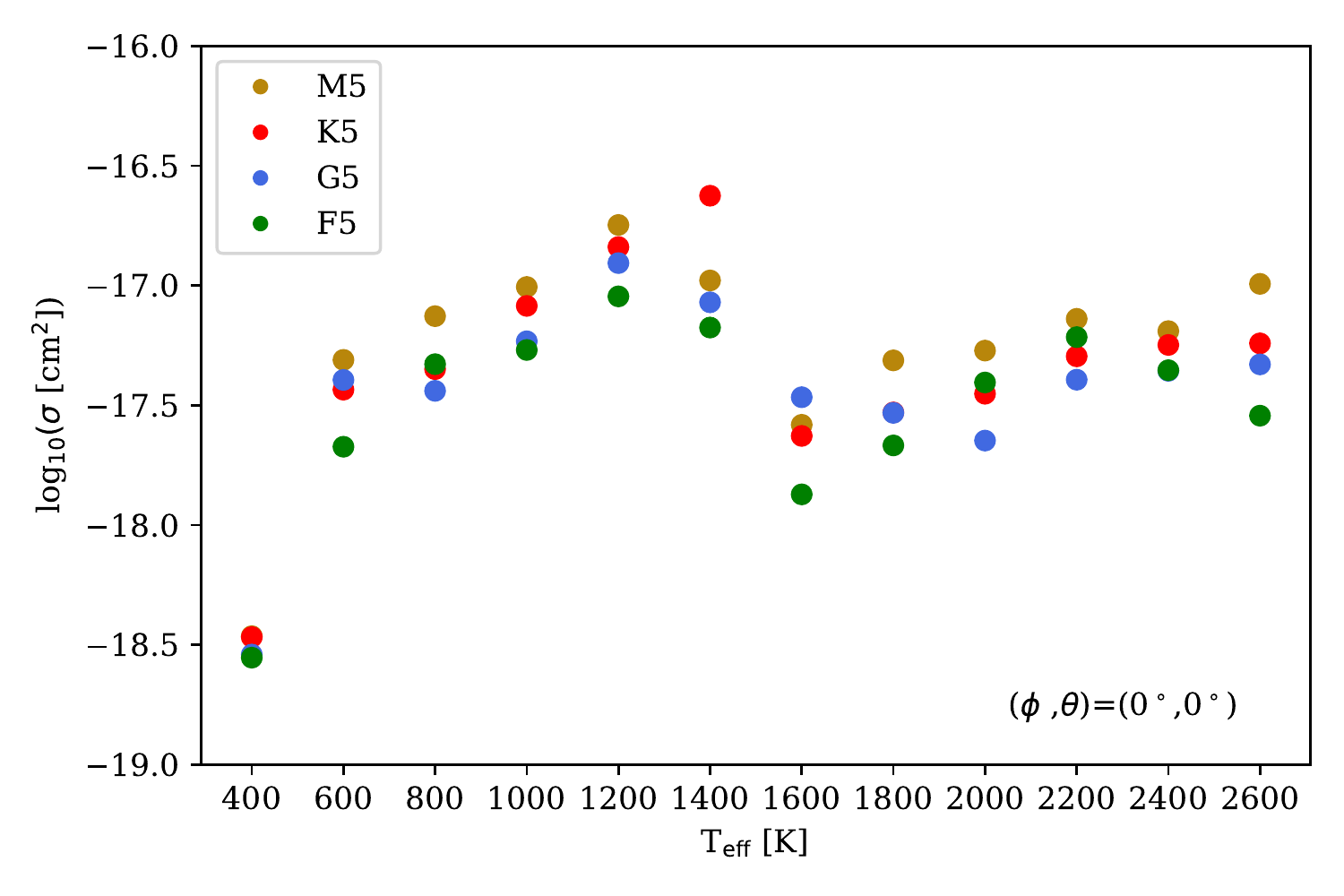}
    \includegraphics[width=20pc]{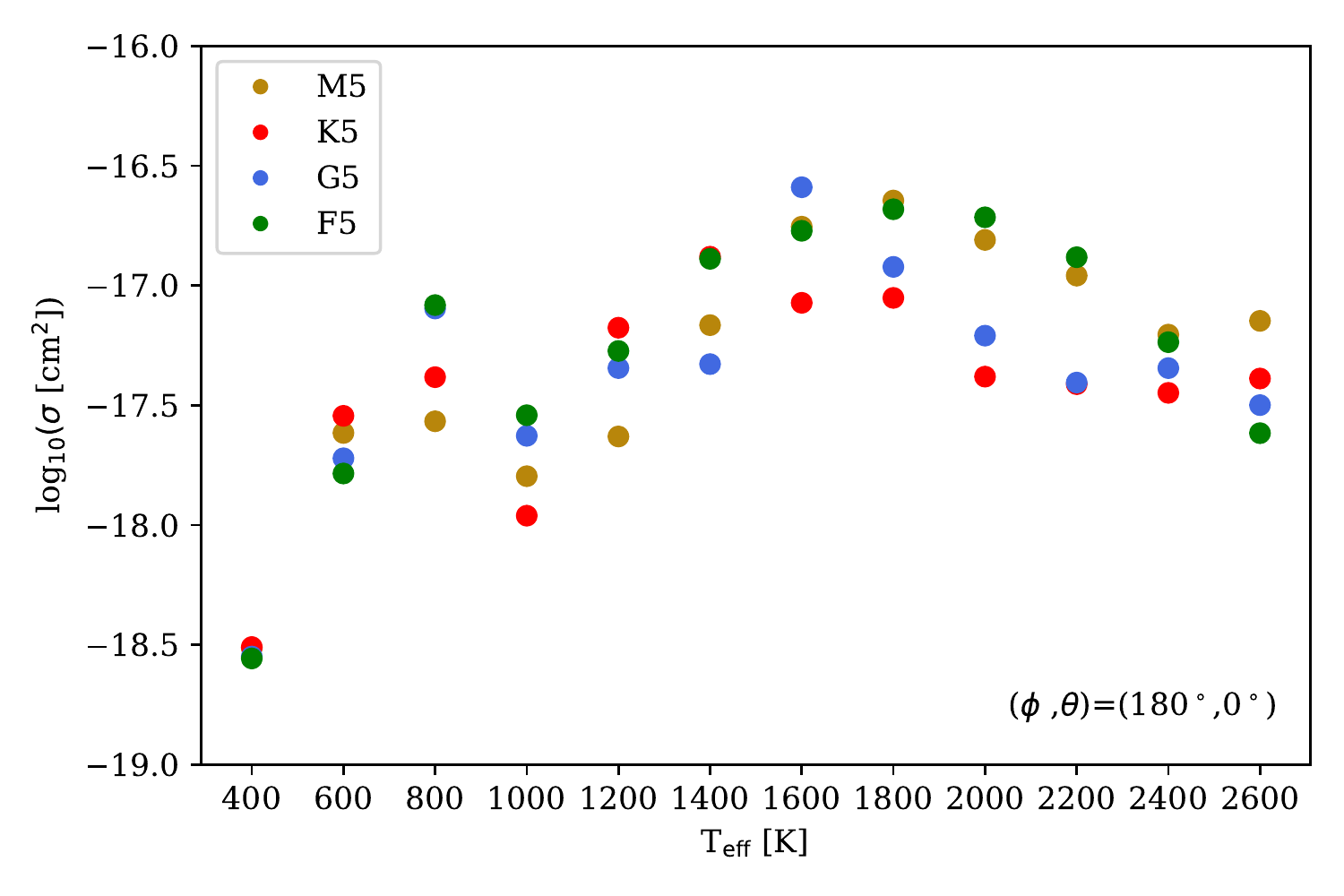}\\
    \includegraphics[width=20pc]{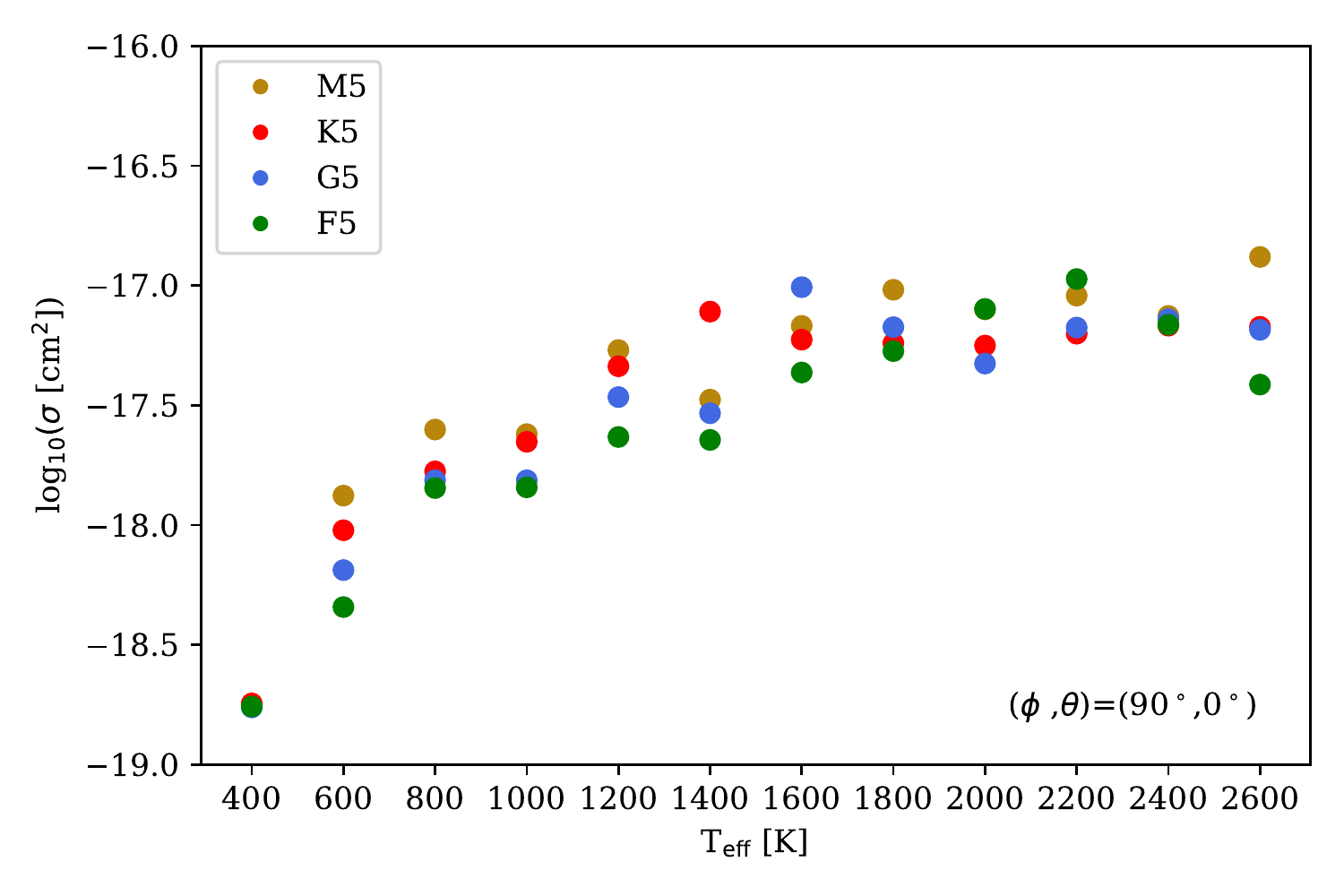}
    \includegraphics[width=20pc]{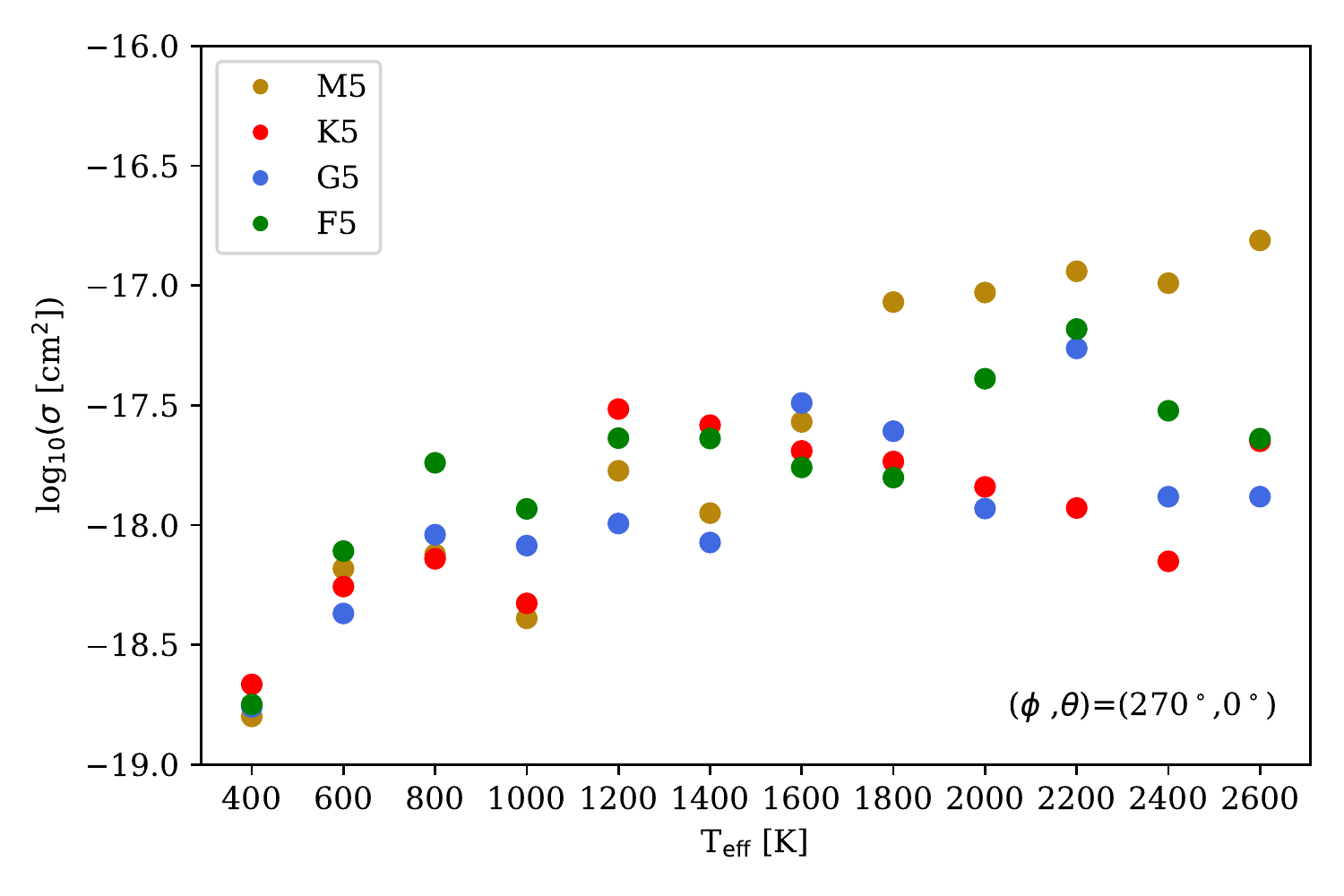}
    
    \caption{Cross sectional area of the gas particles for which the gas is collision dominated throughout the entire pressure regime as a function of planetary effective temperature for each stellar type. Plots are shown for the sub-stellar, anti-stellar, evening terminator and morning terminator points. All models have log(g)=3 [cgs]. Note that this figure is for the uppermost data point only. }
    
   \label{fig:cross_sectional_area_plots}
\end{figure*}

Furthermore, we include the collision cross-sectional area ($\sigma=10^{-11}/T^2$) as a function of pressure in figure \ref{fig:cross_sectional_vs_pressure_plots}.

\section{Diffusive mixing}\label{s:diffmix}

\begin{figure}
\centering
\includegraphics[width=7cm]{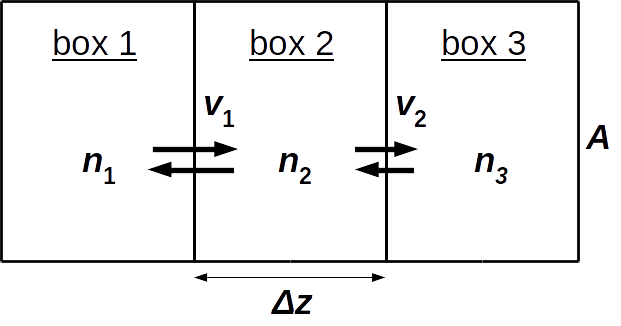}\\[-2mm]
\caption{Diffusive mixing between three boxes.}
\label{3boxes}
\end{figure}

This appendix motivates a new approach how to measure the mass exchange timescale $\tau_{\rm mix}(z)$ from the vertical component of a given velocity field $v_z(\vec{r},t)$, where $\vec{r}$ is the 3D position and $t$ the time. This approach has been used in all {\sc StaticWeather} models presented in this paper.

\subsection{Mixing and diffusion}

Let's assume we have two identical boxes of length $\Delta z$ with cross section $A$ touching each other, see Fig.~\ref{3boxes}. The total number of a certain kind of molecule in one of those boxes is
\begin{equation}
  N = n\,A\,\Delta z
\end{equation} 
where $n\rm\ [cm^{-3}]$ is the molecular particle density. From the 3D hydro model, we observe that matter moves up and down with some average (e.g.\ root-mean-square) velocity $\rm[cm/s]$ as
\begin{equation}
  v = v_{z,\rm rms} = \sqrt{\langle v_z^2 \rangle_t} 
                   = \sqrt{\langle v_z^2 \rangle_{\rm vol}}
  \label{eq:vmix}                   
\end{equation}
which either involves a long-term average over a suitably long time $t$ or a spatial average over a suitably large volume. Since we are interested in the stochastic part of the velocity field, we assume that there is no bulk motion here, i.e. $\langle v_z \rangle_t = \langle v_z \rangle_{\rm vol} = 0$. In real application to a given hydrodynamic structure, this means that we first need to subtract the bulk motion before we can apply Eq.\,(\ref{eq:vmix}).

Because of the random mixing motions, molecules will go from box~1 to box~2 and vice versa. The associated mean particle fluxes $\rm[cm^{-2}s^{-1}]$ through the contact area $A$ are
\begin{eqnarray}
  j_1 = n_1\,v &\quad\quad& \mbox{rightwards,}\\ 
  j_2 = n_2\,v &\quad\quad& \mbox{leftwards.}
\end{eqnarray}
The change of the total number of molecules $N_1$ in the left box is
\begin{align}
  dN_1 =& -j_1\,A\,dt + j_2\,A\,dt = (n_2-n_1)\,v\,A\,dt   \label{eq:dN1}\\
\Rightarrow
  \frac{dn_1}{dt} =& \frac{n_2-n_1}{\Delta z}\,v
                 \;\;\to\;\; -\frac{\partial n}{\partial z}\,v
\end{align}

\paragraph{Diffusion with rate equations:}
The problem can be re-formulated with rate constants
$R=v/\Delta z \rm\;[1/s]$, like a chemist would do
\begin{eqnarray}
  \frac{dn_1}{dt} &=& -n_1 R + n_2 R \label{dn1}\\
  \frac{dn_2}{dt} &=& -n_2 R + n_1 R
\end{eqnarray}

\paragraph{The mixing timescale:}
Let's assume box~1 is full, and box~2 initially has none of those molecules. How long would it take to empty box~1? From Eq.~(\ref{dn1}), with $n_2\to 0$, we find $n_1(t) = n_1(0) \exp(-t/\tau_{\rm mix})$ where
\begin{equation}  
  \tau_{\rm mix} = \frac{1}{R} = \frac{\Delta z}{v}
  \label{tmix1}
\end{equation}
The same result is obtained when considering $dN_2$ in Eq.\,(\ref{eq:dN1}) for the right box 
\begin{equation}
  \frac{dn_2}{dt} = \frac{n_1-n_2}{\tau_{\rm mix}}
  \label{ansatz}
\end{equation}
where now index 1 refers to the ``full'' box, which ultimately provides the supply of fresh condensible material at some distance. In fact, solving the mixing ansatz Eq.\,(\ref{ansatz}) for the mixing timescale results in
\begin{equation}
  {\tau_{\rm mix}} = \frac{n_1-n_2}{\frac{dn_2}{dt}}
                  = \frac{n_1-n_2}{-n_2 R + n_1 R}
                  = \frac{1}{R}
\end{equation}
for any $n_1$ and $n_2$.

\subsection{A linear chain of boxes}

\noindent Let us now repeat the same thought experiment for 3 boxes in
a row as sketched in Fig.~\ref{3boxes}. The rate equations in this case, with
$R_1=v_1/\Delta z$ and $R_2=v_2/\Delta z$ are
\begin{eqnarray}
  \frac{dn_1}{dt} &=& -n_1 R_1 + n_2 R_1 \\
  \frac{dn_2}{dt} &=&  n_1 R_1 - n_2 (R_1+R_2) + n_3 R_2\label{dn2}\\ 
  \frac{dn_3}{dt} &=& -n_3 R_2 + n_2 R_2
\end{eqnarray}
Closer inspection of Eq.~(\ref{dn2}) shows the analogy to Fick's laws
\begin{eqnarray}
  \frac{dn_2}{dt} &=& \frac{n_1-n_2}{\Delta z}\,v_1 
                    + \frac{n_3-n_2}{\Delta z}\,v_2\\
  &\to& \frac{\partial}{\partial z}
      \left(\frac{\partial n}{\partial z}\,v\right)\Delta z
  \;=\; \frac{\partial}{\partial z}
      \left(D\,\frac{\partial n}{\partial z}\right) \ ,
\end{eqnarray}
where we find the diffusion constant $\rm[cm^2/s]$ (velocity $\times$
length) to be
\begin{equation}
  D = v\,\Delta z
  \label{D}
\end{equation}
The meaning of $\Delta z$ is a bit special in Eq.~(\ref{D}). The mixing motions typically have a certain intrinsic range $\ell$, before the incoming particles actually have an effect on the concentration in the box.  When $\Delta z\ll\ell$, those particles simply rush through, there is no time to mix with the ambient gas in the box. In the opposite case, when $\Delta z\gg\ell$, those particles only enrich the regions close to the surface $A$, but not in the entire box, the concentration gradient on the box is substantial, and that local enhancement at the surface should actually be taken into account when we determine the flux backwards to the originating cell, which we do
not. Therefore, the only box thickness where the local Eq.~(\ref{D}) actually works fine is when
\begin{equation} 
  \mbox{Diffusion:}\quad\quad \Delta z \approx \ell \ .
\end{equation}
Indeed, when determining diffusion constants, we must always make some assumption about $\ell$, for example that $\ell$ is the mean free path for gas-kinetic diffusion, or $\ell$ is the scale height $H_p$ for convective mixing.

To find the mixing timescale $\tau_{\rm mix}$ for the 3-box experiment, we assume that the concentration $n_2$ in the sandwich box 2 adjusts quickly to $n_1$ and $n_3$. Setting the time derivative in Eq.~(\ref{dn2}) to zero we find
\begin{equation}
  n_2 = \frac{n_1 R_1 + n_3 R_2}{R_1+R_2} 
  \label{n_2}
\end{equation}
Generalising the derivation of $\tau_{\rm mix}$ from the 2-box experiment, 
\begin{equation}
  \tau_{\rm mix}^{-1} = -\frac{1}{n_1}\frac{dn_1}{dt} 
                    = -\frac{1}{n_1}\Big(-n_1 R_1 + n_2 R_1\Big)
\end{equation}
and using Eq.~(\ref{n_2}) we find
\begin{equation}
  \tau_{\rm mix} = \frac{R_1 + R_2}{R_1 R_2}
                  \;\frac{1}{1-n_3/n_1}  \ ,
\end{equation}
which, in the limiting case of $n_3\to 0$, results in
\begin{equation}
  \tau_{\rm mix} \;\to\; \frac{1}{R_1} + \frac{1}{R_2}
  \;=\; \frac{\Delta z}{v_1} + \frac{\Delta z}{v_2} \ .
  \label{tmix2}
\end{equation}
Again, the same result is obtained when considering the right box (index 3) and using Eq.~(\ref{n_2})
\begin{equation}
  {\tau_{\rm mix}} = \frac{n_1-n_3}{\frac{dn_3}{dt}}
                  = \frac{n_1-n_3}{-n_3 R_2 + n_2 R_2}
                  = \frac{\Delta z}{v_1} + \frac{\Delta z}{v_2} \ .
\end{equation}
This thought experiment can be extended to a linear chain of boxes of arbitrary length $K$. For each chain length, we consider the boundary particle densities $n_1$ and $n_K$ to be given and assume that $n_2\,...\,n_{K-1}$ can be calculated in their stationary limits. This is similar to the Maxwell daemon in nucleation theory, who would always collect the large clusters, break them up into monomers, and return them this way back to the gas phase. Here, we need a daemon who makes sure that $n_K$ stays small, and $n_1$ stays large. That daemon would quickly transport the molecules arriving in the right box back to the left box, to create a stationary problem with constant diffusive fluxes through all interface areas. In our case, the daemon is dust formation and settling, causing a stationary situation.

Assuming $n_K \to 0$, the result is
\begin{equation}
  \tau_{\rm mix} \;=\; 
  \frac{1}{R_1} + \frac{1}{R_2} + ... + \frac{1}{R_{K-1}}
  ~=~ 
  \frac{\Delta z}{v_1} + \frac{\Delta z}{v_2} + ... + \frac{\Delta z}{v_{K-1}}
  \label{tmix3}
\end{equation}
The same result is obtained for the mixing timescale of the right box
\begin{equation}
  {\tau_{\rm mix}} = \frac{n_1-n_K}{\frac{dn_K}{dt}}
   = \ldots 
   = \frac{\Delta z}{v_1} + \frac{\Delta z}{v_2} + 
     ... + \frac{\Delta z}{v_{K-1}} \ .
\end{equation}
In the limiting case $\Delta z\to 0$, the final result is
\begin{equation}
  \tau_{\rm mix}(z) \;=\, \int_0^z \frac{1}{v(z')}\;dz'
  \label{final}
\end{equation}
which shows that the result is independent of the choice of $\Delta z$ (disregarding here the uncertainties in the actual numerical computation of that integral). To summarise:
\begin{itemize}
\setlength\itemsep{0.2mm}
\item Equation (\ref{final}) states an expression for the replenishment timescale $\tau_{\rm mix}$ in consideration of a distant supply. 
\item The replenishment timescale is monotonic increasing with $z$, i.e.\ it always takes longer to replenish an atmospheric layer which is higher above the ground.
\item There can be a bottleneck. If there is a layer between 0 and $z$ where $v(z')$ is particularly slow, all regions above that layer should indeed receive very little mixing supply.  
\item If $v=\rm const$, Eq.~(\ref{tmix3}) agrees with the 2-box result (Eq.~\ref{tmix1}) and the 3-box result (Eq.~\ref{tmix2}), namely $\tau_{\rm mix}(z)=z/v$ which had been used previously in {\sc StaticWeather} (case $\beta=1$).
\end{itemize}

\section{Miscellaneous Figures}
\begin{figure*}
    \centering
    \includegraphics[width=21pc]{figures/OpticalDepthPlots/OpticalDepthFstar_Multirun_JWST_surf.pdf}
    \includegraphics[width=21pc]{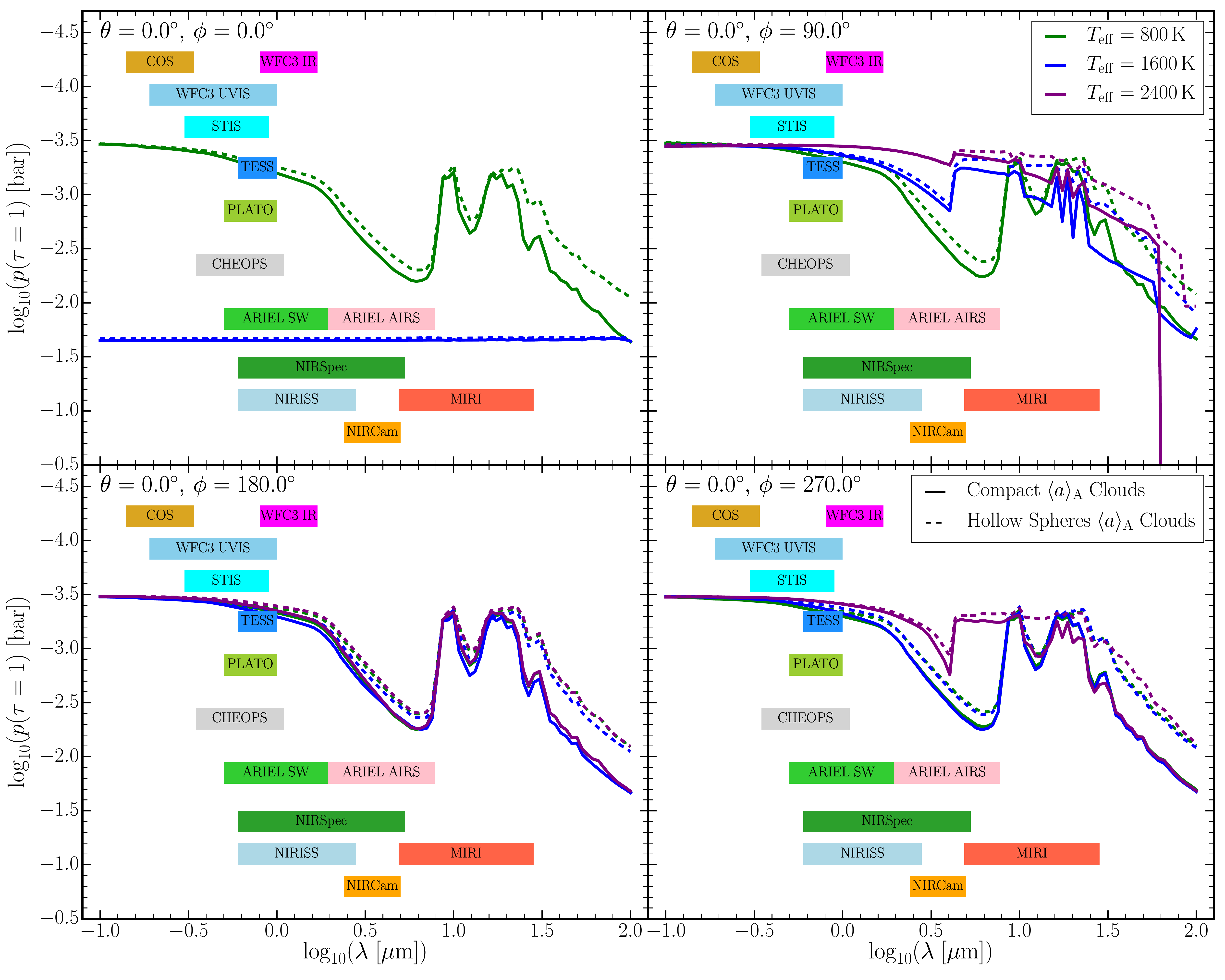}
    \includegraphics[width=21pc]{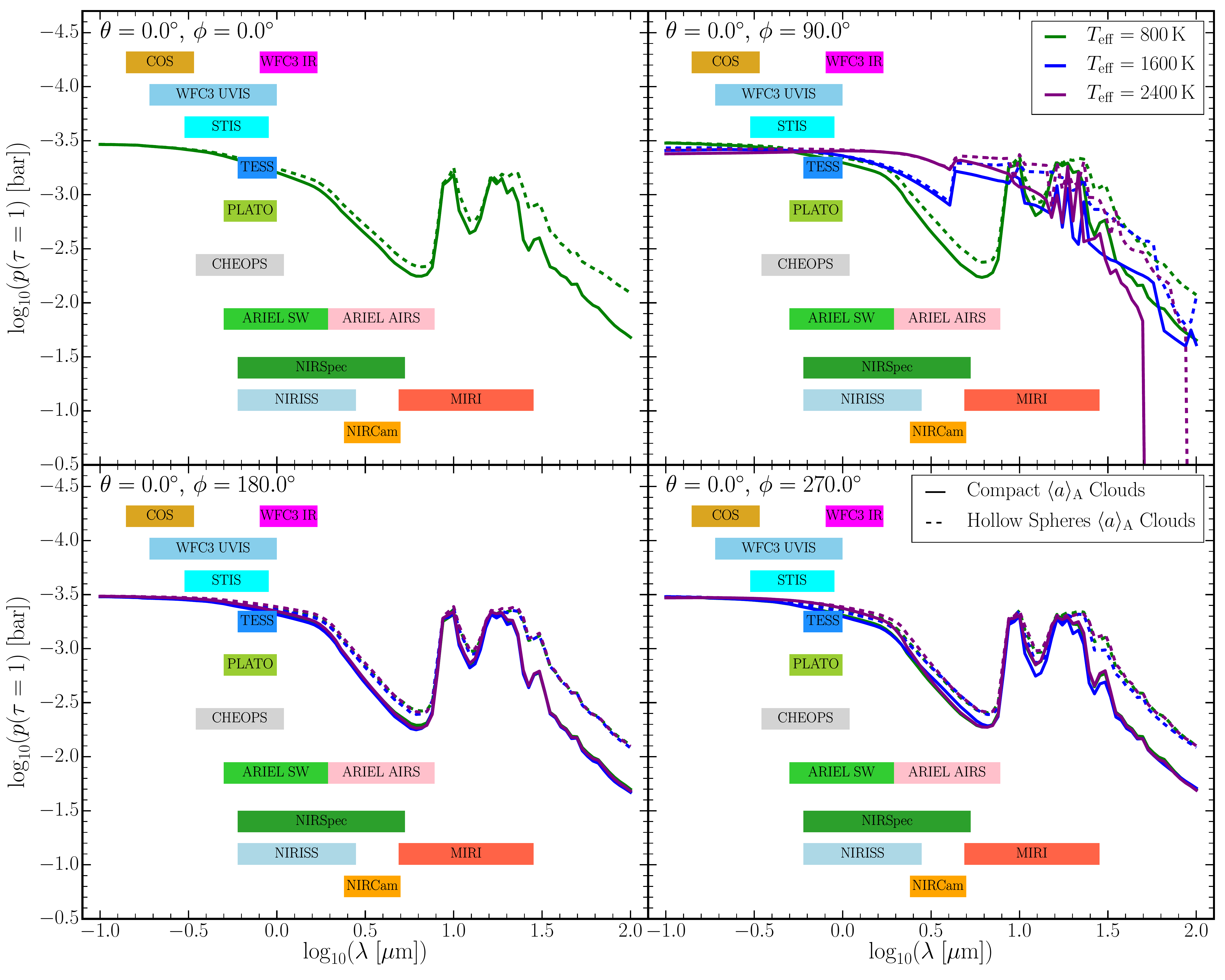}
    \includegraphics[width=21pc]{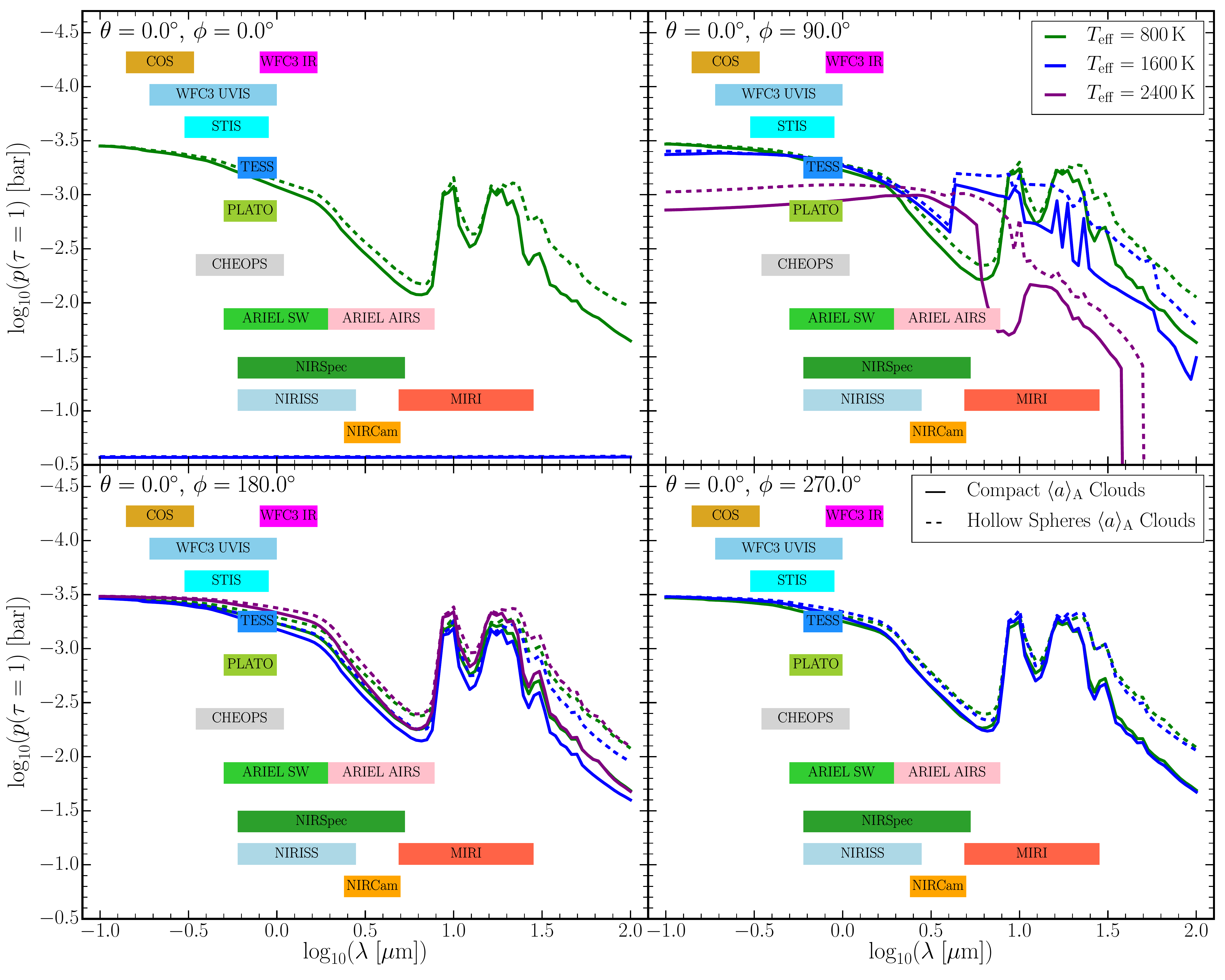}
    \caption{Wavelength-dependent pressure level at which the optical depth due to cloud particles becomes unity, $\tau=1 $, i.e $p_{\rm gas} = p({  \tau_{\lambda} = 1})$, for a grid model planet with \textbf{Upper Left:} an F-type host star, \textbf{Upper Right:} a G-type host star, \textbf{Lower Left:} a K-type host star, and \textbf{Lower Right:} an M-type host star, each at T$_{\rm eff, p}=800, 1600$ and 2400 K, $\log(g) = 3$ [cgs]}
    \label{fig:Opt_Depth_for_multiple_temps}
\end{figure*}

We provide the optical depth plots for all host star classes for completeness in Figure~\ref{fig:Opt_Depth_for_multiple_temps}. 
Tables~\ref{t:UV1} and ~\ref{t:UV2} list a selection of potentially favourable targets for a UV mission. The targets are selected based on  being gas giant exoplanets with a host star effective temperature close to or hotter than that of the sun. The grid models of this work with host stars of F5 (T$_{\rm eff}$~=~6500~K) or G5 (T$_{\rm eff}$~=~5650~K) type are thus most applicable to such potential future UV missions.

\begin{table*}[]
\tiny  
\caption{Physical and orbital parameters of potentially favourable exoplanet targets for a UV mission. The density, surface gravity and $\log_{10}(g)$ were calculated (alongside their respective errors) for each planet based on the mass and radius. All planets marked with * have Msin(i) 
instead of M. Planets marked with ** had their T$_{\rm eff, P}$ calculated from Equ.~12 from \citet{2021MNRAS.tmp.1277B}, with $A_B$=0 and $f$=2. Table~\ref{t:UV2} list the corresponding host star parameters.}
\renewcommand{\arraystretch}{1.5}
\begin{tabular}{llllllll} 
\hline
Planet & a [AU] & P [days] & T$_{\rm eff, P}$ [K] & M$_{\rm P}$ [M$_{\rm Jup}$] & R$_{\rm P}$ [R$_{\rm Jup}$] & $\rho_{\rm P, bulk}$ [g cm$^{-3}$] & log$_{10}$(g) [cm s$^{-2}$] \\ 
\hline
ups And b & 0.05922166±0.00000020 & 4.617033±0.000023 & 1837$^{+46}_{-63}$ ** & 0.6876±0.0044* &  &  &  \\
tau Boo A b & 0.049±0.003 & 3.3124568±0.0000069 & 1997$^{+186}_{-189}$ ** & 4.32±0.04* &  &  &  \\
61 Vir b & 0.050201±0.000005 & 4.2150±0.0006 & 1397±44 ** & 0.016±0.002* &  &  &  \\
51 Peg b & 0.0527±0.0030 & 4.230785±0.000036 & 1557$^{+149}_{-213}$ ** & 0.472±0.039* &  &  &  \\
HD 179949 b & 0.0443±0.0026 & 3.092514±0.000032 &  & 0.916±0.076* &  &  &  \\
HD 75289 b & 0.050±0.000* & 3.509270±0.000064 & 1260 & 0.49±0.03* &  &  &  \\
KELT-20 b & 0.057±0.006 & 3.474119$^{+0.000005}_{-0.000006}$ & 2260±50 & 17 & 1.83±0.07 &  &  \\
HD 209458 b & 0.04707$^{+0.00045}_{-0.00047}$ & 3.52474859±0.00000038 & 1484±18 & 0.682$^{+0.014}_{-0.015}$ & 1.359$^{+0.016}_{-0.019}$ & 0.3603$^{+0.0104}_{-0.0118}$ & 2.9807$^{+0.0264}_{-0.0296}$ \\
HD 212301 A b & 0.030±0.000 & 2.24571±0.00028 & 2195$^{+217}_{-235}$ ** & 0.51±0.04* & 1.07 &  &  \\
HD 149143 b & 0.0530±0.0029 & 4.07182±0.00001 & 1756±285 ** & 1.33±0.15* &  &  &  \\
HAT-P-7b & 0.03813±0.00036 & 2.204737±0.000017 & 2733±21 & 1.806±0.036 & 1.510±0.020 & 0.6956±0.0211 & 3.3121±0.0274 \\
WASP-18b & 0.02087±0.00068 & 0.9414526$^{+0.0000016}_{-0.0000015}$ & 2413±44 & 10.4 & 1.191±0.038 & 8.163 & 4.279 \\
WASP-103b & 0.01985±0.00021 & 0.9255456±0.0000013 & 2489$^{+66}_{-65}$ & 1.455$^{+0.090}_{-0.091}$ & 1.528$^{+0.073}_{-0.047}$ & 0.5408$^{+0.0559}_{-0.0444}$ & 3.2079$^{+0.0916}_{-0.0762}$ \\
WASP-121b & 0.02544$^{+0.00049}_{-0.00050}$ & 1.27492550$^{+0.00000020}_{-0.00000025}$ & 2720±8 & 1.183$^{+0.064}_{-0.062}$ & 1.865±0.044 & 0.2418$^{+0.0164}_{-0.0161}$ & 2.945$^{+0.0636}_{-0.0621}$
\end{tabular}
{\small  \textbf{References:} \textit{ups And b:} \citet{Curiel2011}, \citet{Stassun2019}, \citet{Fuhrmann1998}; \textit{tau Boo A b:} \citet{Butler1997}, \citet{Stassun2019}, \citet{Borsa2015}; \textit{61 Vir b:} \citet{Vogt2010}; \textit{51 Peg b:} \citet{Butler2006}, \citet{Keenan1989}, \citet{Rosenthal2021};
\textit{HD 179949 b:} \citet{Butler2006}, \citet{Rosenthal2021}; 
\textit{HD 75289 b:} \citet{Stassun2017}, \citet{Udry2000}; \textit{KELT-20 b:} \citet{Talens2018}; \textit{HD 209458 b:} \citet{Bonomo2017}, \citet{Evans2015}, \citet{Stassun2017}; \textit{HD 212301 A b:} \citet{Stassun2017}; \textit{HD 149143 b:} \citet{Ment2018}; 
\textit{HAT-P-7 b:} \citet{Bonomo2017}, \citet{Stassun2017}, \citet{Berger2018}, \citet{Morris2013}; \textit{WASP-18b:} \citet{Shporer2019}, \citet{Salz2015}, \citet{Southworth2012}; \textit{WASP-103b:} \citet{Gillon14}, \citet{Bonomo2017}, \citet{Southworth2016}; \textit{WASP-121b:} \citet{Delrez2016}, \citet{MikalEvans2019}.}
\label{t:UV1}
\end{table*}

\begin{table*}[]
\tiny  
\caption{Physical parameters of the exoplanet host stars of potentially favourable targets for a UV mission.}
\renewcommand{\arraystretch}{1.5}
\begin{tabular}{lllllll} 
\hline
Star & T$_{\rm eff}$ [K] & M$_*$ [M$_{\odot}$] & R$_*$ [R$_{\odot}$] & Spectral Type & {[}Fe/H] & Planet \\ 
\hline
HD 9826 & 6105.510$^{+127.253}_{-151.085}$ & 1.150000$^{+0.164999}_{-0.144399}$ & 1.6364900$^{+0.1059680}_{-0.0580015}$ & F8 V & 0.09±0.06 & ups And b \\
HD 120136 & 6466.2700$^{+115.2650}_{-96.8038}$ & 1.320000$^{+0.243739}_{-0.184934}$ & 1.4258800$^{+0.0642849}_{-0.0504688}$ & F7 V & 0.2642300±0.0199902 & tau Boo A b \\
HD 115617 & 5577±33 & 0.942 +0.034-0.029 & 0.963±0.011 & G5 V & -0.01 & 61 Vir b \\
HD 217014 & 5758.000$^{+101.623}_{-119.624}$ & 1.0300000$^{+0.1666990}_{-0.0854185}$ & 1.1756100$^{+0.0673608}_{-0.0353276}$ & G2IV & 0.2057±0.0598 & 51 Peg b \\
HD 179949 & 6168 & 1.21 & 1.2202±0.0375 & F8 V & 0.137 & HD 179949 b \\
HD 75289 & 6117±16 & 1.29±0.10 & 1.23±0.02 & G0 & 0.26 & HD 75289 b \\
HD 185603 & 8980$^{+90}_{-130}$ & 1.89$^{+0.06}_{-0.05}$ & 1.60±0.06 & A2 V & -0.02±0.07 & KELT-20 b \\
HD 209458 & 6065±50 & 1.119±0.033 & 1.155$^{+0.014}_{-0.016}$ & G0 V & 0.01 & HD 209458 b \\
HD 212301 A & 6239±24 & 1.55±0.16 & 1.16±0.02 & F8V & 0.18 & 
HD 212301 A b \\
HD 149143 & 5856 & 1.20±0.20 & 1.44±0.08 & G0 & 0.29 & HD 149143 b \\
HAT-P-7 & 6449±129 & 1.510$^{+0.040}_{-0.050}$ & 1.991$^{+0.084}_{-0.080}$ & F8 & 0.260±0.080 & HAT-P-7b \\
WASP-18 & 6431±48 & 1.46±0.29 & 1.26±0.04 & F6 IV-V & 0.11±0.08 & WASP-18b \\
WASP-103 & 6110±160 & 1.220$^{+0.039}_{-0.036}$ & 1.436$^{+0.052}_{-0.031}$ & F8 V & 4.22$^{+0.12}_{-0.05}$ & WASP-103b \\
WASP-121 & 6459±140 & 1.353$^{+0.080}_{-0.079}$ & 1.458±0.030 & F6 V & 0.13±0.09 & WASP-121b
\end{tabular}
\newline
{\small \textbf{References:} \textit{ups And b:} \citet{Curiel2011}, \citet{Stassun2019}, \citet{Fuhrmann1998}; \textit{tau Boo A b:} \citet{Butler1997}, \citet{Stassun2019}, \citet{Borsa2015}; \textit{61 Vir b:} \citet{Vogt2010}; \textit{51 Peg b:} \citet{Butler2006}, \citet{Keenan1989}, \citet{Rosenthal2021}, \citet{Stassun2019}; 
\textit{HD 179949 b:} \citet{Butler2006}, \citet{Rosenthal2021}; 
\textit{HD 75289 b:} \citet{Stassun2017}, \citet{Udry2000}; \textit{KELT-20 b:} \citet{Talens2018}, \citet{Lund2017}; \textit{HD 209458 b:} \citet{Bonomo2017}, \citet{Evans2015}, \citet{Stassun2017}, \citet{Stassun2019}; \textit{HD 212301 A b:} \citet{Stassun2017}, \citet{Locurto2006}; \textit{HD 149143 b:} \citet{Ment2018}; 
\textit{HAT-P-7 b:} \citet{Bonomo2017}, \citet{Stassun2017}, \citet{Berger2018}, \citet{Morris2013}, \citet{Stassun2019}; \textit{WASP-18b:} \citet{Shporer2019}, \citet{Salz2015}, \citet{Southworth2012}, \citet{Stassun2019}; \textit{WASP-103b:} \citet{Gillon14}, \citet{Bonomo2017}, \citet{Southworth2016}; \textit{WASP-121b:} \citet{Delrez2016}, \citet{MikalEvans2019}, \citet{Stassun2019}.}
\label{t:UV2}
\end{table*}


\end{appendix}

\end{document}